\documentclass[journal, draftcls, onecolumn]{IEEEtran}

\usepackage[pdftex]{graphicx}
\usepackage[cmex10, tbtags]{amsmath}
\usepackage{hhline}
\usepackage[table]{xcolor}
\usepackage{amssymb}
\usepackage[utf8]{inputenc}
\usepackage[english]{babel}
\usepackage{url}
\usepackage[noadjust]{cite}
\usepackage[normalem]{ulem}
\usepackage{epsfig,epsf,subfigure,graphicx,graphics}
\usepackage{tikz}
\usepackage{multirow}
\usepackage{cmap}\usepackage[T1]{fontenc}
\usepackage{enumerate}
\usepackage{xcolor}
\usepackage{soul}

\soulregister\cite7
\soulregister\ref7

\newtheorem{theorem}{Theorem}
\newtheorem{lemma}{Lemma}
\newtheorem{proposition}{Proposition}
\newtheorem{fact}{Fact}
\newtheorem{remark}{Remark}

\addto\captionsenglish{}
\graphicspath{{./fig/}}

\newcommand{\abs}[1]{\left| {#1} \right|}
\newcommand{\ET}[3]{\left\langle {#1}, {#2}, {#3} \right\rangle}
\newcommand{\ETS}[3]{\left\langle {#1}, {#2}, {#3} \right\rangle_{\mathcal P_S}}
\newcommand{\mpelabel}[1]{\stepcounter{equation} \tag{\theequation} \label{#1}}
\renewcommand\hl[1]{#1}

\allowdisplaybreaks

\begin{document}
\title{Degrees of Freedom of the Bursty MIMO X Channel without Feedback}

\author{Shih-Yi~Yeh and I-Hsiang~Wang% <-this % stops a space
\thanks{The material in this paper was presented in part at the IEEE International Symposium on Information Theory, Barcelona, Spain, July 2016.}
\thanks{S.-Y. Yeh is with the Graduate Institute of Communication Engineering, 
National Taiwan University, Taipei 10617, Taiwan (email: steven0416@gmail.com).}% <-this % stops a space
\thanks{I.-H. Wang is with the Department of Electrical Engineering, 
National Taiwan University, Taipei 10617, Taiwan (email: ihwang@ntu.edu.tw).}}% <-this % stops a space

\maketitle

% As a general rule, do not put math, special symbols or citations in the abstract or keywords.
\begin{abstract}
We study the sum degrees of freedom (DoF) of the bursty MIMO X channel without feedback, where the four transmitter-receiver links are intermittently on-and-off, controlled by four Bernoulli random sequences which may be arbitrarily correlated, subject to a symmetry assumption:  The two direct-links have the same level of burstiness, modeled by $\mathrm{Ber}(p_d)$, and so do the cross-links, modeled by $\mathrm{Ber}(p_c)$.  The sum DoF is fully characterized in the regime where $\frac{p_c}{p_d}$ is small, i.e. below a certain threshold, and is partially characterized in the other regime where $\frac{p_c}{p_d}$ is above the threshold.  The achievability is proved with a combination of Han-Kobayashi strategy and interference alignment, which can achieve strictly higher DoF than interference alignment alone.  The converse proof employs a channel-state-sequence pairing technique.  \hl{We highlight that burstiness of the channel disrupts the network topology, turning the MIMO X channel into a network with time-varying topology.  This fundamental difference has striking ramifications.}  In particular, various interference alignment schemes that achieve the DoF of non-bursty MIMO X channels become suboptimal on the bursty channels.  The reciprocity between the forward and the reverse links is lost, and the sum DoF does not saturate when the ratio of the transmitter and the receiver antennas exceeds $\frac{2}{3}$.
\end{abstract}

%\begin{IEEEkeywords}
%Degrees of freedom, MIMO X channels, interference alignment, Han-Kobayashi scheme, interference channels, $<$TBD$>$.
%\end{IEEEkeywords}

% \fi
\IEEEpeerreviewmaketitle

\section{Introduction} \label{Intro}
%auto-ignore

\newcommand{\cy}{\cellcolor{yellow!40}}
\newcommand{\cb}{\cellcolor{blue!20}}

\begin{figure}[h]
\centering
\includegraphics[width=12cm]{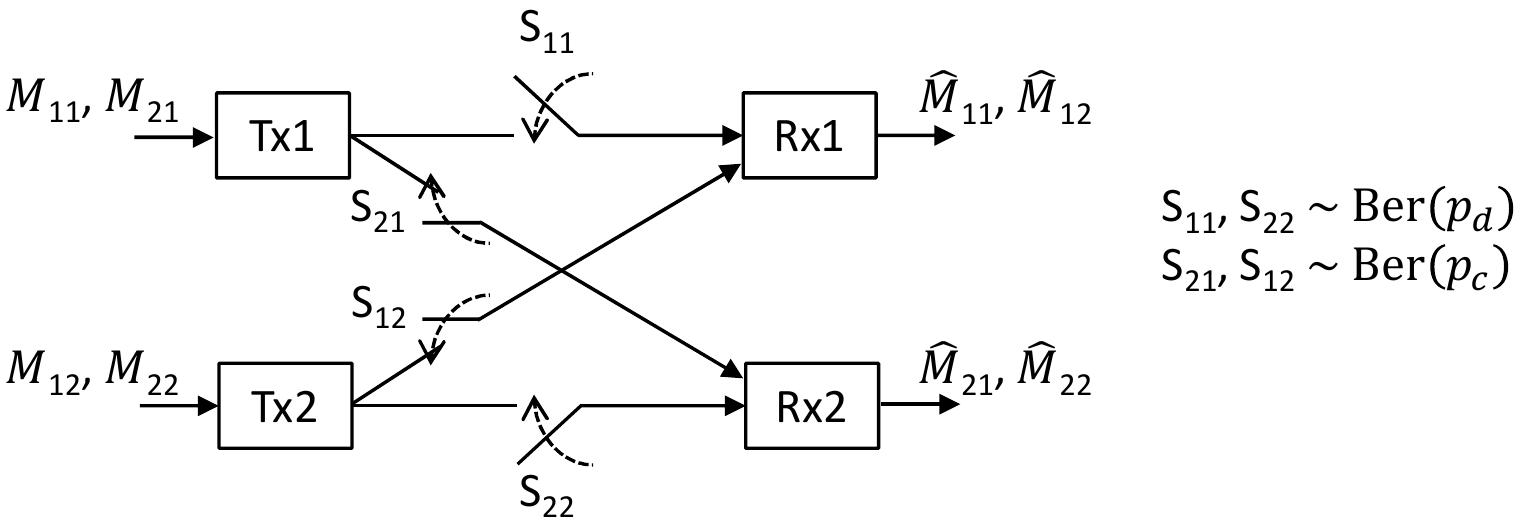}
\caption{Bursty MIMO X Channel ($M_{ji}$: message transmitted from Tx$i$ to Rx$j$, $\hat M_{ji}$: decoded message)}
\label{fig:Bursty_XC}
\end{figure}

With the rapid deployment of wireless networks and the ever-increasing number of devices connected to them, interference becomes more and more prevalent and is often one of the key factors limiting the capacity of a network.  For example, the sum degrees of freedom (DoF) of a two-user-pair MIMO interference channel (IC) with $M$ antennas at all nodes is only $M$ \cite{JF07}, instead of $2M$ as it would be without interference.  Allowing cross-link messaging (i.e. Tx1 to Rx2 and Tx2 to Rx1 as in Fig.~\ref{fig:Bursty_XC}) in such an interference network as proposed in \cite{MMK06} may serve as a simple remedy.  It boosts the sum DoF from $M$ to $4M/3$, and quite remarkably this gain can be realized with simple interference alignment (IA) schemes.  The two-user-pair IC with cross-link messaging allowed is coined X channel (XC) in the literature, and its DoF is characterized in \cite{JS08}--\hspace{1sp}\cite{MOMK14}.

However, the DoF characterization of the MIMO X channel above assumes that the four transmitter-receiver (Tx-Rx) links are always present throughout the transmission, which neglects some phenomena that may be important in real scenarios.   For example, fading is a critical issue for wireless networks and will result in time-varying channel matrices for the four Tx-Rx links.  As a first step toward understanding \hl{how shadowing impacts the channel capacity, we may model it with four randomly on-and-off Tx-Rx links.}  This may also serve as a simple abstraction of the co-channel interference effects when there are other wireless networks operated in the vicinity (\hspace{1sp}\cite{VMA14a},\cite{VMA17}), which can be more commonplace in the era of internet of things (IoT).  Machine-type communications envisioned for the IoT era can also lead to short bursts of interruption to the regular MIMO X channel traffic, and their effects can be captured with four intermittent Tx-Rx links too.  Lastly, if frequency hopping is allowed, the two pairs of users may hop to two different frequency bands and enjoy higher sum rate, even without multi-carrier capability in the physical layer.  In this case, the cross-links may be intermittently off.

This motivates us to investigate the fundamental limit of the \emph{bursty} MIMO X channel, where each of the four Tx-Rx links can be intermittently on-and-off as illustrated in Fig.~\ref{fig:Bursty_XC} ($S_{ij}=1$ : the link is on, $i, j\in \{1,2\}.$)  The burstiness of the four Tx-Rx links is governed by four Bernoulli random sequences which can be arbitrarily correlated at each time instant, subject to the following channel symmetry assumption:  The direct-links (i.e. Tx1 to Rx1 and Tx2 to Rx2) have the same level of burstiness, modeled by $\mathrm{Ber}(p_d)$, and so do the cross-links, modeled by $\mathrm{Ber}(p_c)$, and the direct-links also have the same level of \emph{conditional burstiness}, i.e. $\mathcal P(S_{11}=1|S_{12}=1)=\mathcal P(S_{22}=1|S_{21}=1)=p_{d|c}$.  As we will see in later sections, this symmetry simplifies the analysis of the DoF.  Note that this model is very similar to the one in \cite{VMA17}, except that correlation among the four Bernoulli random sequences are allowed and it is applied to the MIMO X channel where each transmitter is equipped with $M$ antennas and each receiver with $N$ antennas.  We study the sum DoF of this bursty MIMO X channel without feedback.  (For clarity, we shall refer to the usual MIMO X channel whose links are always present as the \emph{non-bursty} channel.)
\begin{figure}[h]
\centering
\includegraphics[width=14.5cm]{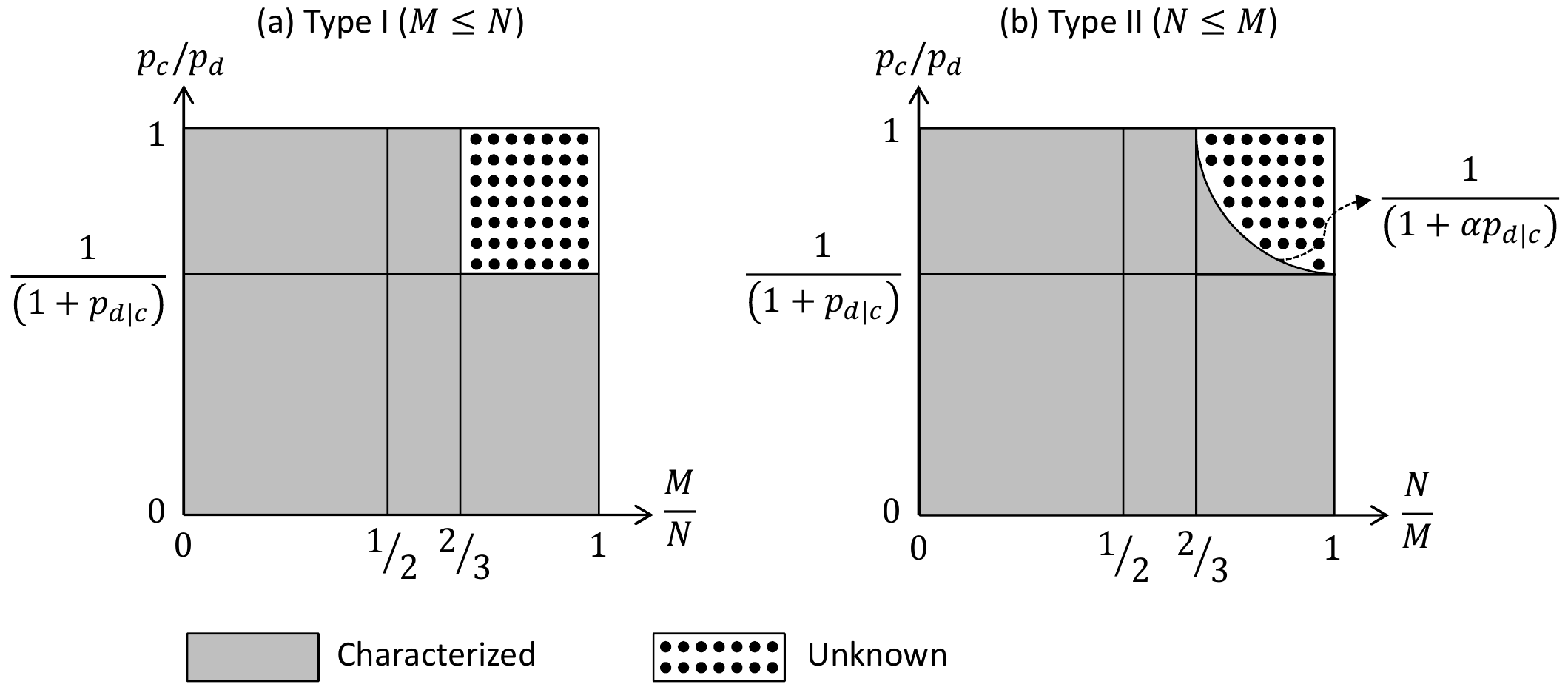}
\caption{Status of sum DoF characterization ($\alpha=\frac{3N-2M}{2N-M}$)}
\label{fig:DoF_Char_Status}
\end{figure}

\begin{figure}[h]
\centering
\begin{tikzpicture}
\node[anchor=south west,inner sep=0] at (0,0) {\includegraphics[width=5.85cm]{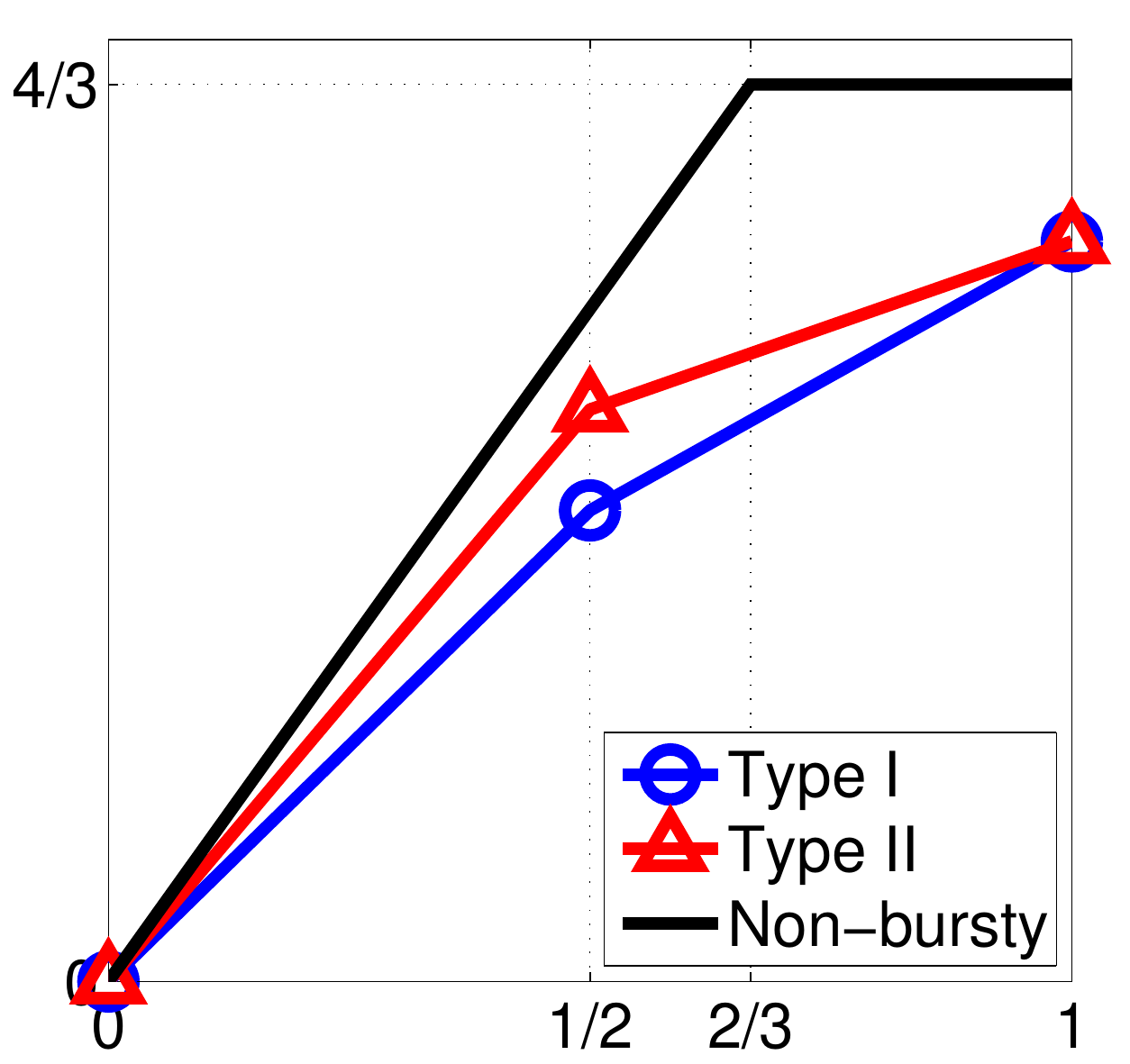}};
\node[anchor=south west,inner sep=0] at (7.5,0) {\includegraphics[width=5.85cm]{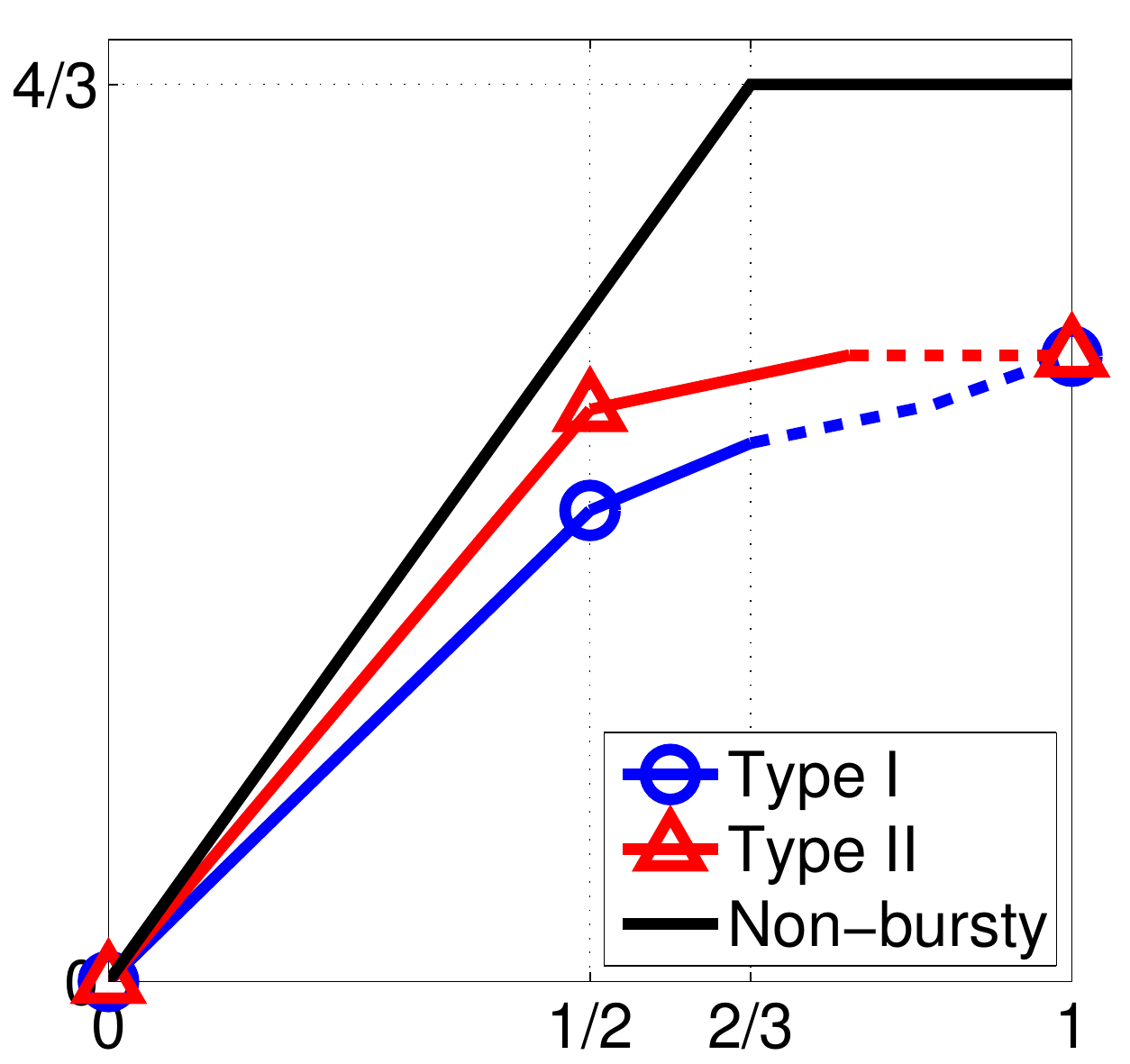}};
\node [right] at (5.6,0.5) {$\frac{\min(M,N)}{\max(M,N)}$};
\node [above] at (0.7,5.3) {$\frac{\eta}{\max(M,N)}$};
\node [above] at (3.1, 5.9) {(a) $p_d=0.7, p_c=0.3, p_{d|c}=0.5$};
\node [right] at (13.1,0.5) {$\frac{\min(M,N)}{\max(M,N)}$};
\node [above] at (8.2,5.3) {$\frac{\eta}{\max(M,N)}$};
\node [above] at (10.6, 5.9) {(b) $p_d=0.7, p_c=0.5, p_{d|c}=0.7$};
\end{tikzpicture}
\caption{Example DoF plots (Dashed lines: lower bounds.)}
\label{fig:DoF_Example_Plots}
\end{figure}
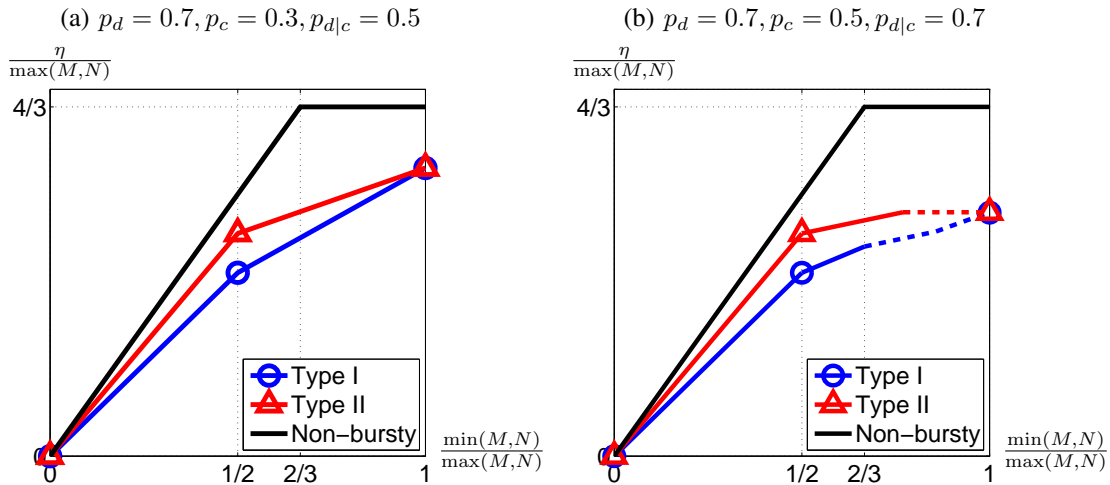

Our main contribution is the characterization of the sum DoF of the bursty MIMO X channel when its $M,N,p_c, p_d$ and $p_{d|c}$ parameters fall in the solid regions of Fig.~\ref{fig:DoF_Char_Status}.  The DoF in the dotted regions remains unknown.  The channels are categorized into two types according to Tx and Rx antenna numbers - \emph{Type I} has $M \le N$ and \emph{Type II} has $N \le M$.  For each channel type, we distinguish two operating regimes, similar to \cite{VMA17}.  The channel is said to be in \emph{Regime 1} when $p_c/p_d \le 1/(1+p_{d\mid c})$, and its sum DoF is fully characterized in this regime.  On the other hand, if $1/(1+p_{d\mid c}) < p_c/p_d \le 1$, then the channel is operated in \emph{Regime 2}, where the sum DoF is only partially known, and as can be seen in Fig.~\ref{fig:DoF_Char_Status}, we know a little more about the DoF of Type II than Type I.  (It will become clear shortly that it suffices to characterize the channel for $p_c \le p_d$ due to channel symmetry.)  The sum DoF $(\eta)$ of the channel in Regime 1 and 2 is illustrated in Fig.~\ref{fig:DoF_Example_Plots}(a) and  \ref{fig:DoF_Example_Plots}(b), respectively.

\hl{We highlight that burstiness of the channel disrupts the network topology, turning the MIMO X channel into a new network with time-varying topology.  This fundamental difference has striking ramifications.}  For example, various IA schemes that achieve the DoF of non-bursty MIMO X channels are found to be sub-optimal when the channels become bursty.  New coding schemes are needed for these channels even when the burstiness of the channel (CSI) is not known to the transmitters.  In particular, we discovered a combination of the Han-Kobayashi (HK) scheme and IA scheme that can outperform IA for the bursty MIMO X channels.  In addition, the non-bursty MIMO X channels are known to have the following properties:
\begin{enumerate}[(i)]  
\item The Type I and the Type II channel are reciprocal, i.e. if we swap $M$ and $N$, the sum DoF is unchanged.
\item For the Type I channel ($M \le N$), the sum DoF saturates at $4N/3$ when $M$ grows beyond $2N/3$.  Similarly the DoF of the Type II channel saturates at $4M/3$ when $N$ reaches $2M/3$.
\item The presence of cross-links does not increase the sum DoF of the channel.  
\end{enumerate}
However, when the channels become bursty, \emph{none} of these are true in general.  The contrasts to (i) and (ii) are evident in Fig.~\ref{fig:DoF_Example_Plots}.  Other contrasts will be shown in later paragraphs.

For the converse proof of the DoF, we extend the state-sequence pairing technique developed in \cite{WSDV13} and derive a DoF upper bound.  However, unlike \cite{WSDV13}, where the channel burstiness is controlled by a single Bernoulli random sequence, our channel now has four Bernoulli random sequences.  So the converse proof becomes much more complicated, as we have 16 channel states at each time instant, as opposed to only two previously.  This difficulty, however, can be overcome with the so-called ``contracted channel" state reduction technique in \cite{VMA17}, which reduces the channel states to only five.  We then use Gaussian distributions to bound a number of differential entropies, and arrive at a DoF upper bound that is tight everywhere in the solid regions of Fig.~\ref{fig:DoF_Char_Status}.

The DoF achievability is proved with schemes that are motivated from the linear IA schemes (\hspace{1sp}\cite{MMK08}, \cite{JS08}) and the HK scheme (\hspace{1sp}\cite{VMA17}, \cite{ETW08}, \cite{HK81}).  The linear IA schemes in \cite{MMK08} can be called \emph{interference-nulling beamforming (INBF)} schemes, as they use beamforming to null out the interference, thereby orthogonalizing the channel. INBF schemes suit our needs particularly well.  However, due to the new channel states induced by the burstiness of the channel, more sophisticated INBF schemes are needed to achieve the DoF in many cases.  Furthermore, for the solid region of Fig.~\ref{fig:DoF_Char_Status}(b) where $1/2 <N/M\le 1$ and $p_c/p_d > 1/(1+p_{d|c})$, the INBF schemes no longer achieve the DoF of the channel.  In this case, superimposing public messages on top of the INBF private messages and decoding them successively similar to the celebrated HK scheme in \cite{ETW08} can yield strictly higher DoF than using the INBF alone, and in fact achieve the DoF of the channel.  For brevity, we name this scheme \emph{HKIA}, since it combines the HK strategy and the IA scheme.  As for the dotted regions of the Fig.~\ref{fig:DoF_Char_Status}, the achievable DoF of our HKIA scheme does not coincide with the upper bound, so the DoF remains unknown there.  However, the HKIA can again achieve higher DoF than the INBF, so the INBF schemes are not DoF-optimal for these channel configurations either.

\subsection*{Related Works}

\hl{In addition to the DoF characterization of the X channel in \cite{MMK06}--{\hspace{1sp}}\cite{MOMK14}, which assume fixed channel matrices throughout the transmission, much work has been done to study the DoF of the time-varying X channel.  Most noticeably this includes a series of works which considers the DoF of the fading X channel and MIMO networks in general with varying level of channel state information at the transmitters (CSIT).  For example, the DoF regions of fading MIMO interference networks without CSIT are reported in \cite{HJSV12}--{\hspace{1sp}}\cite{VV12} and are found to be significantly smaller than those with perfect CSIT.  Instead of not assuming CSIT at all, \cite{GMK11}--{\hspace{1sp}}\cite{KA17} consider the DoF of the X channel with delayed CSIT, following the remarkable findings in \cite{MAT12} that completely stale CSIT is still very useful.  \cite{DJ16} and \cite{TJSP13} look into the interesting DoF problems of MIMO networks with finite-precision CSIT and alternating CSIT, respectively.  Though enlightening, these works on fading channels do not address the possibilities of bursty channels, where each Tx-Rx link may be broken occasionally with non-zero probability.}

\hl{Another line of research sheds light on this and focuses on the burstiness element of the channel (without fading).}  \cite{KPV09} and \cite{MWD13} study the two-user-pair IC where interference (cross-links) may be absent from time to time with a degraded message set formulation.  \cite{WSDV13} and \cite{MWD14} consider the capacity of the bursty IC when delayed feedback was available.  A similar delayed-feedback problem is treated in \cite{VMA14c}, where both direct-links and cross-links can be intermittently on-and-off in a binary fading IC.  \cite{VMA17} then investigates the symmetric binary fading IC with no feedback in the system.  A simplified version of our work presented in \cite{YW16} is the first to consider the DoF of the bursty MIMO X channel without feedback.  It assumes a simple channel model where only the cross-links can be intermittent, controlled by a single Bernoulli random sequence.

\hl{Note also that \cite{SGJ13} and \cite{LKA16} study the capacity (DoF) of the fading interference networks with time-varying topology, which can be viewed as fading \emph{and} bursty channels.  Clearly, our work is most closely related to \cite{MMK08}, \cite{VMA17} and \cite{LKA16}.  We introduce burstiness to the MIMO X channel and evaluate its impacts on the INBF schemes propounded in \cite{MMK08}.  Our work differs from \cite{VMA17} in that we consider MIMO X channel with arbitrarily correlated symmetric burstiness.  Lastly, compare to \cite{LKA16}, which considers the fading SISO X channel with perfect burstiness knowledge at the transmitters but without CSIT, we investigate another extreme --- the non-fading MIMO X channel without burstiness knowledge at the transmitters but with perfect CSIT.}

The rest of this paper is structured as follows.  The formal problem formulation is given in Section~\ref{Prob}.  The main results are summarized in Section~\ref{Main}.  The fundamental differences of the bursty MIMO X channel, compared to the non-bursty channel, are presented in Section~\ref{Fund}.  We then derive the sum DoF upper bound in Section~\ref{Upper}, and show its achievability in Section~\ref{Thm2}.  Section~\ref{Thm3} provides a lower bound of the sum DoF where the DoF remains unknown.  Section~\ref{Disc} addresses some open and subtle issues. Finally, we conclude this paper with the remarks in Section~\ref{Conc}.

\section{Problem Formulation} \label{Prob}
%auto-ignore

\begin{figure}[b]
\centering
\includegraphics[width=9.8cm]{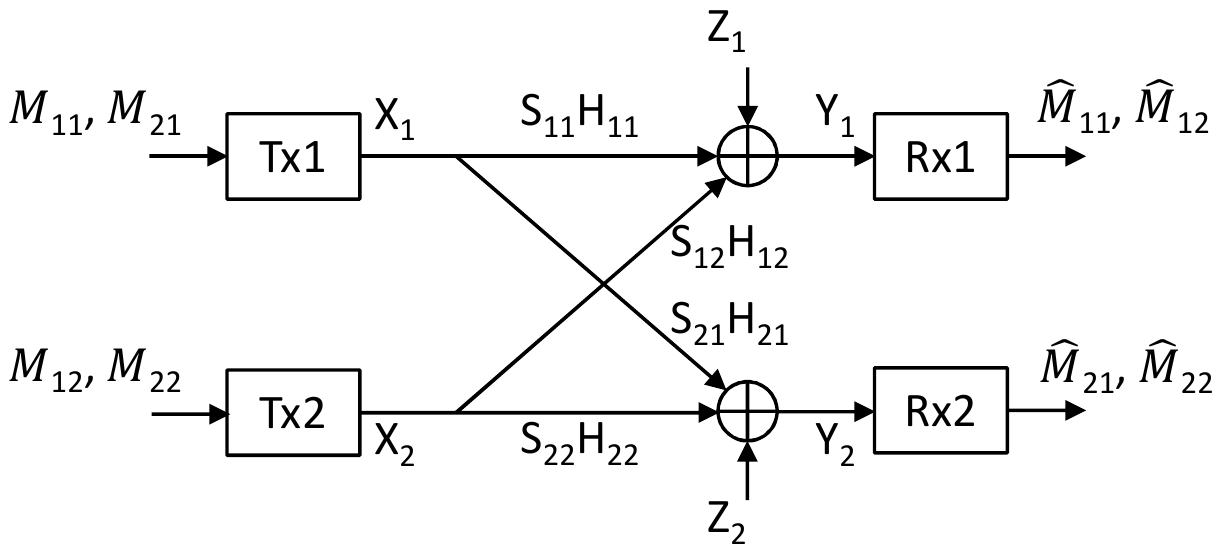}
\caption{Model of the bursty MIMO X channel}
\label{fig:Bursty_XC_Model}
\end{figure}

The detailed system model of the bursty MIMO X channel considered in this work is depicted in Figure~\ref{fig:Bursty_XC_Model}.  There are two transmitters and two receivers in the system, denoted by Tx$i$ and Rx$j$, respectively, for $i, j\in\{1, 2\}$.  Each transmitter is equipped with $M$ antennas, while each receiver has $N$ antennas.  $M_{ji}\sim\mathrm{Unif}\{1,2,\ldots,2^{nR_{ji}}\}$ denotes the message from Tx$i$ to Rx$j$, \hl{encoded over a code block of $n$ symbols with code rate $R_{ji}$}, and $\hat M_{ji}$ is the decoded message at Rx$j$.  $X_i$ represents the signal transmitted by Tx$i$ and $Y_j$ is the received signal at Rx$j$.  Each transmitter has an average transmit power constraint $P$, i.e. $\frac{1}{n}\sum_{k=1}^n{\|X_i[k]\|^2} \le P, i\in \{1, 2\}$, where $X_i[k]$ denotes the $k$-th transmitted symbol of Tx$i$.  $H_{ji}$ models the $N\times M$ channel matrix from Tx$i$ to Rx$j$.  The channel matrices are drawn randomly from a continuous distribution with i.i.d. elements, but are fixed during the transmission.  Each transmitter or receiver is assumed to have perfect knowledge of all channel matrices.  $Z_j$ is the additive Gaussian noise at Rx$j$ with zero mean and unit variance, i.i.d. in time.

The four Tx-Rx links can be intermittently on and off, controlled by four Bernoulli random sequences, $S_{11}[k]$, $S_{12}[k]$, $S_{21}[k]$, and $S_{22}[k],$ which can be arbitrarily correlated at each time instant and are i.i.d. in time.  The link from Tx$i$ to Rx$j$ is on at the $k$-th time instant when $S_{ji}[k]=1,$ and it is off when $S_{ji}[k]=0$.  For brevity of notation, we may drop the dummy time index ($k$) herein and abbreviate $S_{ji}[k]$ as $S_{ji}$ when there is no confusion.  As mentioned earlier, in this paper we consider the symmetric case where $(S_{11}, S_{12})$ has the same distribution as $(S_{21}, S_{22})$, with $S_{11}\sim \mathrm{Ber}(p_d), S_{12}\sim \mathrm{Ber}(p_c), \mathcal P(S_{11}=1|S_{12}=1)=p_{d|c},$ and $\mathcal P(S_{11}=1, S_{12}=1)=p_{cd}=p_c p_{d|c}$.  Each receiver has perfect knowledge of the current burstiness of the two incoming links, e.g. Rx$1$ knows $S_{11}$ and $S_{12}$.  There is no feedback in the system and transmitters do not know any $S_{ij}$ variables.

Let $R_{ji}$ be the code rate from Tx$i$ to Rx$j$, $i, j\in\{1,2\}$.  A rate tuple $(R_{11}, R_{12}, R_{21}, R_{22})$ is said to be achievable on the bursty MIMO X channel if there exists a sequence of codes such that $\mathcal P\{\hat M_{ji} \ne M_{ji}, \textrm{ for some } i, j\in\{1,2\}\}$ converges to zero as the block length of the codes tends to infinity.  The capacity region of the channel is the set of all achievable rate tuples $(R_{11}, R_{12}, R_{21}, R_{22})$\hl{, and the sum capacity of the channel, $C_\mathrm{sum}$, is the supremum of the achievable sum rates $(R_{11} + R_{12}+ R_{21}+ R_{22})$}.  The sum DoF, $\eta$, of the channel follows conventional definition, i.e.
\begin{equation} \label{eq:2_1}
\eta \triangleq \lim_{P\to\infty} \frac{C_\mathrm{sum}}{(\frac{1}{2})\log(P)}.
\end{equation}
Note that in this paper we evaluate the sum DoF of the channel in the almost surely (a.s.) sense, since the channel matrices are drawn from a continuous probability distribution as in \cite{JS08}.

The following notations are used herein to simplify the DoF derivations and discussions.
\begin{enumerate}[(i)]
\item $\langle \eta_{cd}, \eta_{\overline cd}, \eta_{c\overline d} \rangle$:  For the DoF analysis in this paper, it is often natural or convenient to express the DoF as a linear combination of $p_{cd}, p_d-p_{cd},$ and $p_c-p_{cd}$, which are the probabilities of $(S_{ij}, S_{ii})$ being $(1,1), (0,1),$ and $(1,0)$, respectively, $i \ne j$.   So we define $p_{\overline cd}\triangleq p_d-p_{cd}$, and $p_{c \overline d}\triangleq p_c-p_{cd}$, and introduce $\langle \eta_{cd}, \eta_{\overline cd}, \eta_{c\overline d} \rangle$, or the \emph{$\eta$-triple}, to denote $\eta_{cd}p_{cd}+\eta_{\overline cd}p_{\overline cd}+\eta_{c \overline d}p_{c \overline d}$.   For example $\langle2,1,1\rangle=2p_{cd}+1p_{\overline cd}+1p_{c\overline d}$.  Note that the $\eta$-triple is a \emph{scalar} value, not a vector.  When it helps, we may also add subscript $\mathcal P_S$ to the $\eta$-triple (i.e. $\ETS{\eta_{cd}}{\eta_{\overline cd}}{\eta_{c\overline d}}$) to emphasize its dependence on the distribution of the state variables, $\{S_{ij}\mid i, j\in\{1,2\}\}$.
\item $\mathcal D\{x\}$:  Quite often the DoF computation involves entropies and mutual informations, so we define $\mathcal D\{x\} \triangleq \lim_{P\to\infty} \frac{x}{(1/2)\log(P)}$, and use it in places like $\mathcal D\{h(Y_1\mid S_{11},S_{12})\}, \mathcal D\{I(X_1;Y_1\mid S_{11},S_{12})\}$, et al.
\item $X^n$:  We adopt the convention of using $X^n$ to represent a sequence (vector) of random variables, $X[1], X[2], \allowbreak ..., X[n]$, where the number inside $[\ ]$ is the time index.  For instance, $Y_1^n$ denotes the random vector consisting of $n$ received symbols at Rx$1$, i.e. $Y_1[1], Y_1[2], ..., Y_1[n]$.  Similarly, $(S_{11},S_{12})^n$ indicates the following set of derived random variables $\{(S_{11}[k], S_{12}[k])\mid k=1,2,...,n\}$, and $(S_{12}H_{12}X_2+Z_1)^n=\{S_{12}[k]H_{12}X_2[k]+Z_1[k]\mid k=1,2,...,n\}$.  (Recall that the channel matrices are fixed throughout the transmission.) 
\item $M \times N$ channel: We refer to a MIMO channel with $M$ antennas at the transmitter and $N$ antennas at the receiver as an $M \times N$ channel, which can be bursty or non-bursty.
\end{enumerate}

\section{Main Results} \label{Main}
%auto-ignore

We summarize our main results in this section.  First of all, as mentioned in Section~\ref{Intro}, the sum DoF characterization status of the bursty MIMO X channel is shown in Fig.~\ref{fig:DoF_Char_Status}, where the DoF is characterized in the solid regions and remains unknown in the dotted regions.  The sum DoF in the solid regions is established by Theorem~\ref{thm:UB} and \ref{thm:UB_Tightness}.

\begin{theorem} \label{thm:UB}
The sum DoF of the bursty MIMO X channel is upper bounded by $\eta_\mathrm{ub}$ (a.s.), with
\begin{equation*}
\eta_\mathrm{ub} \triangleq \left\{ \begin{aligned}[rl]
2[p_{cd}\min(2M,N)+p_c\min(M,2N)+(p_d-p_c-2p_{cd})\min(M,N)],& \quad \frac{p_c}{p_d} \le \frac{1}{1+p_{d|c}} \\
2[(p_d-p_c)\min(2M,N)+(p_d-p_{cd})\min(M,2N)+(p_{cd}-2p_d+2p_c)\min(M,N)],& \quad \frac{p_c}{p_d} > \frac{1}{1+p_{d|c}}. \\
\end{aligned} \right.
\end{equation*}
\end{theorem}

\begin{theorem} \label{thm:UB_Tightness}
The upper bound, $\eta_\mathrm{ub}$, in Theorem~\ref{thm:UB} is tight except when
\begin{enumerate}[(i)]
\item $2/3 < M/N \le 1 \textrm{ and } \frac{1}{1+p_{d|c}}<\frac{p_c}{p_d} \le 1, \textrm{ or}$
\item $2/3 < N/M \le 1 \textrm{ and } \frac{1}{1+\alpha p_{d|c}}<\frac{p_c}{p_d} \le 1, $
\end{enumerate}
where $\alpha \triangleq \frac{3N-2M}{2N-M}$.
\end{theorem}

Note that $0 < \alpha \le 1$, when $2/3 < N/M \le 1$, so the region where the DoF is unknown is smaller for Type II than for Type I. When $\eta_\mathrm{ub}$ is not tight, the following theorem provides a lower bound of the sum DoF, which can be strictly higher than the DoF achieved by the INBF scheme.  This lower bound is more conveniently expressed in the $\eta$-triple notation.

\begin{theorem} \label{thm:LB}
If $2/3 < M/N \le 1$ and $1/(1+ p_{d|c}) < p_c/p_d \le 1$, then, with $N=3k+q$ where $k$ is an integer and $q\in\{0, 1, 2\}$, the sum DoF is lower bounded by $\eta_\mathrm{lb,1}$, defined as
\begin{equation*}
\eta_\mathrm{lb,1} \triangleq \left\{ \begin{aligned}[rl]
\ET{4k+q}{2k+q}{2M-2k-q},  & \quad \frac{p_c}{p_d} \le  \frac 1{1+\beta_1\:p_{d|c}}\\
\ET{N}{M}{M}+\ET{k}{\left\lfloor \frac{q}{2} \right\rfloor}{-\left\lfloor \frac{q}{2} \right\rfloor}\frac{\ET{2M-N}{-M}{M}}{\ET{2M-N}{-2k-\lceil \frac{q}{2} \rceil}{2k+\lceil \frac{q}{2} \rceil}},  & \quad \frac{p_c}{p_d} > \frac 1{1+\beta_1\:p_{d|c}}, \\
\end{aligned} \right. 
\end{equation*}
where $\beta_1\triangleq \frac{2M-4k-q}{2k+q}.$  On the other hand, if $2/3 < N/M \le 1$ and $1/(1+\alpha p_{d|c}) < p_c/p_d \le 1,$ then, with $M=3k+q$ where $k$ is an integer and $q\in\{0, 1, 2\}$, the sum DoF is lower bounded by $\eta_\mathrm{lb,2}$, defined as
\begin{equation*}
\eta_\mathrm{lb,2} \triangleq \left\{ \begin{aligned}[rl]
M\left( \ET{1}{1}{1}+\ET{1}{0}{0} \frac{\ET{1}{-1}{1}}{\ET{3}{-2}{2}} \right),  & \quad \frac{p_c}{p_d} \le  \frac 1{1+\beta_2\:p_{d|c}}\\
M\ET{1}{1}{1}+k\ET{1}{0}{0},  & \quad \frac{p_c}{p_d} > \frac 1{1+\beta_2\:p_{d|c}}, \\
\end{aligned} \right.
\end{equation*}
where $\beta_2 \triangleq \frac{q}{k+q}$.
\end{theorem}

Sharp contrasts exist between the DoF of the bursty and non-bursty MIMO X channels, including those listed in Table~\ref{tab:Contrasts}.  (For C1, $\max(M,N)$ is assumed to be a multiple of 3 when $\frac{\min(M,N)}{\max(M,N)} > \frac{2}{3}$ and $M\ne N$.)
\vspace{-1ex}
%\begin{table}[htbp]
\begin{table}[h]
\renewcommand{\arraystretch}{1.0}
\centering
\caption{Contrasts between Bursty and Non-bursty MIMO X Channels}
\label{tab:Contrasts}
\vspace{-3ex}
\begin{tabular}{|c|l|c|c|}
	\hline
	  \bfseries Contrast & \bfseries Description & \bfseries Non-bursty & \bfseries Bursty \\ 
	\hline 
	& & & \\ [-3.5ex]
	\hline
	C1 & Does IA always achieve the optimal DoF? & Yes & No \\
	\hline
	C2 & Are Type I and Type II reciprocal? & Yes & No \\
	\hline
	C3 & Does the DoF saturate at $\frac{\min(M,N)}{\max(M,N)}=2/3$? & Yes & No \\
	\hline
	C4 & Can crosslinks increase the DoF? & No & Yes \\
	\hline
\end{tabular}
\end{table}

\section{Fundamental Difference and Its Ramifications} \label{Fund}
%auto-ignore

\hl{A fundamental difference of the bursty MIMO X channel, compared to the non-bursty channel, is that the channel is no longer a MIMO X channel, but in fact a new network with time-varying topology.  This difference has important ramifications, including the sharp contrasts in Table~\ref{tab:Contrasts}, which we will look into in this section to gain insight and intuition.}  We do so with the help of a few concrete and simple channels.  

\subsection{IA Does Not Always Achieve the Optimal DoF} \label{Fund_A}

\begin{figure}[h]
\centering
\includegraphics[width=8.3cm]{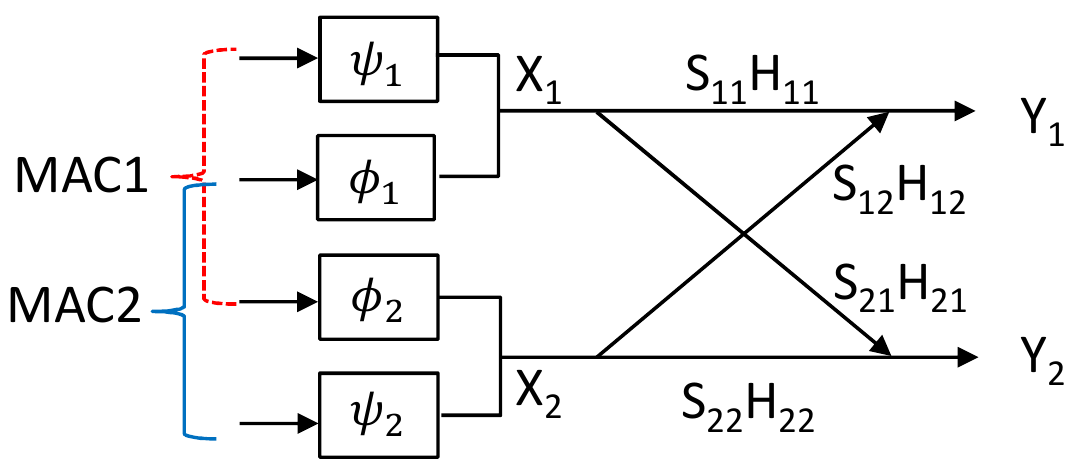}
\caption{The INBF scheme for the $3\times 2$ channel}
\label{fig:INBF3x2}
\end{figure}

\begin{figure}[h]
\centering
\includegraphics[width=9cm]{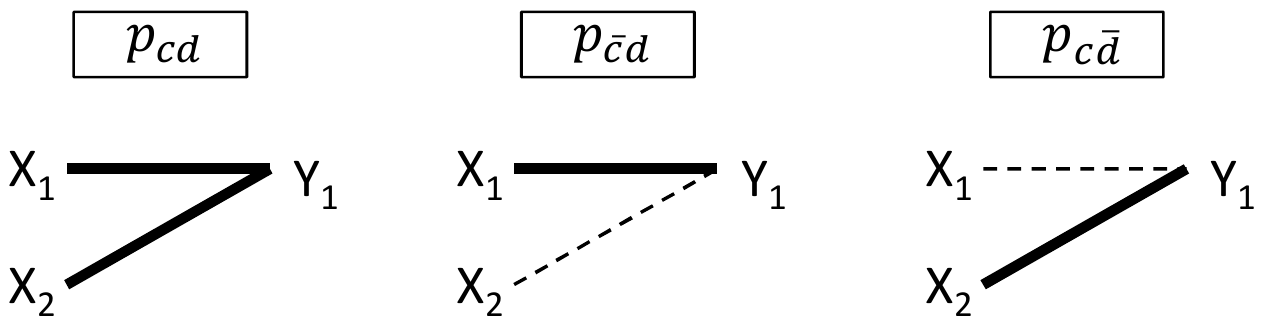}
\caption{The channel states for Rx1 (The states for Rx2 are defined symmetrically.)}
\label{fig:Rx1_State}
\end{figure}

\hl{Interference alignment (IA) is DoF-optimal for the MIMO X channel.  However, burstiness of the channel disrupts the network topology, so it is no longer an X channel.  Instead we now have a network with time-varying topology, which may effectively be an X channel, Z channel, mulitple access channel (MAC), broadcast channel, or point-to-point channel at each time instant.  Intuitively interference alignment need not be optimal in this case.

More precisely, burstiness of the channel creates new channel states which may be underutilized by IA schemes.}  To see this, let us consider the simple $3\times2$ MIMO X channel ($M=3, N=2$).  When this channel is non-bursty, it is well-known that the sum DoF is 4 (a.s.), and can be achieved by the INBF scheme shown in Fig.~\ref{fig:INBF3x2}, which decomposes the X channel into two orthogonal multiple-access channels (MACs) with the four $3 \times 1$ beamforming unit vectors satisfying $H_{21}\psi_1=H_{11}\phi_1=H_{22}\phi_2=H_{12}\psi_2=0$.  When the channel becomes bursty, it is easily verified that the DoF achieved by this scheme is $2(p_c+p_d)$, or $2\langle 2,1,1 \rangle$ (a.s.).  The sub-optimality of this scheme is evident upon an examination of the received signal at Rx1.  (The situation at Rx2 is the same.)  Specifically, consider the channel states for Rx1 illustrated in Fig. \ref{fig:Rx1_State}, where a solid line indicates that the link is on, a dashed line indicates an off link and each state is named by its probability, $p_{cd}, p_{\overline cd},$ or $p_{c\overline d}$.  The dimensions of the received signal at Rx1 in these three channel states are $2, 1,$ and $1$ (a.s.), respectively.  However, we have two antennas at Rx1 and three antennas at each transmitter, so it is clear that we under-utilize the available dimensions of the $p_{\overline cd}$ and $p_{c \overline d}$ channel states.  (Note that we omit the fourth channel state where both links to Rx1 are off and no data is communicated.)

This suggests that by taking advantage of the unused dimensions in the $p_{\overline cd}$ and $p_{c \overline d}$ channel state, we may achieve higher DoF.  One possible way is to introduce public messages.  Let $M_{01}$ and $M_{02}$ be the public messages sent from Tx1 and Tx2, respectively, to be decoded at both receivers, and consider the following random coding scheme:
\begin{equation*}X_1^n(M_{01}, M_{11}, M_{21})=U_1^n(M_{01})+[\psi_1 \: \phi_1]\left[\begin{array}{c}D_1^n(M_{11}) \\
C_1^n(M_{21}) \end{array} \right],
\end{equation*}
where $U_1 \sim \mathcal{N}(0, \frac{P}{6}I_3), C_1 \& D_1 \sim \mathcal{N}(0,  \frac{P}{4}),$ and $[\psi_1 \: \phi_1]\left[\begin{array}{c}D_1^n(M_{11}) \\ C_1^n(M_{21}) \end{array} \right]$ is a shorthand notation for multiplying each of the $n$ components of $\left[\begin{array}{c}D_1^n(M_{11}) \\ C_1^n(M_{21}) \end{array} \right]$ by $[\psi_1 \: \phi_1]$.  $X_2^n(M_{02}, M_{12}, M_{22})$ is encoded symmetrically.  At each receiver, we decode the public messages first, remove them, and then decode the private messages, similar to the HK scheme in \cite{ETW08}.  With this decoding order, clearly the sum DoF achieved by the private messages remains unchanged, i.e. $2\langle 2,1,1 \rangle$.  So the key question is: What DoF is achievable on the public messages?  Since the public messages are decoded at two MACs, it follows that $\eta_{01}$ and $\eta_{02}$ (DoF) satisfying the following inequalities are achievable on $M_{01}$ and $M_{02}$, respectively \hl{$(S\triangleq(S_{11}, S_{12}, S_{21}, S_{22}))$}:
\begin{equation*} \begin{aligned}
\eta_{01} &\le \mathcal D\{\min[I(U_1; Y_1\mid S,U_2), I(U_1; Y_2\mid S,U_2)]\} \\
\eta_{02} &\le \mathcal D\{\min[I(U_2; Y_1\mid S,U_1), I(U_2; Y_2\mid S,U_1)]\} \\
\eta_{01}+\eta_{02} &\le \mathcal D\{\min[I(U_1, U_2; Y_1\mid S), I(U_1, U_2; Y_2\mid S)]\}.
\end{aligned} \end{equation*}

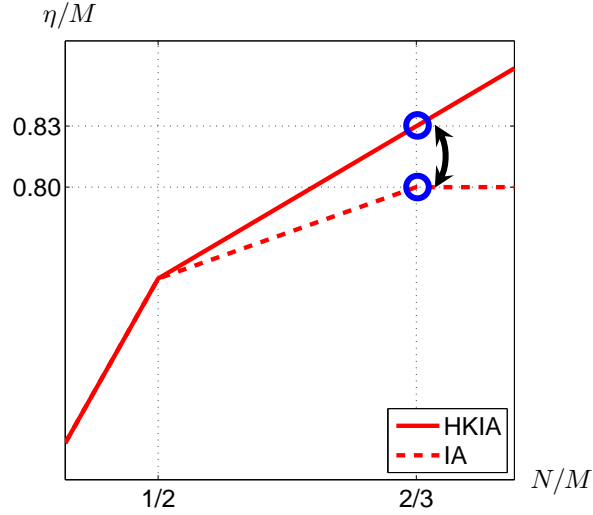
\begin{figure}[h]
\centering
\begin{tikzpicture}
\node[anchor=south west,inner sep=0] at (0,0) {\includegraphics[width=7cm]{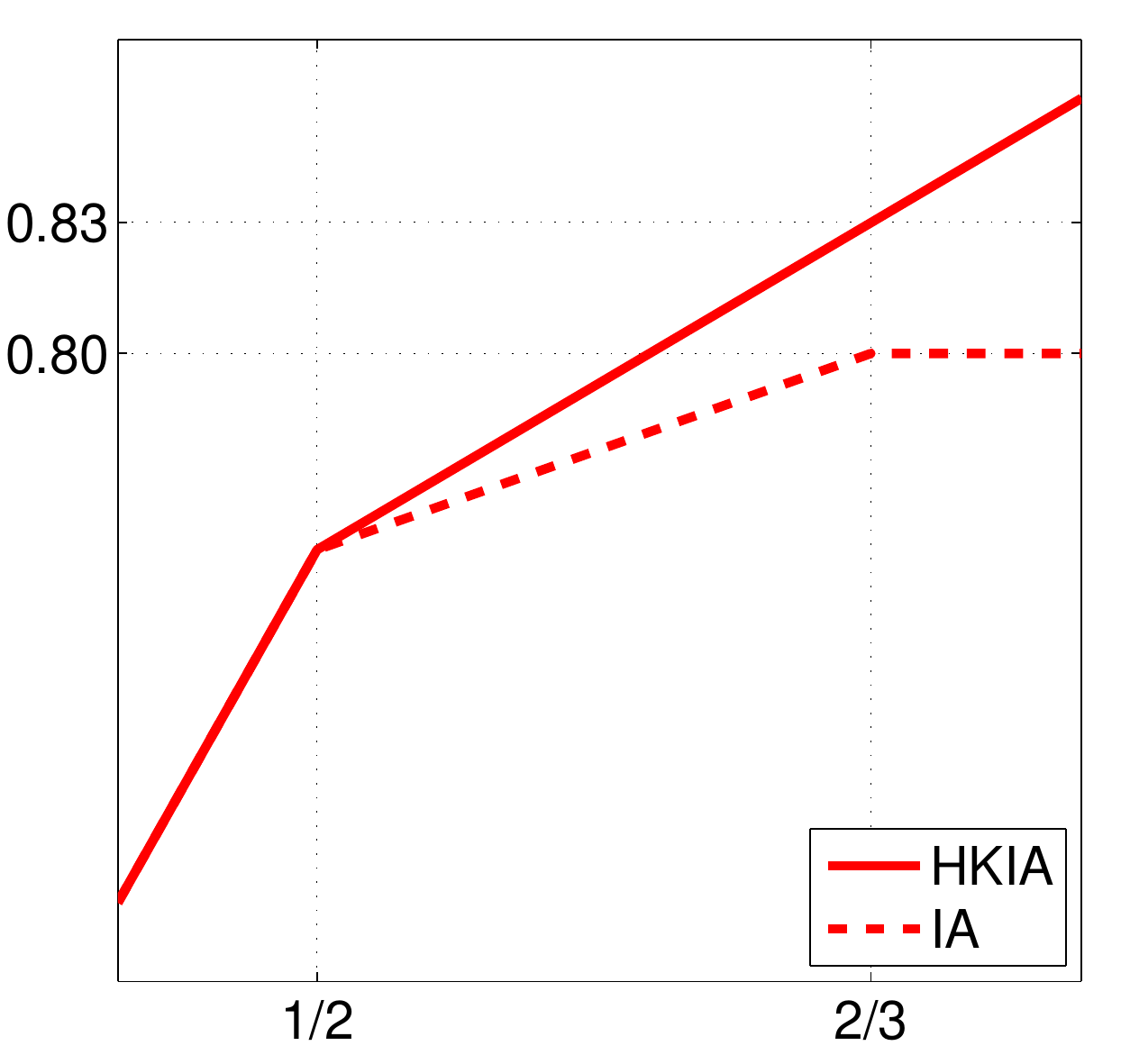}};
\node [right] at (6.8,0.5) {$N/M$};
\node [above] at (0.8,6.4) {$\eta/M$};
\draw [blue, line width=0.8mm] (5.42, 5.23) ellipse (1.5mm and 1.5mm);
\draw [blue, line width=0.8mm] (5.42, 4.41) ellipse (1.5mm and 1.5mm);
\draw [<->,>=stealth, line width=0.8mm] (5.65, 4.40) to [out=60, in=-60] (5.65, 5.24);
\end{tikzpicture}
\caption{DoF achieved by IA and HKIA on the $3\times 2$ bursty channel with $p_d=0.7, p_c=0.5,$ and $p_{d|c}=0.9$ }
\label{fig:3x2_IA_vs_HKIA}
\end{figure}

Following the steps in Section \ref{Thm2}, it can be shown that $\eta_{01}=\eta_{02}=\langle 0,0,1 \rangle$, or $p_c-p_{cd}$, is achievable, which is \emph{non-zero} whenever $p_{cd} < p_c$!  The plausibility of this result can also be seen intuitively as follows.  Recall that the dimensions of the private messages at each receiver are 2, 1, and 1 in the three channel states (Fig. \ref{fig:Rx1_State}), so $\eta_{01}+\eta_{02}$ is upper bounded by $\langle 0,1,1 \rangle$, since there is one free dimension only in the 2nd and 3rd states.  However, each public message also has to be decoded at both receivers.  For $M_{01}$ to be decoded at Rx1, the $\eta_{01}$ cannot be greater than $\langle 0,1,0 \rangle$, because the link from Tx1 to Rx1 exists only in the first two states.  Similarly, for $M_{01}$ to be decoded at Rx2, we need $\eta_{01} \le \langle 0,0,1 \rangle.$  Since $p_c \le p_d$, it follows that $\langle 0,0,1 \rangle \le \langle 0,1,0 \rangle,$ and hence $\eta_{01} \le \langle 0,0,1 \rangle.$ Finally, noting that the DoF of $M_{02}$ is symmetric to that of $M_{01}$, we conclude that $\eta_{01}=\eta_{02}=\langle 0,0,1 \rangle$ is achievable, and the total DoF achievable is $2\langle 2,1,2 \rangle$.  And this is in fact the sum DoF of this channel!  This coding scheme is named HKIA, as it incorporates the ideas of the HK strategy and the IA scheme.  The DoF achieved by the INBF (IA) and the HKIA schemes on this channel ($N/M=2/3$) is illustrated in Fig. \ref{fig:3x2_IA_vs_HKIA}.

So, we have seen that for the $3 \times 2$ bursty MIMO X channel, the earlier INBF scheme no longer achieves the DoF of the channel, because the $p_{\overline cd}$ and $p_{c \overline d}$ states are under-utilized.  The HKIA scheme takes advantage of them by superimposing public messages on the private messages and achieves the DoF of the channel.

\subsection{Type I and II Are Not Reciprocal} \label{Fund_B}

The loss of reciprocity between Type I and Type II bursty channels has been seen in Fig. \ref{fig:DoF_Example_Plots}, and \emph{some} intuition can be gained by considering two simple channels, the $1 \times 3$ and $3 \times 1$ bursty X channels.  The DoF of the bursty $3 \times 1$ channel can be shown to be $2(p_d+p_c-p_{cd})$, or $2\langle 1,1,1 \rangle$, and is achievable by the exactly same INBF scheme as shown in Fig. \ref{fig:INBF3x2}, since each effective MAC clearly achieves a DoF of $p_d+p_c-p_{cd}$.  

\begin{figure}[h]
\centering
\includegraphics[width=7.5cm]{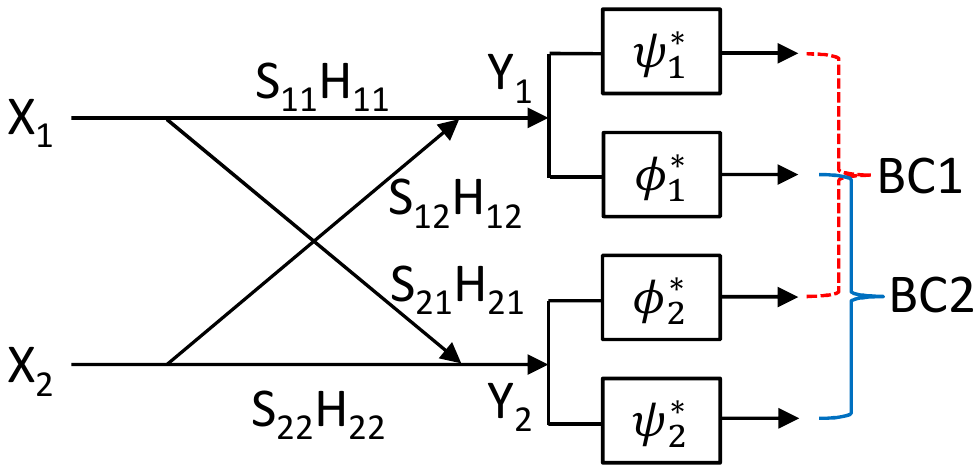}
\caption{The INBF scheme for the $1\times 3$ channel}
\label{fig:INBF1x3}
\end{figure}

However, if we consider the reciprocal INBF scheme for the $1 \times 3$ bursty X channel as depicted in Fig. \ref{fig:INBF1x3}, does it achieve achieve the same DoF?  The answer is \emph{no}, and can be seen as follows.  This INBF scheme decomposes the $1 \times 3$ bursty X channel into two orthogonal broadcast channels (BCs) with the four $3 \times 1$ beamforming vectors satisfying $\psi_1^*H_{12}=\phi_1^*H_{11}=\phi_2^*H_{22}=\psi_2^*H_{21}=0$.
So each effective broadcast channel is a $1\times 1$ broadcast channel, and its DoF is the same as that of a \emph{degraded} $1 \times 1$ broadcast channel, i.e. $p_d$, following similar lines of arguments in \cite{JF07}.  Consequently this INBF scheme achieves only $2p_d$ DoF, which is in fact the DoF of the bursty $1 \times 3$ channel and is strictly less than $2(p_d+p_c-p_{cd})$ whenever $p_c > p_{cd}$.

Therefore in light of the fundamental difference between the bursty BC and MAC, it is understandable that bursty Type I and Type II channels are not reciprocal in general, and a bursty Type II channel can have higher DoF than the corresponding Type I channel.

\subsection{Cross-links Can Increase the DoF}

The bursty $3 \times 1$ channel considered in the previous subsection also sheds light on why the presence of the cross-links can increase the DoF of the MIMO X channel when the channel becomes bursty.  Recall that with the INBF scheme in Fig. \ref{fig:INBF3x2}, the DoF of each MAC is $p_d+p_c-p_{cd}$, or $\langle 1,1,1 \rangle$.  On the other hand, in the absence of the cross-links (i.e. $p_c = 0$), the channel reduces to two orthogonal $3 \times 1$ point-to-point bursty MISO channels, each of which simply has a DoF of $p_d$, or $\langle 1,1,0 \rangle$.  Therefore we see that the cross-links increases the DoF of the bursty $3 \times 1$ X channel because of the presence of the $p_{c \overline d}$ state at each MAC.  Since $p_{c\overline d}=p_c-p_{cd}$ and $p_{cd}=p_c$ when $p_d=1$, this state is suppressed for the non-bursty MIMO X channel, so the presence of the cross-links does not help the DoF.  Only when the channel becomes bursty, can the $p_{c\overline d}$ state emerge and present an opportunity for data transmission, leading to higher DoF.  This explains why cross-links can increase the DoF for the bursty MIMO X channels, but not for the non-bursty channels.

\subsection{DoF Does Not Saturate at $\frac{\min(M,N)}{\max(M,N)} = \frac{2}{3}$}

\begin{figure}[h]
\centering
\begin{tikzpicture}
\node[anchor=south west,inner sep=0] at (0,0) {\includegraphics[width=7cm]{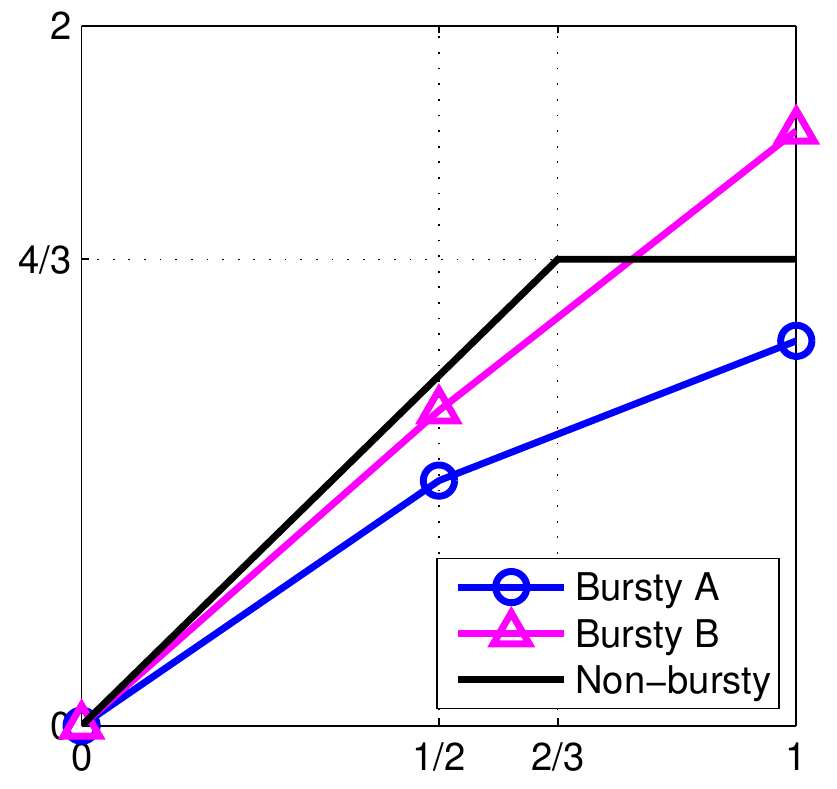}};
\node [right] at (6.9,0.6) {$M/N$};
\node [above] at (0.8,6.55) {$\eta/N$};
\end{tikzpicture}
\caption{DoF of the Type I channel in Regime 1 (Bursty A: $p_d=0.7, p_c=0.3, p_{d|c}=0.5$, Bursty B: $p_d=0.9, p_c=0.1, p_{d|c}=0.5$)}
\label{fig:Type_I_DoF_No_Sat}
\end{figure}

\hl{The key problem limiting the DoF of the non-bursty MIMO X channel is interference.  When the disparity between $M$ and $N$ is great, we have sufficiently large null spaces of the channel matrices where we can beamform and eliminate (hide) interference.  This explains why, for instance, when $\frac{M}{N}\le\frac{2}{3}$, the DoF grows linearly with $M$.  However, when $M$ and $N$ becomes comparable, e.g. $\frac{2}{3}<\frac{M}{N}\le 1$, the dimensions of the null spaces shrink to such a point that interference cannot be hidden in them perfectly any more, so further DoF growth is impeded.

This changes for the bursty X channel, for the network topology is disrupted, and interference is no longer constantly there at each receiver.  For example, when $p_c$ is small and $p_d$ is large, interference is present for only a small fraction of time, or is mostly eliminated ``by nature,'' and we effectively have two orthogonal point-to-point channels most of the time.  This clearly opens up more opportunities for data transmission, so DoF does not saturate as in the non-bursty channel.}  Two bursty Type I DoF curves are illustrated in Fig. \ref{fig:Type_I_DoF_No_Sat}.

\begin{remark}
\hl{
It is clear from the discussions above that the time-varying topology caused by burstiness of the channel is the fundamental reason underlying all of the above contrasts between the bursty and non-bursty X channel.  However, we suspect that the lack of feedback is also key to the loss of reciprocity as discussed in Section \ref{Fund_B}.  This conjecture is based on two facts: (1) The apparent asymmetry of channel topology information at the transmitters and the receivers. (2) The bursty MAC and bursty BC considered in Section \ref{Fund_B} do become reciprocal when the transmitters are aware of the channel topology too.
}
\end{remark}

\section{The Upper Bound} \label{Upper}
%auto-ignore

We now prove the upper bound of the sum DoF of the bursty MIMO X channel given by Theorem \ref{thm:UB}.  The basic idea of the proof is to extend the state-sequence pairing technique in \cite{WSDV13} to the more general bursty MIMO X channel. We do this in three major steps.  First we provide extra side information to the two receivers, so that the entropy terms of the rate bounds at the two receivers can be combined for subsequent simplification.  We then delicately choose a pair of state sequence realizations that can bring the difference of the combined entropy terms to a computable form.  Finally, we obtain the upper bound by resorting to the fact that Gaussian distributions maximize differential entropies and applying the law of large numbers to $\{S_{ij}\mid i, j\in\{1,2\}\}$.  Before we start, however, we first present a few simple lemmas and facts that will prove useful in bounding the sum DoF.

\subsection{Preliminaries} 

The following lemma extends Lemma 2 of \cite{VMA17} to the bursty MIMO X channel, and can be proved along similar lines.  Alternatively a simple proof following the arguments for the broadcast channel \cite{EGK11} is shown below.
\begin{lemma} \label {lemma_1}
The capacity region of the bursty MIMO X channel without feedback depends on distribution of ($S_{11}, S_{12}, S_{21}, S_{22}$) only through the marginal distributions of ($S_{11}, S_{12}$) and ($S_{21}, S_{22}$).
\end{lemma}
\begin{IEEEproof}
Let $P_e\triangleq \mathcal \{M_{ji}\ne \hat M_{ji}, \textrm{for some}\: i, j\in\{1,2\}\}$, and $P_{ej}\triangleq \mathcal \{M_{ji}\ne \hat M_{ji}, \textrm{for some}\: i\in\{1,2\}\}, j=1,2.$  It is easily verified that, for any sequence of coding and decoding schemes, $P_e$ tends to zero if and only if $P_{e1}$ and $P_{e2}$ both tend to zero.  But $P_{e1}$ and $P_{e2}$ depends only on the marginal distribution of $Y_1$ and $Y_2$, respectively, and hence in turn only on the marginal distributions of ($S_{11}, S_{12}$) and ($S_{21}, S_{22}$), respectively, for any given set of channel matrices (c.f. Fig. \ref{fig:Bursty_XC_Model}).  This establishes the lemma.
\end{IEEEproof}

\begin{figure}[h]
\centering
\includegraphics[width=16cm]{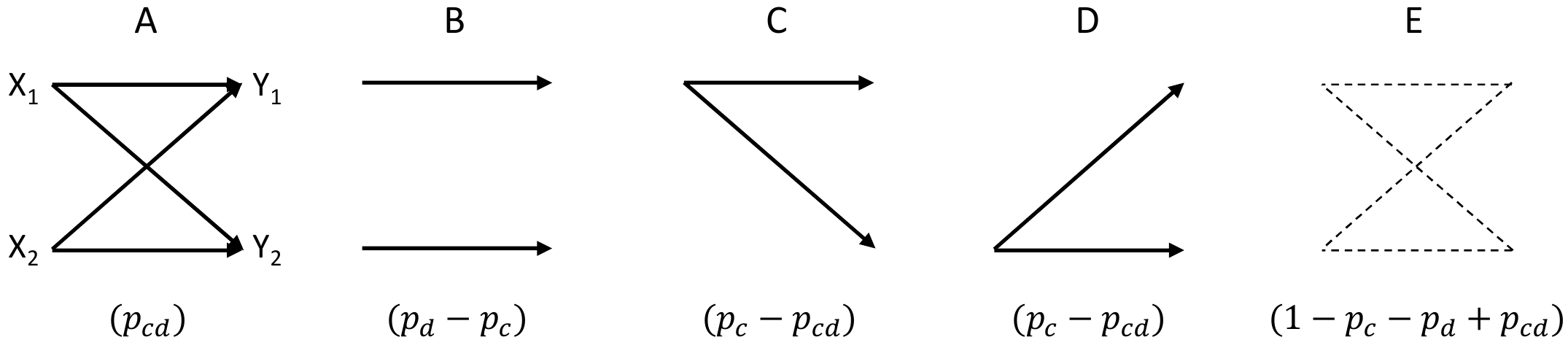}
\caption{The contracted channel}
\label{fig:Contracted_Channel}
\end{figure}

A simple and useful consequence of Lemma \ref{lemma_1} is that it allows us to reduce channel states \cite{VMA17}.  Specifically, for any bursty MIMO X channel with $p_c \le p_d$, it is easily checked that its marginal distributions of ($S_{11}, S_{12}$) and ($S_{21}, S_{22}$) are identical to those of the contracted channel shown in Fig. \ref{fig:Contracted_Channel} with only five channel states, where a solid line indicates that the link is on and the probability of each state is given in the parentheses, which will be denoted as $P_A, P_B, ..., P_E$ hereafter.  \hl{(Recall that the marginal distribution of ($S_{ij}, S_{ii}$) is determined by $p_{cd}, p_{\bar cd},$ and $p_{c \bar d}$ as formulated in Section \ref{Prob}.  Refer to the illustration in Fig. \ref{fig:Rx1_State} too.)}  Therefore, we shall consider this contracted channel, when deriving the upper bound of the sum DoF.

Let $\mathcal C(p_c, p_d, p_{cd})$ denote the capacity region of the bursty MIMO X channel without feedback.  We have the following lemma as a result of the symmetry of the channel.  It follows from this lemma that swapping the values of $p_c$ and $p_d$ does not change the sum capacity of the channel, so it suffices to characterize the sum DoF of the channel for $p_c \le p_d$.

\begin{lemma} \label {lemma_2}
$(R_{11}, R_{12}, R_{21}, R_{22})\in \mathcal C(p_0, p_1, p_2)$ if and only if $(R_{21}, R_{22}, R_{11}, R_{12})\in \mathcal C(p_1, p_0, p_2)$ for any legitimate set of probabilities $p_0, p_1,$ and $p_2$.
\end{lemma}
\begin{IEEEproof}
Due to the channel symmetry, the proof is immediate by swapping/relabeling Rx1 and Rx2. 
\end{IEEEproof}

Let $\mathrm{tr}(A)$ denotes the trace of a matrix $A$.  We collect some well-known or easily-proved facts below, which will facilitate the bounding of the differential entropies.  (i) and (ii) are simple; (iii) and (iv) can be proved with techniques to bound conditional differential entropies of jointly Gaussian random vectors, e.g. in \cite{AV09}.

\begin{fact} \label {fact_dof_ub}
Let $X$ and $Y$ be two (possibly correlated) $M$-dimensional Gaussian random vectors whose correlation matrices are $R_X$ and $R_Y$, respectively, with $\mathrm{tr}(R_X) \le kP$ and $\mathrm{tr}(R_Y) \le kP$ for some positive constant $k$. Let $Z$ and $W$ be two independent $N$-dimensional Gaussian random vectors, each distributed according to $\mathcal N(0, I_N)$, independent of $X$ and $Y$.  Let $A, B$ and $C$ be three $N \times M$ random matrices with i.i.d. elements drawn from a continuous distribution, independent of $X, Y, Z$ and $W$.  Then, with probability 1,
\begin{enumerate}[(i)]
\item $\mathcal D\{h(AX+Z)\} \le \min(M,N),$
\item $\mathcal D\{h(AX+BY+Z)\} \le \min(2M,N),$
\item $\mathcal D\{h(AX+Z\mid CX+W)\} \le \min(M,2N) - \min(M, N), \text{and}$
\item $\mathcal D\{h(AX+BY+Z\mid CY+W)\} \le \min(M, N).$
\end{enumerate}
\end{fact}

\subsection{Providing Extra Side Information to the Receivers}

First of all, letting $S_1 \triangleq (S_{11}, S_{12}), S_2 \triangleq (S_{21}, S_{22}),$ and $S\triangleq (S_1, S_2)$, the sum rate at Rx1 can be bounded as follows:
\begin{align*} 
n&(R_{11} + R_{12} - \epsilon_n) \\
& \leq I(M_{11},M_{12} ; Y_1^n,S_1^n) \\
& \overset{(a)}{\leq} I(M_{11},M_{12} ; Y_1^n,S^n, M_{21})\\
& \overset{(b)}{=} I(M_{11},M_{12} ; Y_1^n \mid S^n, M_{21})\\
& =  h(Y_1^n \mid S^n, M_{21}) - h(Y_1^n \mid S^n, M_{21}, M_{11}, M_{12}) \\
& \overset{(c)}{=} h(Y_1^n \mid S^n, M_{21}) - h((S_{12}H_{12}X_2+Z_1)^n \mid S^n,  M_{12}) \\
& = \sum_{u^n}{\mathcal P\{S^n=u^n\}[h(Y_1^n \mid S^n=u^n, M_{21}) 
	 				- h((S_{12}H_{12}X_2+Z_1)^n \mid S^n=u^n,  M_{12})]} \mpelabel{eq:5_1}
\end{align*}
where side information ($S_2^n, M_{21}$) is provided to Rx1 in (a), independence between $(M_{11}, M_{12})$ and $(S^n, M_{21})$ leads to (b), and (c) follows because $X_1^n$ becomes deterministic when $M_{11}$ and $M_{21}$ are given.

$R_{21} + R_{22}$ is bounded symmetrically, so we also have
\begin{align} \begin{split} \label{eq:5_2}
n&(R_{21} + R_{22} - \epsilon_n) \\
& \leq \sum_{v^n}{\mathcal P\{S^n=v^n\}[h(Y_2^n \mid S^n=v^n, M_{12}) 
					   - h((S_{21}H_{21}X_1+Z_2)^n \mid S^n=v^n,  M_{21})] }.
\end{split} \end{align}

Rearranging the terms, the sum rate is hence bounded from above by
\begin{align} \begin{split} \label{eq:5_3}
n&(R_{11}+R_{12}+R_{21} + R_{22} - \epsilon_n) \\
& \le \left[\sum_{u^n}{\mathcal P\{S^n=u^n\}h(Y_1^n \mid S^n=u^n, M_{21})}
	-\sum_{v^n}{\mathcal P\{S^n=v^n\}h((S_{21}H_{21}X_1+Z_2)^n \mid S^n=v^n,  M_{21})}\right]\\
& \quad + \left[\sum_{v^n}{\mathcal P\{S^n=v^n\}h(Y_2^n \mid S^n=v^n, M_{12})}
	- \sum_{u^n}{\mathcal P\{S^n=u^n\}h((S_{12}H_{12}X_2+Z_1)^n \mid S^n=u^n,  M_{12})}\right].
\end{split} \end{align}
Next we bound the two differences inside the brackets separately on the contracted channel in Fig. \ref{fig:Contracted_Channel}.

\subsection{Pairing State Sequence Realizations}

\begin{figure}[h]
\centering
\begin{tikzpicture}
\node[anchor=south west,inner sep=0] at (0,0) {\includegraphics[width=12cm]{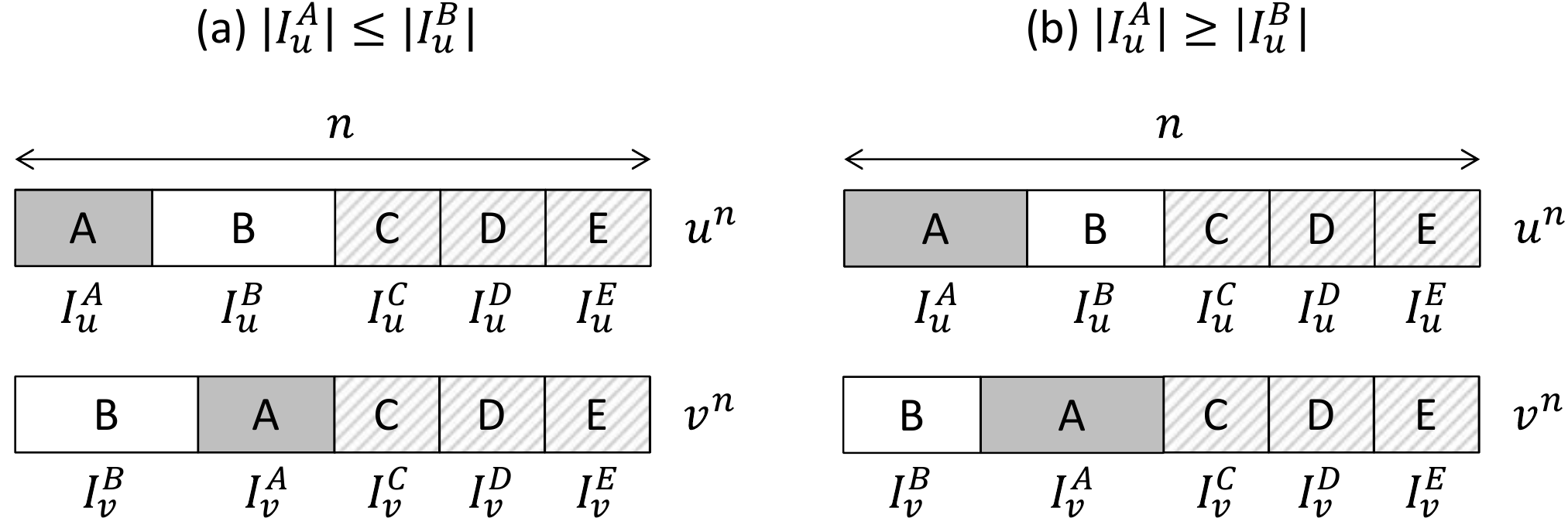}};

\path [fill=white] (0,0.0) rectangle (12,0.5);
\path [fill=white] (0,1.4) rectangle (12,1.9);
\path [fill=white] (1.5,3.5) rectangle (4,4.1);
\path [fill=white] (7.8,3.5) rectangle (10.4,4.1);

\node [right] at (0.6,0.2) {$\mathcal I_v^B \qquad \mathcal I_v^A \quad\:\: \mathcal I_v^C \quad \mathcal I_v^D \quad \mathcal I_v^E$};
\node [right] at (6.75,0.2) {$\mathcal I_v^B \qquad \mathcal I_v^A \qquad \mathcal I_v^C \quad \mathcal I_v^D \quad \mathcal I_v^E$};
\node [right] at (0.4,1.65) {$\mathcal I_u^A \qquad \mathcal I_u^B \qquad \mathcal I_u^C \quad \mathcal I_u^D \quad \mathcal I_u^E$};
\node [right] at (6.95,1.65) {$\mathcal I_u^A \qquad \mathcal I_u^B \quad\:\: \mathcal I_u^C \quad \mathcal I_u^D \quad \mathcal I_u^E$};
\node [right] at (1.2,3.8) {\fontsize{11}{11}\selectfont (a) \normalsize{$\left| \mathcal I_u^A\right| \le \left| \mathcal I_u^B \right|$}};
\node [right] at (7.5,3.8) {\fontsize{11}{11}\selectfont (b) \normalsize{$\left| \mathcal I_u^A\right| \ge \left| \mathcal I_u^B \right|$}};
\end{tikzpicture}
\caption{Pairing of state sequence realizations}
\label{fig:State_Sequence_Pairing}
\end{figure}
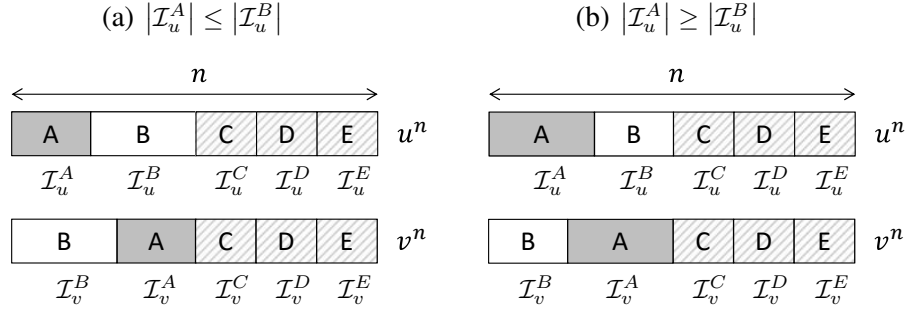

For each difference in (\ref{eq:5_3}), we expand the two summations and pair up $\mathcal P\{S^n=u^n\}h(Y_1^n \mid S^n=u^n, M_{21}) - \mathcal P\{S^n=v^n\}h((S_{21}H_{21}X_1+Z_2)^n \mid S^n=v^n,  M_{21})$ with $u^n$ and $v^n$ chosen as follows:  State C, D and E occur in the same places in $u^n$ and $v^n$, and State A has the same number of occurrences in $u^n$ and $v^n$ but has no or minimum overlap, as illustrated in Fig. \ref{fig:State_Sequence_Pairing}.  Note also that in the figure $\mathcal I_u^A$ and $\mathcal I_v^A$ denote the set of time indexes of State A in $u^n$ and $v^n$, respectively.  Similar notations are used for other channel states.  To keep the expressions in this and next subsection simple, the following notations are also employed:
\begin{enumerate}[(i)]
\item Usual set algebra notations: $|\:|$ : cardinality, $\setminus$ : set minus.  E.g. $\abs{\mathcal I_u^A}$ , $\mathcal I_u^B\setminus \mathcal I_v^A$
\item $\mathcal I_u^{AB} \triangleq \mathcal I_u^A \cup \mathcal I_u^B$ and $\mathcal I_u^{ABC} = \mathcal I_u^A \cup \mathcal I_u^B \cup \mathcal I_u^C,$ and so on.
\item $X^{I_u^A} \triangleq \left\{X[k]\mid k\in \mathcal I_u^A\right\}.$  E.g  $(H_{11}X_1+Z_1)^{\mathcal I_u^A}=\left\{(H_{11}X_1[k]+Z_1[k])\mid k\in \mathcal I_u^A\right\}$
\end{enumerate}

With the above pairing strategy, clearly $\mathcal P\{S^n=u^n\}=\mathcal P\{S^n=v^n\}$, and we can bound $h(Y_1^n\mid S^n=u^n, M_{21})-h((S_{21}H_{21}X_1+Z_2)^n \mid S^n=v^n,  M_{21})$ as follows. First note that
$$h(Y_1^n\mid S^n=u^n, M_{21})=h\left((H_{11}X_1+H_{12}X_2+Z_1)^{\mathcal I_u^A},(H_{11}X_1+Z_1)^{\mathcal I_u^{BC}},(H_{12}X_2+Z_1)^{\mathcal I_u^D}, Z_1^{\mathcal I_u^E} \mid M_{21}\right) \textrm{, and}$$
$$h((S_{21}H_{21}X_1+Z_2)^n \mid S^n=v^n,  M_{21})=h\left((H_{21}X_1+Z_2)^{\mathcal I_v^{AC}}, Z_2^{\mathcal I_v^{BDE}}\mid M_{21}\right),$$
where $\{S^n=u^n\}$ and $\{S^n=v^n\}$ are dropped because they are independent of $X_1^n$ and $X_2^n$ when there is no feedback in the system.  Letting $\Omega_1 \triangleq \left((H_{11}X_1+H_{12}X_2+Z_1)^{\mathcal I_u^A},(H_{11}X_1+Z_1)^{\mathcal I_u^{BC}},(H_{12}X_2+Z_1)^{\mathcal I_u^D}, Z_1^{\mathcal I_u^E}\right)$, and $\Omega_2 \triangleq \Big((H_{21}X_1+Z_2)^{\mathcal I_v^{AC}}, Z_2^{\mathcal I_v^{BDE}}\Big)$, we then have
\begin{align*}
h(\Omega_1\mid M_{21}) &= I(\Omega_1; X_1^n,X_2^n\mid M_{21})+h(Z_1^n) \\
& \le I(\Omega_1, \Omega_2; X_1^n,X_2^n\mid M_{21})+h(Z_1^n) \\
& = h(\Omega_1, \Omega_2 \mid M_{21}) - h(Z_2^n) \\
& = h(\Omega_2 \mid M_{21}) + h(\Omega_1\mid M_{21}, \Omega_2) - h(Z_2^n),
\end{align*}
and hence
\begin{equation} \label {eq:5_4}
h(Y_1^n\mid S^n=u^n, M_{21})-h((S_{21}H_{21}X_1+Z_2)^n \mid S^n=v^n,  M_{21})  \le h(\Omega_1\mid M_{21}, \Omega_2) - h(Z_2^n).
\end{equation}

For $\abs{\mathcal I_u^A} \le \abs{\mathcal I_u^B}$, (\ref{eq:5_4}) can be further bounded by
\begin{align*} 
h&(Y_1^n\mid S^n=u^n, M_{21})-h((S_{21}H_{21}X_1+Z_2)^n \mid S^n=v^n,  M_{21}) \\
& \le  h(\Omega_1\mid \Omega_2) - h(Z_2^n) \\
& \overset{(a)}\le h\left((H_{11}X_1+H_{12}X_2+Z_1)^{\mathcal I_u^A}\right) + h\left((H_{11}X_1+Z_1)^{\mathcal I_u^B\setminus \mathcal I_v^A}\right) + h\left((H_{11}X_1+Z_1)^{\mathcal I_v^{AC}}\mid (H_{21}X_1+Z_2)^{\mathcal I_v^{AC}}\right) \\
& \quad + h\left((H_{12}X_2+Z_1)^{\mathcal I_u^D}\right),  \mpelabel{eq:5_5}
\end{align*}
where (a) follows from the chain rule and the fact that conditioning reduces entropy.

Due to symmetry, we clearly also have
\begin{align} \begin{split} \label{eq:5_6}
h&(Y_2^n \mid S^n=v^n, M_{12})-h((S_{12}H_{12}X_2+Z_1)^n \mid S^n=u^n,  M_{12}) \\
& \le h\left((H_{21}X_1+H_{22}X_2+Z_2)^{\mathcal I_v^A}\right) + h\left((H_{22}X_2+Z_2)^{\mathcal I_v^B\setminus \mathcal I_u^A}\right) + h\left((H_{22}X_2+Z_2)^{\mathcal I_u^{AC}}\mid (H_{12}X_2+Z_1)^{\mathcal I_u^{AC}}\right) \\
& \quad + h\left((H_{21}X_1+Z_2)^{\mathcal I_v^D}\right).
\end{split} \end{align}

The case of $\abs{\mathcal I_u^A} \ge \abs{\mathcal I_u^B}$ is treated similarly in Appendix \ref{App_A}.

\subsection{Bounding Differential Entropies} \label {sec_5D}

Each of the entropy terms in (\ref{eq:5_5}) can in turn be bounded as follows.
\begin{align*}
h\left((H_{11}X_1+H_{12}X_2+Z_1)^{\mathcal I_u^A}\right) & \le \sum_{k\in \mathcal I_u^A}{h(H_{11}X_1[k]+H_{12}X_2[k]+Z_1[k])} \\
	& = \left|\mathcal I_u^A\right|h(H_{11}X_{1Q_\alpha}+H_{12}X_{2Q_\alpha}+Z_1\mid Q_{\alpha}) \\
	& \le \left|\mathcal I_u^A\right|h(H_{11}X_{1Q_\alpha}+H_{12}X_{2Q_\alpha}+Z_1) \\
	& \le \left|\mathcal I_u^A\right|h(H_{11}X_{1G_\alpha}+H_{12}X_{2G_\alpha}+Z_1),  \mpelabel{eq:5_7}
\end{align*}
where $Q_\alpha$ is the time-sharing random variable uniformly distributed on $\mathcal I_u^A$, and given $Q_\alpha=k$, $(X_{1Q_\alpha}, X_{2Q_\alpha})$ has the same distribution as $(X_1[k], X_2[k])$ for any $k\in \mathcal I_u^A$.  $(X_{1G_\alpha}, X_{2G_\alpha})$ are jointly Gaussian with the same mean and covariance matrix as those of $(X_{1Q_\alpha}, X_{2Q_\alpha})$.  The other entropy terms in (\ref{eq:5_5}) can be similarly bounded for $P_A \le P_B$, and it is straightforward to show
\begin{align} \begin{split} \label{eq:5_9}
h\left((H_{11}X_1+Z_1)^{\mathcal I_u^B\setminus \mathcal I_v^A}\right) &\le \left(\abs{\mathcal I_u^B}-\abs{\mathcal I_v^A}\right)h(H_{11}X_{1G_\beta}+Z_1), \\
h\left((H_{11}X_1+Z_1)^{\mathcal I_v^{AC}}\mid (H_{21}X_1+Z_2)^{\mathcal I_v^{AC}}\right) &\le \left(\abs{\mathcal I_v^A} +\abs{\mathcal I_v^C}\right)  h((H_{11}X_{1G_\gamma}+Z_1)\mid (H_{21}X_{1G_\gamma}+Z_2)\textrm{, and} \\ 
h\left((H_{12}X_2+Z_1)^{\mathcal I_u^D}\right) &\le \abs{\mathcal I_u^D}h(H_{12}X_{2G_\delta}+Z_1), 
\end{split} \end{align}
where $X_{1G_\beta}, X_{1G_\gamma},$ and $X_{2G_\delta}$ apparently parallel the definitions in (\ref{eq:5_7}).  Note that we have used the fact that Gaussian distributions maximize (conditional) differential entropies \cite[Ch2]{EGK11}, \cite[Lemma 1]{T87}.

Now let $\mathcal T_\epsilon^{(n)}$ be the set of typical sequences of $S^n$, i.e. those $u^n$ with $|\:|\mathcal I_u^s|/n-P_s|<\epsilon$ for all $s\in\{A,B,C,D,E\}$.  Then $\mathcal P\big\{S^n \notin \mathcal T_\epsilon^{(n)}\big\}=\delta_n \to 0$ as $n \to \infty$, for any $\epsilon > 0$, due to the law of large numbers.  As a result, together with Fact \ref{fact_dof_ub}, it is not hard to verify that, for $P_A \le P_B$, we have
\begin{align} \begin{split} \label {eq:5_10}
\frac{1}{n}&\mathcal D\left\{ h(Y_1^n \mid S^n, M_{21}) - h((SH_{21}X_1+Z_2)^n \mid S^n,  M_{21}) \right\} \\
& \le (P_A+\epsilon)\min(2M,N) + (P_B-P_A+2\epsilon)\min(M,N)+(P_A+P_C+2\epsilon)[\min(M,2N)-\min(M,N)] \\
& \qquad +(P_D+\epsilon)\min(M,N) + \delta_n K_2 \\
& = P_A\min(2M,N)+(P_A+P_C)\min(M,2N)+(P_B-2P_A)\min(M,N) + \epsilon K_1 + \delta_n K_2 \quad \textrm{(a.s.),}
\end{split} \end{align}
where $K_1$ and $K_2$ are some finite constants. ($\delta_n K_2$ bounds the contribution of the non-typical sequences.)   Since $\epsilon$ and $\delta_n$ can both be made arbitrarily small as $n \to \infty$, this implies that
\begin{align} \begin{split} \label {eq:5_10a}
\frac{1}{n}&\mathcal D\left\{ h(Y_1^n \mid S^n, M_{21}) - h((SH_{21}X_1+Z_2)^n \mid S^n,  M_{21}) \right\} \\
& \le P_A\min(2M,N)+(P_A+P_C)\min(M,2N)+(P_B-2P_A)\min(M,N) \\
& = p_{cd}\min(2M,N)+p_c\min(M,2N)+(p_d-p_c-2p_{cd})\min(M,N) \quad \textrm{(a.s.).}
\end{split} \end{align}

Finally, recalling the symmetry between (\ref{eq:5_5}) and (\ref{eq:5_6}), we have proved Theorem \ref{thm:UB} for Regime 1 $\left(p_c\le \frac{p_d}{1+p_{d|c}}\right)$, which is equivalent to $P_A \le P_B$.  The proof of Theorem \ref{thm:UB} for Regime 2 $\left(p_c > \frac{p_d}{1+p_{d|c}}\right)$ is obtained by considering $P_A > P_B$, following the same lines, and is relegated to Appendix \ref{App_A}.

\section{Proof of Theorem 2 : Achievability schemes} \label{Thm2}
%auto-ignore

\vspace{-1ex}
\begin{table}[b]
\renewcommand{\arraystretch}{1.0}
\centering
\caption{Sum DoF Upper Bound in the $\eta$-triple notation}
\label{tab:DoF_UB}
\vspace{-3ex}
\begin{tabular}{|c|c|c|c|}
	\hline
	  \bfseries Type & $\boldsymbol{\frac{M}{N} \textrm{ or } \frac{N}{M}}$ & \bfseries Regime 1 & \bfseries Regime 2 \\ 
	\hline 
	& & & \\ [-3.5ex]
	\hline
	\multirow{2}{*}{I} &  $0 < \frac{M}{N} \le \frac{1}{2}$ & $2\langle M,M,0 \rangle$ & $2\langle M,M,0 \rangle$ \\
	\hhline{~---}
	 & $\frac{1}{2} < \frac{M}{N} \le 1$ & $2\langle N-M,M,0 \rangle$ & $2\langle M,N-M,2M-N \rangle$ \\
	\hline
	\multirow{2}{*}{II} &  $0 < \frac{N}{M} \le \frac{1}{2}$ & $2\langle N,N,N \rangle$ & $2\langle N,N,N \rangle$ \\
	\hhline{~---}
	& $\frac{1}{2} < \frac{N}{M} \le 1$ & $2\langle M-N,N,M-N \rangle$ & $2\langle N,M-N,N \rangle$ \\
	\hline
\end{tabular}
\end{table}

In this section, we show that the upper bound derived in the previous section is in fact tight in Regime 1 and partially tight in Regime 2 (c.f. Fig. \ref{fig:DoF_Char_Status}) with four closely related coding schemes, thereby establishing Theorem \ref{thm:UB_Tightness}.  Not surprisingly the INBF schemes in \cite{MMK08} that achieve the best-known DoF on the non-bursty MIMO X channel serve as our baseline schemes, and it turns out that they still work well for the Type I channel in Regime 2 even when the channel becomes bursty.  However, new beamforming schemes are needed to achieve the sum DoF of the bursty MIMO X channel in Regime 1, where $p_c$ is usually small.  Moreover, for the Type II channel operated in Regime 2, the INBF schemes alone no longer achieve the DoF, when $N/M > 1/2$.  Further rate splitting and introducing public messages can offer higher DoF and, in some cases, achieve the DoF of the Type II channel in Regime 2.  

Our proof in this section makes use of the simple facts in Fact \ref{fact_dof_pp_mac}.  It is easy to see from Fact \ref{fact_dof_pp_mac} that when the channel state $S$ takes the form of Fig. \ref{fig:Rx1_State}, the achievable DoF can be expressed in the $\eta$-triple notation as $\ETS{\mathrm{rank}\{H(p_{cd})\}}{\mathrm{rank}\{H(p_{\overline cd})\}}{\mathrm{rank}\{H(p_{c\overline d})\}}$.  To facilitate the proof, it helps to express the upper bound also in the $\eta$-triple notation, as shown in Table \ref{tab:DoF_UB}.

\begin{fact} \label{fact_dof_pp_mac}
For a point-to-point MIMO Gaussian channel, modeled by $Y=H(S)X+Z$, where the channel matrix $H(S)$ varies with a discrete channel state $S$ known to the receiver, the following DoF is achievable:  $\sum_{s}\mathcal P_S(s)\mathrm{rank}\{H(s)\}$.  Similarly, for a MIMO Gaussian MAC, modeled by $Y=H_1(S)X_1+H_2(S)X_2+Z$, where the channel state $S$ is known to the receiver, if we let $H(S)=[H_1(S) \: H_2(S)]$ and $X=\left[\begin{array}{c}X_1 \\ X_2\end{array}\right]$, then exactly the same form of sum DoF, i.e.  $\sum_{s}\mathcal P_S(s)\mathrm{rank}\{H(s)\}$, is also achievable on the MAC.
\end{fact}

We now show that the beamforming schemes illustrated in Fig. \ref{fig:Type_I_BF} suffice to achieve the sum DoF upper bound of the Type I bursty MIMO X channel in Regime 1 and part of Regime 2.  Note that details are given only for beamforming filters at Tx1 and Rx1 in Fig. \ref{fig:Type_I_BF}, Fig. \ref{fig:Class_1_BF}, and Fig. \ref{fig:Type_II_BF}.  It is understood that the beamforming at Tx2 and Rx2 is done symmetrically.

\subsection{Type I, Regime 1 : INBF without cross-link messaging} 

\begin{figure}[t]
\centering
\includegraphics[width=15.5cm]{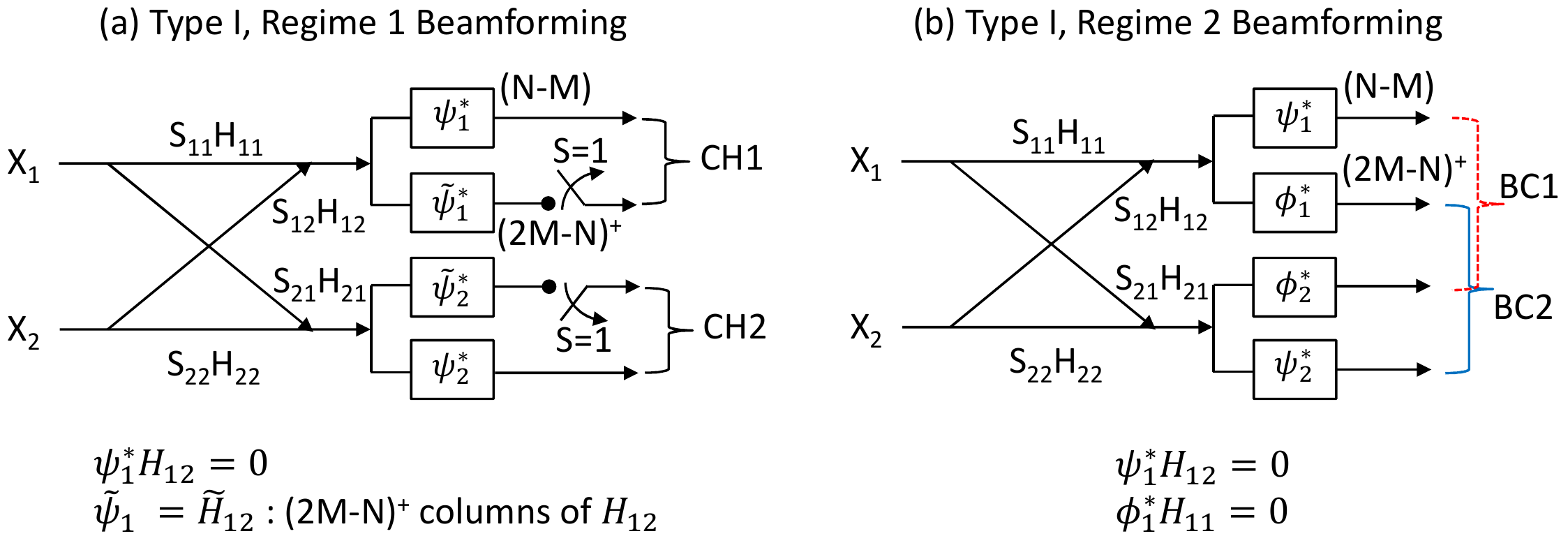}
\caption{Beamforming schemes that achieve the DoF upper bound of the Type I channel ($(x)^+\triangleq \max(x,0)$)}
\label{fig:Type_I_BF}
\end{figure}

Let us consider the beamforming scheme shown in Fig. \ref{fig:Type_I_BF}(a) for the Type I channel operated in Regime 1 with $1/2 < M/N \le 1$ first.  (The achievability scheme for $M/N \le 1/2$ is simpler and can be viewed as a special degenerate case.)  Since $p_c$ is generally small in Regime 1, we essentially have two orthogonal point-to-point channels on the direct-links for a large fraction of time.  So a simple and natural coding strategy in this regime is to treat cross-link interference as erasure and use the channel as two orthogonal point-to-point channels on the direct-links.  This is the basic idea of Fig. \ref{fig:Type_I_BF}(a), where four linear beamforming filters (matrices) are employed.  $\psi_1$ and $\psi_2$ filters null out the cross-link interference at the receivers, resulting in two effective point-to-point $M \times (N-M)$ channels, free of interference.  $\tilde{\psi}_1$ and $\tilde{\psi}_2$ filters, on the other hand, exploit the opportunity for data transmission when the cross-links are off, thereby increasing the dimensions of the effective channels to $M \times M$.  $\tilde{\psi}_1$ and $\tilde{\psi}_2$ outputs are discarded when the cross-links are present.

More precisely, we choose $\psi_1$ to be an $N \times (N-M)$ matrix, satisfying $\psi_1^*H_{12}=0$.  This is certainly feasible since the $\mathrm{rank}(H_{12})=M$ (a.s.), and $\psi_1$ can be made of the $(N-M)$ basis vectors of the orthogonal complement of the column space of $H_{12}$, i.e. $(\mathrm{Col}\: H_{12})^\perp$.    $\tilde\psi_1$ is designed so that $\left[ \psi_1 \: \tilde\psi_1 \right]$ is full rank, and may simply consist of any $(2M-N)$ columns of $H_{12}$.  It is easy to verify that $\left[ \psi_1 \: \tilde\psi_1 \right]^*H_{11}$ is full rank (a.s.), so we effectively have a point-to-point time-varying channel from Tx1 to Rx1, where the rank of the effective channel matrix is $(N-M), M$ and $0$ (a.s.) in the $p_{cd}, p_{\overline cd}$ and $p_{c\overline d}$ states (Fig. \ref{fig:Rx1_State}), respectively.  Therefore, by Fact \ref{fact_dof_pp_mac}, a DoF of $\langle N-M, M, 0 \rangle$ is achievable on the link from Tx1 to Rx1.  The same DoF is clearly achievable on the Tx2 to Rx2 link due to symmetry.  Therefore, we can achieve a total DoF of $2\langle N-M, M, 0 \rangle$, which is exactly the respective upper bound in Table \ref{tab:DoF_UB}!  Note that this scheme maximizes the achievable DoF in the $p_{\overline cd}$ state, while sacrificing some DoF of the $p_{cd}$ state, which is reasonable because $p_{cd} \le p_d - p_c \le p_{\overline cd}$ in Regime 1.

\begin{figure}[b]
\centering
\includegraphics[width=13.7cm]{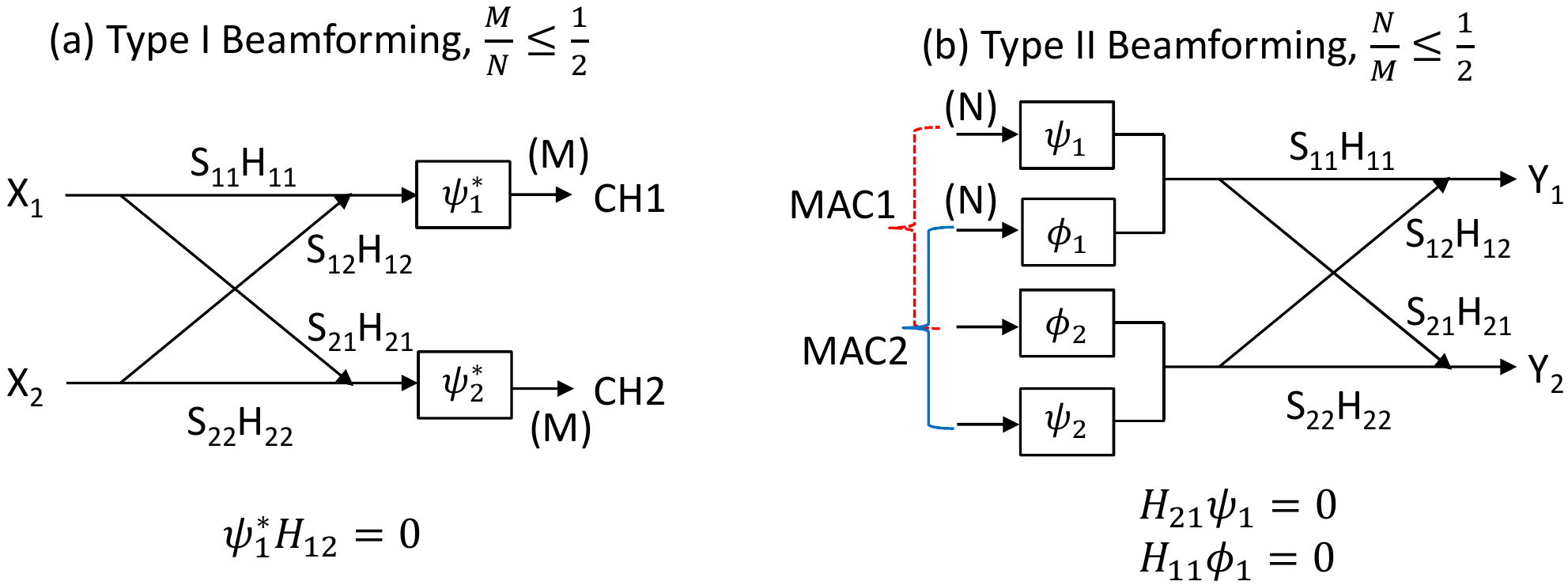}
\caption{Simplified beamforming schemes when $\frac{M}{N} \le \frac{1}{2}$ or $\frac{N}{M} \le \frac{1}{2}$}
\label{fig:Class_1_BF}
\end{figure}

In the case of $M/N \le 1/2$, the $\psi_1$ and $\psi_2$ filters can eliminate the interference coming from the other transmitter when the cross-links are present, while preserving the full $M$ dimensions of the direct-link signals.  So the $\tilde\psi_1$ and $\tilde\psi_2$ filters are unnecessary, and the rank of each effective point-to-point channel is $M$ whenever the direct-link is present, even if we reduce the size of the $\psi_1$ and $\psi_2$ matrices to $N \times M$.  It follows that the achievable DoF is $2\langle M, M, 0 \rangle$ in this case, which matches the upper bound as well.  This simplified beamforming scheme is illustrated in Fig. \ref{fig:Class_1_BF}(a). 

\subsection{Type I, Regime 2 : INBF with cross-link messaging}

For the Type I channel operated in Regime 2, we again consider the $1/2 < M/N \le 2/3$ case first.  Note that in Regime 2, $p_{cd}$ and $p_c$ generally become larger, so the cross-links are present for an increasingly larger fraction of time.  Intuitively, to maximize the sum DoF in this case, the paradigm should be shifted from avoiding the cross-links to exploiting the cross-links.  One way to accomplish this is cross-link messaging with the beamforming scheme depicted in Fig. \ref{fig:Type_I_BF}(b).  Note that this is exactly the same INBF scheme for the non-bursty MIMO X channel as proposed in \cite{MMK08}.

To determine the DoF achievable by this scheme when the channel becomes bursty, recall that this scheme decomposes the channel into two orthogonal broadcast channels, with $\psi_1^*H_{12}=\psi_2^*H_{21}=0$ and $\phi_1^*H_{11}=\phi_2^*H_{22}=0$.  Now consider the effective broadcast channel, BC1, from Tx1 to $\psi_1$ and $\phi_2$.  Since the effective channel matrix $\left[\begin{array}{c}\psi_1^*H_{11} \\ \phi_2^*H_{21}\end{array} \right]$ is full-rank (a.s.), with a simple channel inversion at Tx1 we have effectively two orthogonal point-to-point channels.  It then follows from Fact \ref{fact_dof_pp_mac} that sum DoF of $(N-M)p_d+(2M-N)p_c=\ETS{M}{N-M}{2M-N}$ can be achieved on BC1.  Due to symmetry, the same DoF is also achievable on the other effective broadcast channel (BC2), so the total achievable DoF clearly coincides with the upper bound in Table \ref{tab:DoF_UB}, i.e. $2\langle M, N-M, 2M-N \rangle$.

When $M/N \le 1/2$, the $N-M$ dimensions of the $\psi_1$ filter output can capture all of the $M$ dimensions of the signal coming from Tx1, and so can the $\psi_2$ filter.  In this case, it does not make sense to split the transmitter dimensions and send messages on the cross-links, since $p_c \le p_d$.  So the $\phi_1$ and $\phi_2$ filters can be removed, and by sending messages on the direct-links only, we can clearly achieve a DoF of $2Mp_d = 2\ETS{M}{M}{0},$ which also matches the upper bound.  Moreover,  the $\psi_1$ and $\psi_2$ matrices can be reduced to $N \times M$ matrices as shown in Fig. \ref{fig:Class_1_BF}(a), without affecting the achievable DoF.  

It is interesting to note that the two beamforming schemes in Fig. \ref{fig:Type_I_BF} originating from two different paradigms degenerate into the same scheme in Fig. \ref{fig:Class_1_BF}(a) when $M/N \le 1/2$, because in this case both paradigms reduce to maximizing the utility of the direct links.  This may explain why, when $M/N \le 1/2$, the sum DoF of the channel takes the same form in both regimes, and depends on $p_d$ only, i.e. $2\langle M,M,0 \rangle=2Mp_d.$

\begin{figure}[t]
\centering
\includegraphics[width=16.5cm]{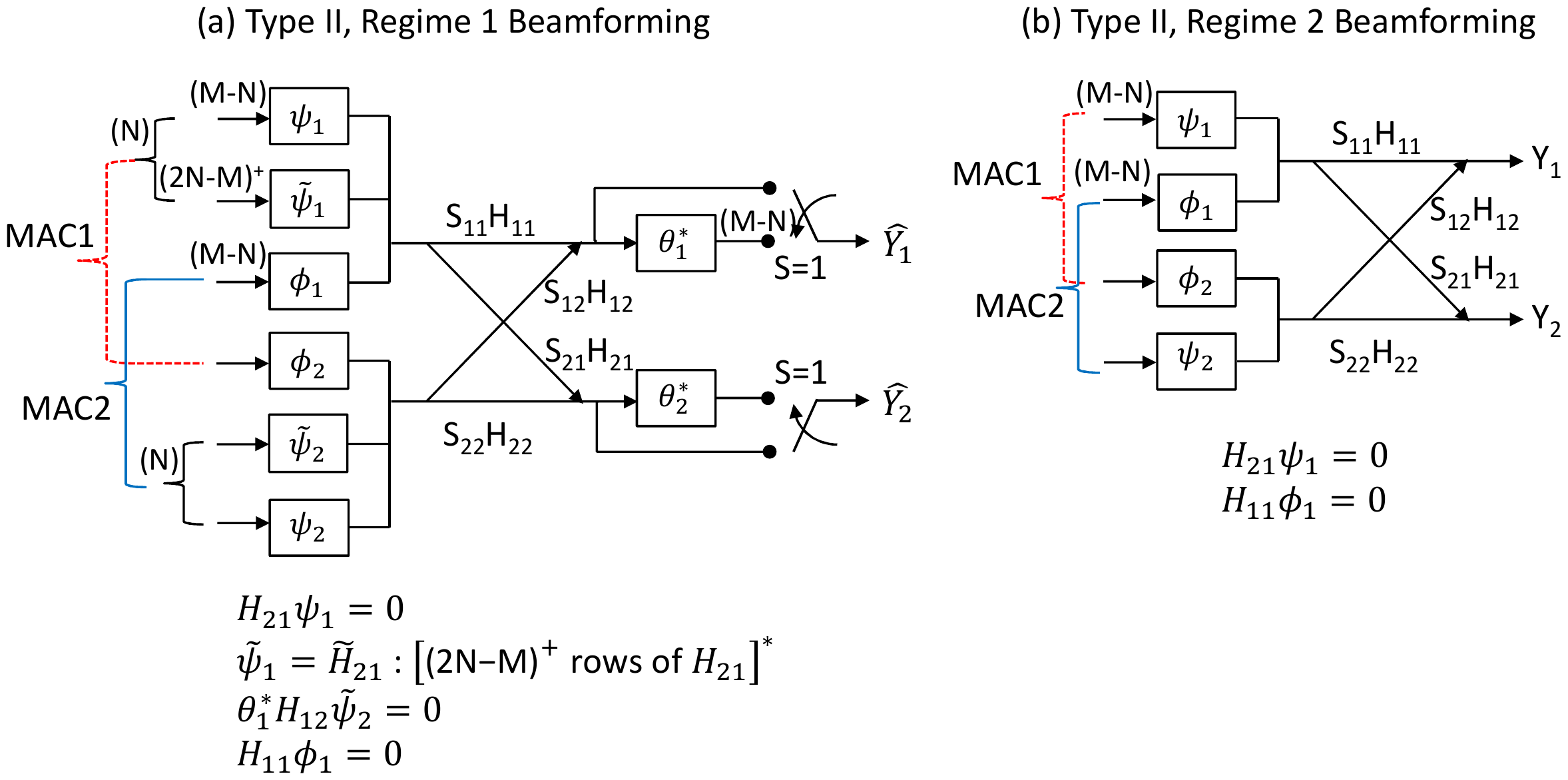}
\caption{Beamforming schemes that, optionally together with public messages, achieve the DoF upper bound of the Type II channel}
\label{fig:Type_II_BF}
\end{figure}

In summary, with the proof in this and the previous subsection, we have shown that the sum DoF upper bound is indeed tight for the Type I channel in the solid regions of Fig. \ref{fig:DoF_Char_Status}(a), and thereby proving the half of Theorem \ref{thm:UB_Tightness}.  Next, let us move on to the Type II channel.  Since the transmitters have more antennas than the receivers on the Type II channel, a natural coding attempt is to employ beamforming filters on the transmitter side reciprocal to those on the receiver side of the Type I channel.  Although this attempt is headed roughly in the right direction, it is not completely correct and optimal for the Type II channel.  As we have seen in Section \ref{Fund}, the Type II bursty channel is not strictly reciprocal to the Type I channel, and neither is its beamforming scheme, due to a fundamental difference between bursty BC and MAC without feedback, i.e. cross-links play a more important role on the MAC and Type II channel.  Therefore, some corrections to the reciprocal scheme are needed, and we will show that with the modifications shown in Fig. \ref{fig:Type_II_BF}, and optionally introducing public messages, we can achieve the sum DoF of the bursty Type II channel.

\subsection{Type II, Regime 1 : INBF with cross-link messaging}

Let us start with Regime 1 and $N/M>1/2$.  Comparing the beamforming scheme in Fig. \ref{fig:Type_II_BF}(a) for the Type II channel to that in Fig. \ref{fig:Type_I_BF} for the Type I channel, we note that they are reciprocal, except for the $\phi_1$ and $\phi_2$ filters in Fig. \ref{fig:Type_II_BF}(a).  Specifically $\psi_1$ is an $M \times (M-N)$ matrix satisfying $H_{21}\psi_1=0$, and $\tilde\psi_1$ of dimensions $M \times (2N-M)$ is chosen to make $\left[ \psi_1\: \tilde\psi_1 \right]$ full-rank (a.s.)  $\psi_2$ and $\tilde\psi_2$ are designed symmetrically at Tx2.  In other words, the $\psi_1$ and $\psi_2$ filters are again designed to null out cross-link interference and effect two point-to-point channels on the direct-links, free of interference.  The $\tilde\psi_1$ and $\tilde\psi_2$ filters enlarge the dimensions of the effective point-to-point links from $M-N$ to $N$ when the cross-link are off.  However, when the cross-links are on, we need $\theta_1$ and $\theta_2$ filters to null out the interference from $\phi_2$ and $\phi_1$, respectively.  This can be done with two $N \times (M-N)$ matrices satisfying $\theta_1^*H_{12}\phi_2=\theta_2^*H_{21}\phi_1=0$.  This extra work is needed on the Type II channel because only the receivers know the state of the channel.  Except for this slight technical difference, $\psi_1, \tilde\psi_1, \psi_2$ and $\tilde\psi_2$ play exactly the same role as in the Type I channel, and present $\langle M-N, N, 0\rangle$ DoF on each direct-link.  ($\theta_1^*H_{11}\left[\begin{array}{c}\psi_1 \\ \tilde\psi_1 \end{array} \right]$ is easily verified to be full-rank almost surely.)

However, more can be done on the Type II channel.  Since a Type II channel can be decomposed into two MACs and there are unused dimensions at each transmitter, we incorporate the $\phi_1$ and $\phi_2$ filters to take advantage of the extra opportunity of communication offered by the cross-links when the direct-links are off.  With $H_{22}\phi_2=0$, the $\phi_2$ filter is transparent to Rx2, but it can offer $(M-N)$ dimensions for data transmission from Tx2 to Rx1 in the $p_{c\overline d}$ state (Fig. \ref{fig:Rx1_State}), due to Fact \ref{fact_dof_pp_mac}.  As such, it is easy to see that the achievable DoF at Rx1 is increased from $\langle M-N, N, 0\rangle$ to $\langle M-N, N, M-N \rangle$.  By symmetry, the same DoF is clearly also achievable at Rx2, so the achievable total DoF is $2\langle M-N, N, M-N \rangle$, which is exactly the upper bound of the Type II channel in Regime 1 when $N/M > 1/2$!  The reader may have noticed a subtle issue here: Is $\theta_1^*H_{12}\phi_2$ full rank (a.s.)?  Although $\theta_1$ and $H_{12}$ are correlated, the answer is still yes, and can be seen in Appendix \ref{App_B}.  Similarly, $\theta_2^*H_{21}\phi_1$ is also full rank (a.s.)

When $N/M \le 1/2$, similarly the $\psi_1$ and $\psi_2$ filters alone can deliver $N$ dimensions of signals to Rx1 and Rx2, respectively, whenever the direct-links are on.  So the $\tilde\psi_1$ and $\tilde\psi_2$ filters again become redundant and can be removed.  Moreover, the dimensions of the $\psi_1, \psi_2, \phi_1$ and $\phi_2$ filters can all be reduced to $M \times N$, and still achieve the DoF upper bound $2\langle N,N,N \rangle$.  This simplified scheme is shown in Fig. \ref{fig:Class_1_BF}(b).

\subsection{Type II, Regime 2 : INBF with cross-link messaging, overlaid with public messages (HKIA)} \label{Achi_D}

\subsubsection{Sub-optimality of INBF} \label{Achi_D_1}

Finally we come to the more intriguing case of the Type II channel in Regime 2, where INBF may not achieve the DoF of the channel.  To see this, let us consider the case of $1/2 < N/M \le 2/3$ first.  Following the philosophy of the paradigm shift on the Type I channel from avoiding the cross-links to exploiting them, we remove the $\tilde\psi_1$ and $\tilde\psi_2$ filters and use the $(2N-M)$ dimensions at each transmitter to beamform (with $\phi_1$ and $\phi_2$) to the cross-link receiver.  ($\theta_1$ and $\theta_2$ filters should therefore also be removed.)  Moreover, the dimensions of the $\phi_1$ and $\phi_2$ filters can be dialed up from $M\times (2N-M)$ to $M \times (M-N)$.  This leads us to the beamforming scheme in Fig. \ref{fig:Type_II_BF}(b).  With Fact \ref{fact_dof_pp_mac}, it is straightforward to verify that this scheme can deliver $2\langle N, M-N, M-N \rangle$ DoF, which is already greater than the reciprocal of the DoF of the corresponding Type I channel, i.e. $\langle N, M-N, 2N-M \rangle$.  The difference is apparently due to the increased dimensions of the $\phi_1$ and $\phi_2$ filters, and is another manifestation of the bursty BC-MAC difference.  

At this point, we have used up the available dimensions for beamforming to orthogonalize the channel.  However, the DoF upper bound is not attained, and an examination of the signal dimensions at the receivers, as was done in Section \ref{Fund}, quickly reveals that we fall short of the potential of the channel:  The received signal dimensions at each receiver are $N, (M-N)$ and $(M-N)$ in the $p_{cd}, p_{\overline cd}$ and $p_{c\overline d}$ states, respectively (c.f. Fig. \ref{fig:Rx1_State}), but there are $N$ Rx antennas.   So the $p_{\overline cd}$ and $p_{c\overline d}$ states are underutilized by the INBF scheme, since $(M-N) < N\le M$ when $1/2 < N/M \le 2/3$.

\subsubsection{DoF-achieving HKIA scheme} \label{Achi_D_2}

One way to make full use of the under-utilized states is the \emph{HKIA} scheme which employs rate splitting, similar to the HK strategy in \cite{ETW08}.  Specifically, we introduce public messages, $M_{01}$ and $M_{02}$, and superimpose them on the INBF signals in Fig. \ref{fig:Type_II_BF}(b) which carry the private messages.  ($M_{01}$ and $M_{02}$ are the messages from Tx1 and Tx2, respectively, to both receivers.)  The transmit power is split evenly between public and private messages.  More precisely, we have the following random coding scheme, generalized from the one used in Section \ref{Fund_A}.  (For completeness, we repeat some of the earlier definitions and arguments below.)
\begin{equation} \label{eq:6_1}
X_1^n(M_{01}, M_{11}, M_{21})=U_1^n(M_{01})+[\psi_1 \: \phi_1]\left[\begin{array}{c}D_1^n(M_{11}) \\
C_1^n(M_{21}) \end{array} \right],
\end{equation}
where $U_1 \sim \mathcal{N}\left(0, \frac{P}{2M}I_M\right), C_1 \& D_1 \sim \mathcal{N}\left(0,  \frac{P}{4(M-N)}I_{M-N}\right),$ and $[\psi_1 \: \phi_1]\left[\begin{array}{c}D_1^n(M_{11}) \\ C_1^n(M_{21}) \end{array} \right]$ is a shorthand notation for multiplying each of the $n$ components of $\left[\begin{array}{c}D_1^n(M_{11}) \\ C_1^n(M_{21}) \end{array} \right]$ by $[\psi_1 \: \phi_1]$.  $X_2^n(M_{02}, M_{12}, M_{22})$ is encoded symmetrically:  
\begin{equation} \label{eq:6_1a}
X_2^n(M_{02}, M_{12}, M_{22})=U_2^n(M_{02})+[\psi_2 \: \phi_2]\left[\begin{array}{c}D_2^n(M_{22}) \\
C_2^n(M_{12}) \end{array} \right],
\end{equation}
where $U_2, C_2,$ and $D_2$ have the same distributions as $U_1, C_1,$ and $D_1$.  Successive decoding is employed at each receiver, where the public messages are decoded and removed first, and then the private messages are  decoded.  Note that although this scheme is motivated from $1/2 < N/M \le 2/3$, there is nothing preventing us from using it for $2/3 < N/M \le 1$.  The achievable sum DoF of this HKIA scheme is given in Proposition \ref{prop_HKIA_dof}. To facilitate the discussion hereafter, we shall use the terms \emph{private DoF} ($\eta_\mathrm{priv}$) and \emph{public DoF} ($\eta_\mathrm{pub}$) to refer to the sum DoF achievable on the private messages and public messages, respectively.

\begin{proposition} \label{prop_HKIA_dof}
The following private and public DoF can be achieved by the HKIA scheme in Section \ref{Achi_D_2}:
\begin{enumerate}[(i)]
\item $\eta_\mathrm{priv} = 2\ET{N}{M-N}{M-N}$ and $\eta_\mathrm{pub} = 2\ET{0}{0}{2N-M}$; when $1/2 < N/M \le 2/3$.
\item $\eta_\mathrm{priv} = 2(M-N)\ET{2}{1}{1}$ and $\eta_\mathrm{pub} = 2\ET{3N-2M}{0}{2N-M}$, when $2/3 < N/M \le 1$ and $p_c \le \frac{p_d}{1+\alpha p_{d|c}}$,
\end{enumerate}
where $\alpha = \frac{3N-2M}{2N-M}$.  In both cases, the HKIA scheme achieves a total DoF of $2\ET{N}{M-N}{N}$.
\end{proposition}

Note that non-zero public DoF is achieved whenever $p_{c \overline d}>0$, and the total achievable DoF matches the upper bound given in Table \ref{tab:DoF_UB}.

\subsubsection{Proof of Proposition \ref{prop_HKIA_dof} for $1/2 < N/M \le 2/3$} \label{Achi_D_3}

First of all, public messages are removed when we decode the private messages, so the achievable private DoF clearly remains the same as in Section \ref{Achi_D_1}, which readily proves that $\eta_\mathrm{priv}=2\langle N, M-N, M-N \rangle$ is achievable.  So the key task is to evaluate the achievable $\eta_\mathrm{pub}$.  To do so, let $\eta_{01}$ and $\eta_{02}$ denote the achievable DoF on $M_{01}$ and $M_{02}$, respectively.  Since the public messages are decoded at both receivers (or effectively two MACs) in the presence of the private messages, $(\eta_{01}, \eta_{02})$ pair satisfying the following set of inequalities are achievable (i.e. the intersection of the DoF regions of the two MACs):
\begin{align} \begin{split} \label{eq:6_2}
\eta_{01} &\le \mathcal D\{\min[I(U_1; Y_1\mid S_1,U_2), I(U_1; Y_2\mid S_2,U_2)]\}, \\
\eta_{02} &\le \mathcal D\{\min[I(U_2; Y_1\mid S_1,U_1), I(U_2; Y_2\mid S_2,U_1)]\}, \textrm{ and}\\
\eta_{01}+\eta_{02} &\le \mathcal D\{\min[I(U_1, U_2; Y_1\mid S_1), I(U_1, U_2; Y_2\mid S_2)]\},
\end{split} \end{align}
Each of the mutual informations in (\ref{eq:6_2}) can be expressed as the difference of two conditional differential entropies, e.g. $I(U_1; Y_1\mid S,U_2)=h(Y_1\mid S,U_2)-h(Y_1\mid S,U_1,U_2).$  So we first compute $\mathcal D\{h(Y_1\mid S_1)\}, \mathcal D\{h(Y_1\mid S_1, U_1)\}, \mathcal D\{h(Y_1\mid S_1, U_2)\},$ and $\mathcal D\{h(Y_1\mid S_1, U_1, U_2)\}$.  The $\eta$-triple notation comes particularly handy in this analysis.  Following the steps given in Appendix \ref{App_C}, it is not hard to show that
\begin{align*} 
\mathcal D\{h(Y_1\mid S_1)\}				&=\langle N, N, N\rangle, \\
\mathcal D\{h(Y_1\mid S_1, U_1)\}			&=\langle N, M-N, N\rangle, \\
\mathcal D\{h(Y_1\mid S_1, U_2)\}			&=\langle N, N, M-N\rangle, \textrm{ and}\\
\mathcal D\{h(Y_1\mid S_1, U_1, U_2)\}	&=\langle N, M-N, M-N\rangle.  \mpelabel{eq:6_3}
\end{align*}
By symmetry, we clearly also have
\begin{align*}
\mathcal D\{h(Y_2\mid S_2)\}				&=\langle N, N, N\rangle, \\
\mathcal D\{h(Y_2\mid S_2, U_2)\}			&=\langle N, M-N, N\rangle, \\
\mathcal D\{h(Y_2\mid S_2, U_1)\}			&=\langle N, N, M-N\rangle, \textrm{ and}\\
\mathcal D\{h(Y_2\mid S_2, U_1, U_2)\}	&=\langle N, M-N, M-N\rangle.  \mpelabel{eq:6_4}
\end{align*}
With (\ref{eq:6_3}) and (\ref{eq:6_4}), it is then straightforward to show that (\ref{eq:6_2}) reduces to
\begin{align} \begin{split} \label{eq:6_5}
\eta_{01}, \: \eta_{02} &\le \langle 0, 0, 2N-M \rangle, \textrm{ and} \\
\eta_{01}+\eta_{02} &\le \langle 0, 2N-M, 2N-M \rangle.
\end{split} \end{align}
From this we conclude that $\eta_\mathrm{pub}=\eta_{01}+\eta_{02}$ satisfying the following inequality is achievable:
\begin{equation} \label{eq:6_5a}
\eta_\mathrm{pub} \le \min\Big(2\langle 0,0,2N-M \rangle, \quad \langle 0, 2N-M, 2N-M \rangle\Big) =  2\langle 0,0,2N-M \rangle,
\end{equation}
where the last equality holds because $p_c \le p_d$.  This establishes Proposition \ref{prop_HKIA_dof} for $1/2 < M/N \le 2/3$.

\subsubsection{Proof of Proposition \ref{prop_HKIA_dof} for $2/3 < N/M \le 1$} \label{Achi_D_4}

Again, it is easy to see that the following private DoF is achievable when $2/3 < N/M \le 1$:
\begin{equation} \label{eq:6_5b}
\eta_\mathrm{priv} = 2(M-N)\ET{2}{1}{1}
\end{equation}
As for the public DoF, it is not hard to verify that (\ref{eq:6_3}) and (\ref{eq:6_4}) remain valid, except for the following modification:
\begin{equation} \label{eq:6_6}
\mathcal D\{h(Y_1\mid S_1, U_1, U_2)\} =	\mathcal D\{h(Y_2\mid S_2, U_1, U_2)\} =\langle 2M-2N, M-N, M-N\rangle.
\end{equation}
As a result, (\ref{eq:6_2}) evaluates to
\begin{align} \begin{split} \label{eq:6_7}
\eta_{01}, \: \eta_{02} &\le \langle 3N-2M, 0, 2N-M \rangle, \textrm{ and} \\
\eta_{01}+\eta_{02} &\le \langle 3N-2M, 2N-M, 2N-M \rangle,
\end{split} \end{align}
which implies that the following public DoF is achievable:
\begin{equation} \label{eq:6_8}
\eta_\mathrm{pub}=\eta_{01}+\eta_{02} \le \min\Big(2\langle 3N-2M, 0, 2N-M \rangle, \quad \langle 3N-2M, 2N-M, 2N-M \rangle\Big).
\end{equation}
Note that $2\ETS{3N-2M}{0}{2N-M} \le \ETS{3N-2M}{2N-M}{2N-M}$ when $(3N-2M)p_{cd} \le (2N-M)(p_d-p_c)$, or equivalently $p_c \le \frac{p_d}{1+\alpha p_{d|c}}$, where $\alpha = \frac{3N-2M}{2N-M}$.  Therefore, $\eta_\mathrm{pub} \le 2\langle 3N-2M, 0, 2N-M \rangle$ is achievable when $p_c \le \frac{p_d}{1+\alpha p_{d|c}}$, which concludes the proof of Proposition \ref{prop_HKIA_dof}.

\subsubsection{Simplified scheme for $N/M \le 1/2$} \label{Achi_D_5}

When $N/M \le 1/2$, note that the beamforming scheme in Fig. \ref{fig:Type_II_BF}(b) has completely used up the $N$ dimensions of the receivers in each of $p_{cd}, p_{\overline cd},$ and $p_{c\overline d}$ states.  Overlaying the private messages with public messages in this case is hence unfruitful and unnecessary.  The private messages alone can deliver a sum DoF of $2\langle N,N,N \rangle$, which already matches the upper bound.  Moreover, the dimensions of the $\psi_1, \phi_1, \psi_2$ and $\phi_2$ filters can all be reduced to $M \times N$, resulting in the simplified scheme in Fig. \ref{fig:Class_1_BF}(b).

Once again, when $N/M \le 1/2$, the two beamforming schemes in Fig. \ref{fig:Type_II_BF} for the Type II channel both reduce to the same scheme in Fig. \ref{fig:Class_1_BF}(b), and the sum DoF of the channel has the same form, i.e. $2\ETS{N}{N}{N}=2Np_d+2Np_{c\overline d}$, in Regime 1 and Regime 2.   Unlike the Type I channel, however, the sum DoF in this case varies with $p_c$ and $p_{cd}$, due to the fundamental difference between the bursty BC and MAC discussed earlier.

To recapitulate this section, we have shown four closely related coding schemes that attain the sum DoF upper bound of the bursty MIMO X channel in the solid regions of Fig. \ref{fig:DoF_Char_Status}, and hence have proved Theorem \ref{thm:UB_Tightness}.

\section{Proof of Theorem 3 : Blending HKIA schemes and fine-tuning power allocation} \label{Thm3}
%auto-ignore

We now turn to the proof of Theorem \ref{thm:LB}, which furnishes a lower bound of the sum DoF of the bursty MIMO X channel where the sum DoF remains unknown, i.e. in the dotted regions of Fig. \ref{fig:DoF_Char_Status}.  This lower bound is obtained by further extending the HKIA scheme shown in Section \ref{Achi_D}, and it surpasses the sum DoF achievable by both the IA scheme and the HKIA scheme in Section \ref{Achi_D}.  To differentiate HKIA schemes with different INBF schemes on a bursty MIMO X channel in this section, we shall use $\left[N_{D_1}, N_{C_1}, N_{C_2}, N_{D_2}\right]_\mathrm{II}$ to refer to an HKIA scheme on the Type II channel where the $D_1, C_1, C_2,$ and $D_2$ random vectors take $N_{D_1}, N_{C_1}, N_{C_2},$ and $N_{D_2}$ dimensions, respectively, c.f. (\ref{eq:6_1}) and (\ref{eq:6_1a}).  Similarly, the $\left[N_{D_1}, N_{C_1}, N_{C_2}, N_{D_2}\right]_\mathrm{I}$ notation shall also be used to distinguish the HKIA schemes on the Type I channel.

\begin{figure}[b]
\centering
\begin{tikzpicture}
\node[anchor=south west,inner sep=0] at (0,0) {\includegraphics[width=7cm]{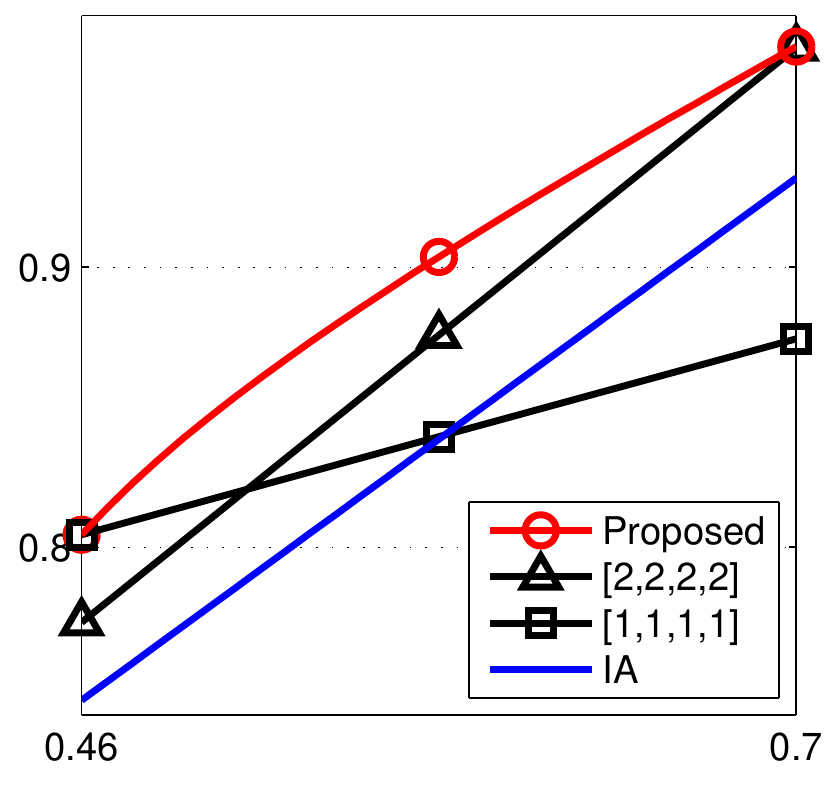}};
\node [right] at (6.7,0.65) {$p_c$};
\node [above] at (0.7,6.55) {$\eta/6$};
\node [right] at (6.1,1.67) {$_\mathrm{II}$};
\node [right] at (6.1,1.27) {$_\mathrm{II}$};
%\node [right] at (5.95,1.45) {$_\mathrm{II}$};
%\node [right] at (5.95,1.1) {$_\mathrm{II}$};
\end{tikzpicture}
\caption{Three HKIA schemes for the 6x5 bursty channel with $p_d=0.7, p_{d|c}=0.9, \frac{p_d}{1+\alpha p_{d|c}}<p_c\le p_d$}
\label{fig:HKIA_LB_6x5}
\end{figure}

\subsection{Motivation}

\begin{figure}[b]
\centering
\includegraphics[width=14.5cm]{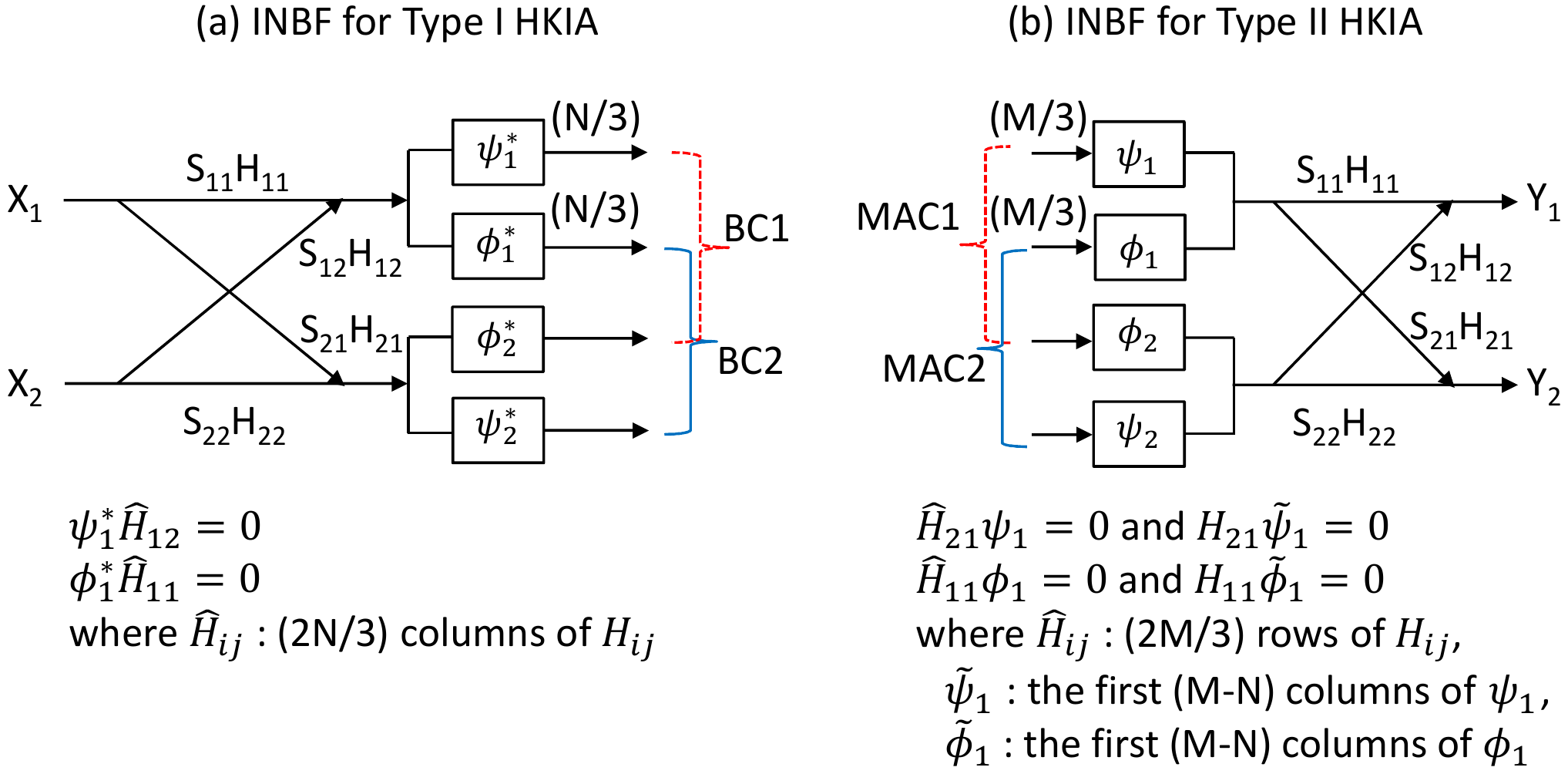}
\caption{INBF schemes for the HKIA DoF lower bounds when $\max(M,N)$ is a multiple of $3$}
\label{fig:HKIA_LB_INBF}
\end{figure}

To motivate the general scheme, let us first consider a concrete and simple example, namely the $6 \times 5$ bursty X channel with $\frac{p_d}{1+\alpha p_{d|c}} < p_c \le p_d$, where $\alpha = \frac{3N-2M}{2N-M}=\frac{3}{4}$.  For this channel, we know from (\ref{eq:6_5b}) and (\ref{eq:6_8}) that the HKIA scheme ($[1,1,1,1]_\mathrm{II}$) in Section \ref{Achi_D} achieves a DoF of $2(M-N)\langle 2,1,1 \rangle + \langle 3N-2M, 2N-M, 2N-M \rangle=\ETS{7}{6}{6}$.  However, a moment of reflection reveals that this clearly cannot be optimal.  For example, when $p_c, p_d$ and $p_{cd}$ are all close to $1$, the achieved sum DoF is $\ETS{7}{6}{6} \simeq 7$.  But in this case we almost have a non-bursty $6 \times 5$ X channel, and by using the usual INBF (IA) scheme (Fig. \ref{fig:HKIA_LB_INBF}(b)) for the non-bursty channel alone we can easily achieve $\ETS{8}{4}{4} \simeq 8$ DoF, which should also be close to being optimal.  So intuitively to enhance the lower bound we would consider an alternative HKIA scheme ($[2,2,2,2]_\mathrm{II}$) with the INBF scheme in Fig. \ref{fig:HKIA_LB_INBF}(b).  This scheme differs from the previous HKIA scheme only in private message signaling, with $D_1, C_1, C_2, D_2 \sim \mathcal N(0, \frac{P}{4M/3}I_{M/3})$ in (\ref{eq:6_1}) and (\ref{eq:6_1a}).  Following the same steps in Section \ref{Achi_D}, it is easy to show that this HKIA achieves a DoF of $\ETS{8}{4}{8}$, which is strictly greater than $\ETS{8}{4}{4}$ whenever $p_{c\overline d}>0$.  These three lower bounds are illustrated in Fig. \ref{fig:HKIA_LB_6x5}.

Rather than just taking the maximum of $\ETS{7}{6}{6}$ and $\ETS{8}{4}{8}$ to be the new lower bound, we consider a simple way to \emph{blend} these two HKIA schemes that yields an even better bound.  Specifically, let each of $D_1, C_1, C_2$ and $D_2$ now takes $\mathcal N\left(0, \left[\begin{array}{cc} \frac{1}{8}P & 0\\ 0 & \frac{1}{8}P^a\end{array}\right]\right)$ distribution, where $a \in [0, 1]$ is a power tuning factor.  Note that, with $a=1$, it is exactly the $[2,2,2,2]_\mathrm{II}$ HKIA scheme, while it yields the $[1,1,1,1]_\mathrm{II}$ HKIA lower bound when $a=0$.  Clearly, with this scheme, tuning up `$a$' would increase the private DoF but decrease the public DoF, and vice versa.  So by optimizing `$a$' for any given $p_c, p_d$ and $p_{cd}$, we can find a good trade-off between private and public DoF and obtain a lower bound that is guaranteed to be no less than $\max(\ETS{7}{6}{6}, \ETS{8}{4}{8})$.  As shown in Appendix \ref{App_D}, this HKIA scheme achieves a DoF of $6\left(\ETS{1}{1}{1}+\ETS{1}{0}{0} \frac{\ETS{1}{-1}{1}}{\ETS{3}{-2}{2}}\right)$ and is illustrated as `Proposed' in Fig. \ref{fig:HKIA_LB_6x5}.  Compared to the $[1,1,1,1]_\mathrm{II}$ HKIA scheme, this scheme performs markedly better because it trades off between the DoF achievable on the public and private messages by exploring two beamforming options, and fine-tuning the power allocation.

\begin{remark}
In the spirit of trading off between public and private DoF, one may want to introduce another power tuning factor `$b$' and consider $\mathcal N\left(0, \left[\begin{array}{cc} \frac{1}{8}P^b & 0\\ 0 & \frac{1}{8}P^a\end{array}\right]\right)$ for $D_1, C_1, C_2$ and $D_2$.  This, however, is unnecessary, because the optimal value of `$b$' is in fact unity, as shown in Appendix \ref{App_D}.
\end{remark}

\begin{remark}
Note also that we fine-tune the power of the private messages, but not the public messages, because reducing the power of the public messages does not help the private DoF for our successive decoding scheme.
\end{remark}

\begin{remark}
For ease of presentation, we shall reverse the usual order and treat the Type II channel first.  The HKIA scheme for the Type I channel will be shown in Section \ref{Thm3_B}.
\end{remark}

\subsection{HKIA Lower Bound for the Type II Channel} \label{Thm3_A}

For $\frac{2}{3} < \frac{N}{M} \le 1$ and $\frac{p_d}{1+\alpha p_{d|c}} < p_c \le p_d$, when $M$ is a multiple of $3$, the general HKIA scheme blends the $\big[(M-N),(M-N),(M-N),(M-N)\big]_\mathrm{II}$ and the $\big[\frac{M}{3},\frac{M}{3},\frac{M}{3},\frac{M}{3}\big]_\mathrm{II}$  INBF schemes, c.f. Fig. \ref{fig:Type_II_BF}(b) and Fig. \ref{fig:HKIA_LB_INBF}(b).  The random coding scheme takes the same form as the one in Section \ref{Achi_D_2}, with only the following changes to the private message encoding: We employ the INBF in Fig. \ref{fig:HKIA_LB_INBF}(b), where the input vector to each beamforming matrix is partitioned into two groups.  The first $(M-N)$ components are allocated with average power of $P^b$ and are beamformed in such a way that they cause no interference to the unintended receiver.   The rest $(N-2M/3)$ components are transmitted with $P^a$ average power, and their beamforming nulls out interference only partially at the unintended receiver.  More specifically, $D_1, C_1, C_2$ and $D_2$ in (\ref{eq:6_1}) and (\ref{eq:6_1a}) now each takes the following distribution:
\begin{equation} \label{eq:7_1}
D_1, C_1, C_2, D_2 \sim \mathcal N\left(0, \left[\begin{array}{cc} \frac{1}{4(M-N)}P^bI_{M-N} & 0\\ 0 & \frac{1}{4(N-2M/3)}P^aI_{N-2M/3}\end{array}\right]\right) \quad (\mathrm{i.i.d.}),
\end{equation}
where $a, b\in [0,1]$.  This HKIA scheme is extended in Appendix \ref{App_D} for arbitrary $M=3k+q$, where $k$ is an integer and $q\in\{0,1,2\}$.  As a result we have the following proposition, which establishes $\eta_\mathrm{lb,2}$ of Theorem \ref{thm:LB}.

\begin{proposition} \label{prop:HKIA_LB_2}
For the HKIA scheme in Section \ref{Thm3_A} and Appendix \ref{App_D}, the optimal value of `$b$' is $1$.  The optimal value of `$a$' and the achievable sum DoF are as follows ($\beta_2=\frac{q}{k+q}$):

\begin{equation*} 
a_\mathrm{opt} = \left\{\begin{aligned}[rl] 
\frac{\ET{3N-2M}{M-2N}{2N-M}}{(k-M+N)\ET{3}{-2}{2}}, & \quad \frac{p_c}{p_d} \le \frac{1}{1+\beta_2\: p_{d\mid c}}, \\
1, & \quad \frac{p_c}{p_d} > \frac{1}{1+\beta_2\: p_{d\mid c}}. 
\end{aligned} \right.
\end{equation*}

\begin{equation*}
\eta_\mathrm{total} \le \left\{ \begin{aligned}[rl]
M\left( \ET{1}{1}{1}+\ET{1}{0}{0}\frac{\ET{1}{-1}{1}}{\ET{3}{-2}{2}} \right),  & \quad \frac{p_c}{p_d} \le  \frac 1{1+\beta_2\: p_{d|c}},\\
M\ET{1}{1}{1}+k\ET{1}{0}{0},  & \quad \frac{p_c}{p_d} > \frac 1{1+\beta_2\: p_{d|c}}.
\end{aligned} \right.
\end{equation*}

\end{proposition}

Though somewhat complicated by the more sophisticated INBF and power optimization, the proof of Proposition \ref{prop:HKIA_LB_2} follows the same lines as in Section \ref{Achi_D_3}, and is relegated to Appendix \ref{App_D}.  The key effort is to re-evaluate the entropies and DoF based on the new distributions and beamforming and to optimize `$a$' and `$b$'.

\begin{remark}
 It is reasonable that the optimal value of `$b$' is unity.  To see this, recall that the first $(M-N)$ components of $D_1, C_1, C_2$ and $D_2$ are allocated with average power $P^b$ and they cause no interference to the unintended receiver, but not so for the rest $(N-2M/3)$ components.  In other words, the first $(M-N)$ components are better beamformed, and it is more effective to send messages on them.  This explains why `$b$' is maximized and assumes a larger value than `$a$'.
\end{remark}

\subsection{HKIA Lower Bound for the Type I Channel} \label{Thm3_B}

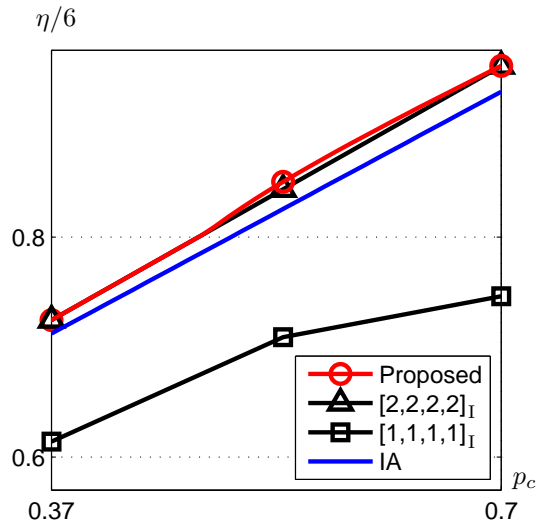
\begin{figure}[b]
\centering
\begin{tikzpicture}
\node[anchor=south west,inner sep=0] at (0,0) {\includegraphics[width=7cm]{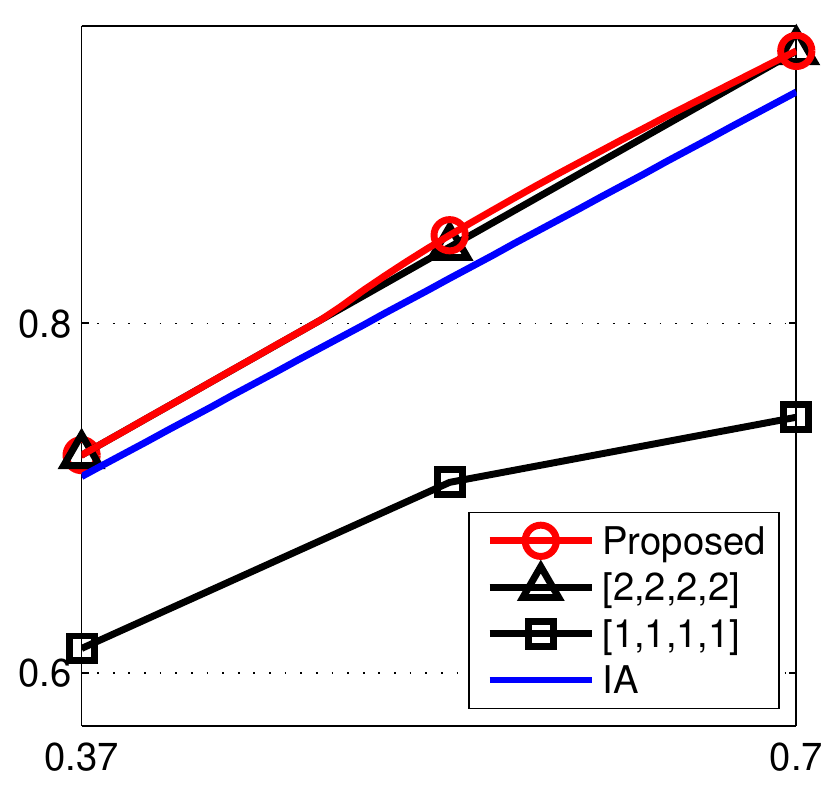}};
\node [right] at (6.7,0.65) {$p_c$};
\node [above] at (0.75,6.5) {$\eta/6$};
\node [right] at (6.1,1.57) {$_\mathrm{I}$};
\node [right] at (6.1,1.17) {$_\mathrm{I}$};
%\node [right] at (5.95,1.37) {$_\mathrm{I}$};
%\node [right] at (5.95,1.0) {$_\mathrm{I}$};
\end{tikzpicture}
\caption{Three HKIA schemes for the 5x6 bursty channel with $p_d=0.7, p_{d|c}=0.9, \frac{p_d}{1+\alpha p_{d|c}}<p_c\le p_d$}
\label{fig:HKIA_LB_5x6}
\end{figure}

\subsubsection{An Example} \label{Thm3_B_1}

To derive a DoF lower bound for the Type I channel with $\frac{2}{3} < \frac{M}{N} \le 1$ and $\frac{p_d}{1+p_{d|c}}<p_c\le p_d$, a natural attempt is to duplicate the schemes on the Type II channel in the previous subsection by constructing HKIA schemes with reciprocal beamforming.  Although this strategy again yields a better lower bound for the Type I channel than the IA scheme, the reciprocity between the HKIA schemes for the Type I and Type II channels is weak, as can be seen by comparing Fig. \ref{fig:HKIA_LB_5x6} to Fig. \ref{fig:HKIA_LB_6x5}.  Here $[1,1,1,1]_\mathrm{I}$ and $[2,2,2,2]_\mathrm{I}$ are the HKIA schemes on the $5 \times 6$ channel with beamforming reciprocal to that for the $[1,1,1,1]_\mathrm{II}$ and $[2,2,2,2]_\mathrm{II}$, respectively, on the $6 \times 5$ channel earlier.  Note that the $[2,2,2,2]_\mathrm{I}$ HKIA again outperforms IA, showing the benefits of introducing public messages, although the gain is not as great as before.  However, somewhat surprisingly the $[1,1,1,1]_\mathrm{I}$ performs strictly worse than the $[2,2,2,2]_\mathrm{I}$, and this is in fact true in general.  Therefore, we base our HKIA scheme for the Type I channel on the $[2,2,2,2]_\mathrm{I}$ HKIA whose beamforming as shown in Fig. \ref{fig:HKIA_LB_INBF}(a), and add power tuning for private messages to it.  (No blending of HKIA schemes here.)  This leads to `Proposed' curve in Fig. \ref{fig:HKIA_LB_5x6} on the $5 \times 6$ channel.  The non-reciprocity between the HKIA schemes on the Type I and Type II channels will be discussed further in Section \ref{Disc}.

\subsubsection{General HKIA Scheme} \label{Thm3_B_2}

The general HKIA scheme for the Type I channel also closely parallels the one for the Type II channel in Section \ref{Achi_D}, with the only difference again being the encoding of the private messages.  Specifically, when $N$ is a multiple of $3$, the $\big[\frac{N}{3},\frac{N}{3},\frac{N}{3},\frac{N}{3}\big]_\mathrm{I}$ INBF scheme shown in Fig. \ref{fig:HKIA_LB_INBF}(a) is employed together with the following random coding scheme:

\begin{equation} \label{eq:7_11}
X_1^n(M_{01}, M_{11}, M_{21})=U_1^n(M_{01})+\left[\begin{array}{c} G_1^{-1} \\ 0 \end{array}\right] \left[\begin{array}{c}D_1^n(M_{11}) \\ C_1^n(M_{21}) \end{array} \right],
\end{equation}
where $U_1 \sim \mathcal{N}\left(0, \frac{P}{2M}I_M\right), C_1 \& D_1 \sim \mathcal{N}\left(0,  \frac{P^a}{4N/3}I_{N/3}\right), a\in [0,1]$ and $G_1=\left[\begin{array}{c} \psi_1^*\hat H_{11} \\ \phi_2^*\hat H_{21} \end{array}\right]$ is the effective channel matrix of the first broadcast channel (i.e. BC1 in Fig. \ref{fig:HKIA_LB_INBF}(a)), with $\hat H_{ji}$ denoting the first $\frac{2}{3}N$ columns of $H_{ji}$.  $X_2^n(M_{02}, M_{12}, M_{22})$ is encoded symmetrically:  
\begin{equation} \label{eq:7_12}
X_2^n(M_{02}, M_{12}, M_{22})=U_2^n(M_{02})+\left[\begin{array}{c} G_2^{-1} \\ 0 \end{array}\right] \left[\begin{array}{c}D_2^n(M_{22}) \\ C_2^n(M_{12}) \end{array} \right],
\end{equation}
where $(U_2, C_2, D_2)$ have the same distributions as $(U_1, C_1,D_1)$, and $G_2$ is the effective channel matrix of BC2.  Appendix \ref{App_E} extends this HKIA scheme for arbitrary $N=3k+q$, where $k$ is an integer and $q\in\{0,1,2\}$, leading to Proposition \ref{prop:HKIA_LB_1}, which establishes $\eta_\mathrm{lb,1}$ of Theorem \ref{thm:LB} and is proved in Appendix \ref{App_E}.

\begin{proposition} \label{prop:HKIA_LB_1}
For the HKIA scheme in Section \ref{Thm3_B} and Appendix \ref{App_E}, the optimal value of `$a$' and the achievable sum DoF are as follows:

\begin{equation*}
a_\mathrm{opt} = \left\{\begin{aligned}[rl] 
1, & \quad \frac{p_c}{p_d} \le  \frac 1{1+\beta_1\:p_{d|c}} \\
\frac{\ET{2M-N}{-M}{M}}{\ET{2M-N}{-2k-\lceil \frac{q}{2} \rceil}{2k+\lceil \frac{q}{2} \rceil}}, & \quad \frac{p_c}{p_d} >  \frac 1{1+\beta_1\:p_{d|c}},
\end{aligned} \right.
\end{equation*}

\begin{equation*}
\eta_\mathrm{total} \le \left\{ \begin{aligned}[rl]
\ET{4k+q}{2k+q}{2M-2k-q},  & \quad \frac{p_c}{p_d} \le  \frac 1{1+\beta_1\:p_{d|c}}\\
\ET{N}{M}{M}+\ET{k}{\left\lfloor \frac{q}{2} \right\rfloor}{-\left\lfloor \frac{q}{2} \right\rfloor}\frac{\ET{2M-N}{-M}{M}}{\ET{2M-N}{-2k-\lceil \frac{q}{2} \rceil}{2k+\lceil \frac{q}{2} \rceil}},  & \quad \frac{p_c}{p_d} > \frac 1{1+\beta_1\:p_{d|c}}, \\
\end{aligned} \right. 
\end{equation*}
where $\beta_1\triangleq \frac{2M-4k-q}{2k+q}.$

\end{proposition}

\section{Discussion} \label{Disc}
%auto-ignore

We discuss some open and subtle issues in this section, starting with the non-reciprocity of the HKIA scheme on the Type I and the Type II channel, as observed in the previous section.

\subsection{The non-reciprocity of the HKIA scheme on the Type I and the Type II channel} \label{Disc_A}

\begin{figure}[h]
\centering
\includegraphics[width=14.5cm]{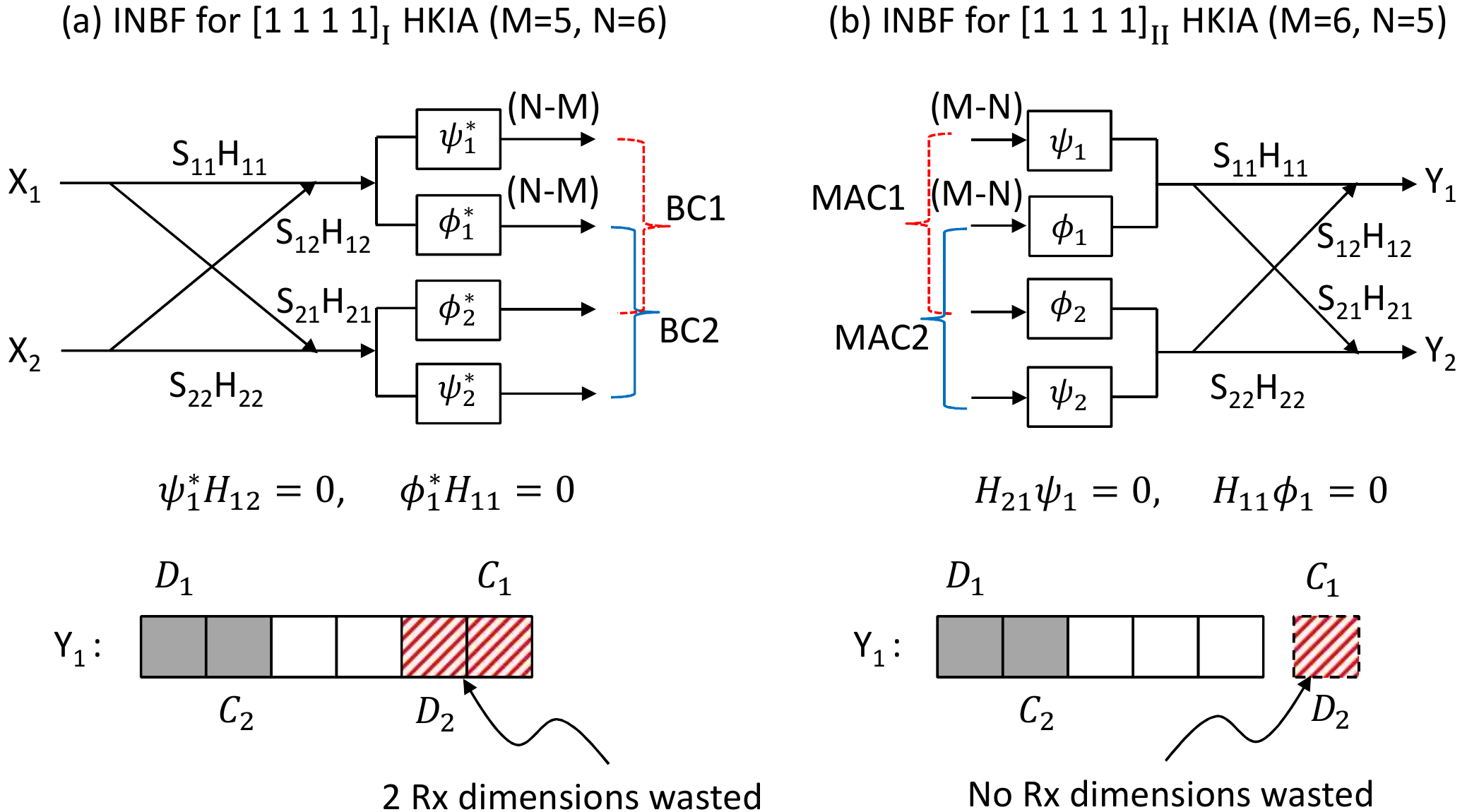}
\caption{The INBF efficiency for the $[1,1,1,1]_\mathrm{I}$ and the $[1,1,1,1]_\mathrm{II}$ HKIA scheme}
\label{fig:HKIA_INBF_Efficiency}
\end{figure}

The non-reciprocity of the HKIA schemes is not due to the fundamental BC-MAC difference as shown in Section \ref{Fund_B}.  Rather, it is because introducing public messages is more effective on the Type II channel than on the Type I channel.  To see this, let us look into the striking difference between the $[1,1,1,1]_\mathrm{II}$ and the $[1,1,1,1]_\mathrm{I}$ HKIA scheme in Fig. \ref{fig:HKIA_LB_6x5} and Fig. \ref{fig:HKIA_LB_5x6} for the $6 \times 5$ and the $5 \times 6$ channel.  First of all, note that the INBF scheme of the $[1,1,1,1]_\mathrm{I}$ and the $[1,1,1,1]_\mathrm{II}$ HKIA as shown in Fig. \ref{fig:HKIA_INBF_Efficiency} are clearly reciprocal, decomposing the channel into two orthogonal MACs and BC, respectively.  Moreover, it is easily verified that they also achieve \emph{identical} private DoF, i.e. $2(p_c+p_d)=2\ETS{2}{1}{1}$!

However, the achievable public DoF is markedly different: $\min(\ETS{6}{0}{8}, \ETS{3}{4}{4})$ is achievable by the $[1,1,1,1]_\mathrm{II}$ but only $\min(\ETS{4}{0}{6}, \ETS{2}{3}{3})$ is achieved by the $[1,1,1,1]_\mathrm{I}$.  The reason for the low public DoF of the $[1,1,1,1]_\mathrm{I}$ HKIA is actually obvious upon an inspection of the received private signal dimensions at each receiver.   Recall that, from (\ref{eq:6_1}), (\ref{eq:6_1a}), (\ref{eq:7_11}) and (\ref{eq:7_12}), $D_1$ and $C_2$ carry the messages to Rx1, while $D_2$ and $C_1$ convey the messages to Rx2.  With the $[1,1,1,1]_\mathrm{II}$ INBF, the desired signals $D_1$ and $C_2$ take two dimensions at Rx1, while undesired $D_2$ and $C_1$ are aligned and nulled, consuming no Rx dimensions, as illustrated in Fig. \ref{fig:HKIA_INBF_Efficiency}(b).  On the other hand, since the $[1,1,1,1]_\mathrm{I}$ INBF is done on the receivers, while desired $D_1$ and $C_2$ signals again occupy two Rx dimensions, the undesired $D_2$ and $C_1$ signals are not nulled at Rx1, when we decode the public messages.  As a result, $D_2$ and $C_1$ present two extra dimensions of interference (noise) to the public messages at Rx1, thus impairing the achievable public DoF, as illustrated in Fig. \ref{fig:HKIA_INBF_Efficiency}(a).  The situation at Rx2 is the same.  Therefore, the INBF scheme works efficiently for the public message decoding on the Type II channel, but not so on the Type I channel.  This explains why the $[1,1,1,1]_\mathrm{II}$ achieves higher DoF than the $[1,1,1,1]_\mathrm{I}$, and is even DoF-optimal on the Type II channel in part of Regime 2 as shown in Section \ref{Achi_D}.

\begin{figure}[h]
\centering
\includegraphics[width=15cm]{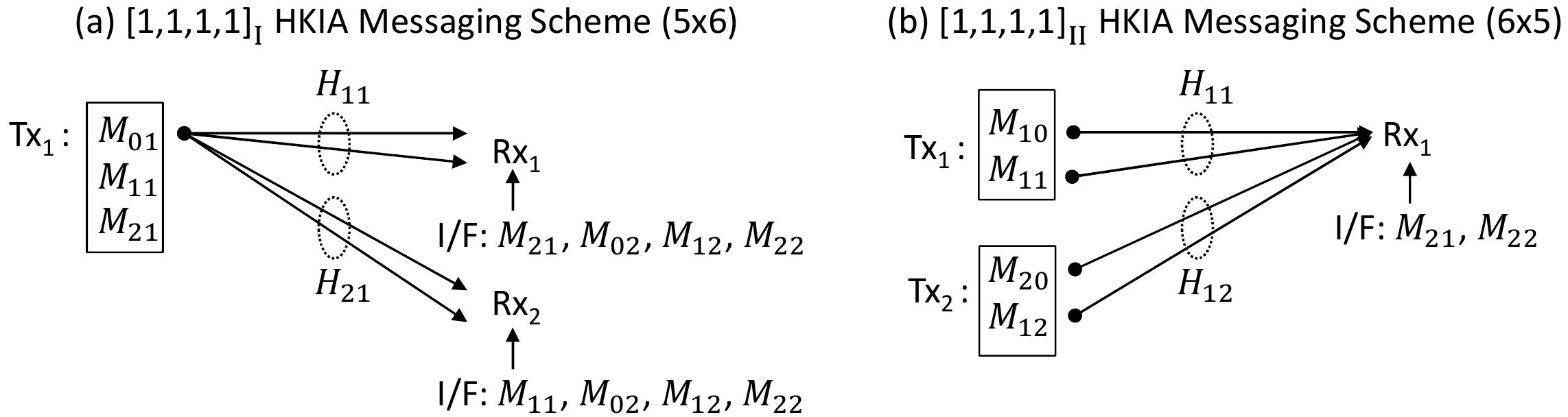}
\caption{Comparison of the $[1,1,1,1]_\mathrm{I}$ and the $[1,1,1,1]_\mathrm{II}$ HKIA messaging scheme (I/F: interference)}
\label{fig:HKIA_Msg_Compare}
\end{figure}

The non-reciprocity of the HKIA schemes can also be seen by comparing the messaging schemes of the $[1,1,1,1]_\mathrm{I}$ and the $[1,1,1,1]_\mathrm{II}$ HKIA, as illustrated in Fig. \ref{fig:HKIA_Msg_Compare}, where we view the $[1,1,1,1]_\mathrm{I}$ messages from a BC perspective and the $[1,1,1,1]_\mathrm{II}$ messages from a MAC perspective.  Recall that the $[1,1,1,1]_\mathrm{II}$ HKIA decodes the public messages ($M_{01}$ and $M_{02}$) first as a MAC, and then decodes the private messages ($M_{11}$ and $M_{12}$) as another MAC.  In terms of the BC-MAC duality, the $[1,1,1,1]_\mathrm{I}$ HKIA is apparently not dual to the $[1,1,1,1]_\mathrm{II}$.  However, from this perspective, it is also unclear how to make corrections to the $[1,1,1,1]_\mathrm{I}$ HKIA to make it dual.

\subsection{The frequency hopping scenario}

There is a subtlety that warrants some attention when we apply our results to applications where frequency hopping is allowed, as mentioned in Section \ref{Intro}.  Recall that our definition of DoF essentially measures the spectral efficiency of the system, c.f. (\ref{eq:2_1}).  Since the MIMO X channel is spectrally more efficient than the respective point-to-point MIMO channel, having the two user pairs hop to two different frequency bands leads to lower spectral efficiency.  So when there are two frequency bands available, we should employ cross-link messaging in both bands, instead of frequency hopping, if higher spectral efficiency is desired.  However, an implicit premise of the above argument is the availability of multi-carrier capability in the system.

For applications or scenarios without multi-carrier capability, frequency hopping still provides a simple way to increase the sum rate of the two user pairs, e.g. in the earlier IEEE 802.11a/b/g wireless LANs.  In this case, frequency hopping may be one of the reasons for the absence of the cross-links in our channel model, and we may lend our results to this scenario.  Nonetheless, it should be understood that the DoF computed now is a special normalized  sum rate over possibly two frequency bands, not the usual spectral efficiency on the two frequency bands.  For example, suppose each transmitter and receiver in the system is designed for 20MHz bandwidth and we have a pool of 20MHz frequency bands for frequency hopping.  Then the DoF ($\eta$) signifies that, at high SNR, the system capacity is approximately the total capacity of $\eta$ AWGN channels, each of 20MHz bandwidth, not 40MHz.

\section{Conclusion} \label{Conc}
%auto-ignore

In this paper, we make progress on extending our knowledge of the sum DoF of the MIMO X channel, by considering a simple bursty channel model without feedback.  The sum DoF of this bursty channel is fully characterized  in Regime 1, and is partially characterized in Regime 2.  The DoF characterization for $\frac{2}{3} < \frac{M}{N} < \frac{\ETS{3}{-2}{2}}{\ETS{2}{-1}{1}}$ in Regime 2 remains a quest.

We also show striking differences between the bursty and the non-bursty MIMO X channel.  In particular, burstiness of the channel creates new channel states, which may be under-utilized by the IA schemes.  Further rate splitting and introducing public messages can achieve strictly higher DoF than the IA schemes, and is even DoF-optimal in some cases.  Moreover, the Type I and the Type II channel are no longer reciprocal under this bursty channel model.  And the DoF of the bursty X channel does not saturate when $\frac{\min(M,N)}{\max(M,N)}$ reaches $\frac{2}{3}$.

While the IA schemes still play an important role on the bursty MIMO X channel, new coding schemes are needed to cope with the extra channel states.  When $p_c$ is small (Regime 1), treating crosslink interference as erasure/noise turns out to be DoF-optimal.  As $p_c$ grows (Regime 2), however, the paradigm shifts from avoiding crosslinks to exploiting them, and the usual IA schemes become useful.  Furthermore, due to the fundamental difference between bursty BC and MAC with the burstiness known only to the receivers, the Type II channel needs more complex beamforming schemes than the Type I channel.  The Type II channel even needs to overlay the IA signals with public messages as in HKIA to make full use of the channel states and achieve the DoF in Regime 2.  When $\frac{M}{N}$ is close to 1 in Regime 2, HKIA also proves to be beneficial on both the Type I and the Type II channel.


\appendices
\section{Proof of Sum DoF Upper Bound for Regime 2} \label {App_A}
%auto-ignore

For $\abs{\mathcal I_u^A} > \abs{\mathcal I_u^B}$, c.f. Fig. \ref{fig:State_Sequence_Pairing}(b), (\ref{eq:5_4}) is similarly bounded by
\begin{align*}
h&(Y_1^n\mid S^n=u^n, M_{21})-h((S_{21}H_{21}X_1+Z_2)^n \mid S^n=v^n,  M_{21}) \\
& \le  h(\Omega_1\mid \Omega_2) - h(Z_2^n) \\
& \overset{(a)}\le h\left((H_{11}X_1+H_{12}X_2+Z_1)^{\mathcal I_v^B}\right) + h\left((H_{11}X_1+H_{12}X_2+Z_1)^{\mathcal I_u^A \setminus \mathcal I_v^B}\mid (H_{21}X_1+Z_2)^{\mathcal I_u^A \setminus \mathcal I_v^B}\right) \\
& \quad + h\left((H_{11}X_1+Z_1)^{\mathcal I_u^{BC}}\mid (H_{21}X_1+Z_2)^{\mathcal I_u^{BC}}\right)  + h\left((H_{12}X_2+Z_1)^{\mathcal I_u^D}\right), \mpelabel{eq:A_1}
\end{align*}
where (a) follows from the chain rule and the fact the conditioning reduces entropy. And again due to symmetry we also have
\begin{align*} 
h&(Y_2^n \mid S^n=v^n, M_{12})-h((S_{12}H_{12}X_2+Z_1)^n \mid S^n=u^n,  M_{12}) \\
& \overset{(a)}\le h\left((H_{21}X_1+H_{22}X_2+Z_2)^{\mathcal I_u^B}\right) + h\left((H_{21}X_1+H_{22}X_2+Z_2)^{\mathcal I_v^A \setminus \mathcal I_u^B}\mid (H_{12}X_2+Z_1)^{\mathcal I_v^A \setminus \mathcal I_u^B}\right) \\
& \quad + h\left((H_{22}X_2+Z_2)^{\mathcal I_v^{BC}}\mid (H_{12}X_2+Z_1)^{\mathcal I_v^{BC}}\right)  + h\left((H_{21}X_1+Z_2)^{\mathcal I_v^D}\right).  \mpelabel{eq:A_2}
\end{align*}

Following the same single-letterization steps in Section \ref{sec_5D}, each of the entropy terms in (\ref{eq:A_1}) can then be bounded by 
\begin{align*}
h&\left((H_{11}X_1+H_{12}X_2+Z_1)^{\mathcal I_v^B}\right) \le \left|\mathcal I_v^B\right|h(H_{11}X_{1G_{\hat\alpha}}+H_{12}X_{2G_{\hat\alpha}}+Z_1), \\
h&\left((H_{11}X_1+H_{12}X_2+Z_1)^{\mathcal I_u^A \setminus \mathcal I_v^B}\mid (H_{21}X_1+Z_2)^{\mathcal I_u^A \setminus \mathcal I_v^B}\right) \\
& \le \left(\left|\mathcal I_u^A\right|-\left|\mathcal I_v^B\right|\right) h(H_{11}X_{1G_{\hat\beta}}+H_{12}X_{2G_{\hat\beta}}+Z_1 \mid H_{21}X_{1G_{\hat\beta}}+Z_2), \\
h&\left((H_{11}X_1+Z_1)^{\mathcal I_u^{BC}}\mid (H_{21}X_1+Z_2)^{\mathcal I_u^{BC}}\right) \le \left(\left|\mathcal I_u^B\right|+\left|\mathcal I_u^C\right|\right)h((H_{11}X_{1G_{\hat\gamma}}+Z_1)\mid (H_{21}X_{1G_{\hat\gamma}}+Z_2), \textrm{ and} \\
h&\left((H_{12}X_2+Z_1)^{\mathcal I_u^D}\right) \le \left|\mathcal I_u^D\right|h((H_{12}X_{2G_{\hat\delta}}+Z_1),  \mpelabel{eq:A_3}
\end{align*}
where $X_{1G_{\hat\alpha}}, X_{2G_{\hat\alpha}}, X_{1G_{\hat\beta}}, X_{2G_{\hat\beta}}, X_{1G_{\hat\gamma}}$, and $X_{2G_{\hat\delta}}$ parallel the respective Gaussian random vectors in Section \ref{sec_5D}.

Therefore, due to the law of large numbers and Fact \ref{fact_dof_ub}, it follows that, for $P_A > P_B$, as $n \rightarrow \infty$
\begin{align} \begin{split} \label {eq:A_4}
\frac{1}{n}&\mathcal D\left[ h(Y_1^n \mid S^n, M_{21}) - h((SH_{21}X_1+Z_2)^n \mid S^n,  M_{21}) \right] \\
& \le P_B\min(2M,N) + (P_A-P_B)\min(M,N)+(P_B+P_C)[\min(M,2N)-\min(M,N)]+P_D\min(M,N) \\
& = P_B\min(2M,N)+(P_B+P_C)\min(M,2N)+(P_A-2P_B)\min(M,N) \\
& = (p_d-p_c)\min(2M,N)+(p_d-p_{cd})\min(M,2N)+(p_{cd}-2p_d+2p_c)\min(M,N) \quad \textrm{(a.s.).}
\end{split} \end{align}
Since (\ref{eq:A_2}) is symmetric to (\ref{eq:A_1}), in light of (\ref{eq:5_3}) we have proved Theorem \ref{thm:UB} for $p_c > \frac{p_d}{1+p_{d|c}}$, which is equivalent to $P_A > P_B$.

\section{The rank of $\left( \theta_1^*H_{12}\phi_2 \right)$ in Type II, Regime 1 Beamforming} \label {App_B}
%auto-ignore

The fact that $\theta_1^*H_{12}\phi_2$ is full-rank (a.s.) can be seen as follows.  First of all, note that $\theta_1^*H_{12}$ is full-rank (a.s.).  Defining $A\triangleq \theta_1^*H_{12}$, it suffices to show that $\mathcal P\{A\phi_2 \textrm{ is singular}\}=0$.  Suppose $\phi_2=\left[u_1\:u_2\: ...\: u_{M-N} \right],$ where $u_1,u_2, ..., u_{M-N}$ are chosen randomly, independently, and uniformly within the unit ball of the null space of $H_{22}$.  (Note that $\phi_2$ is a function of $H_{22}$, independent of $A$, and is full-rank (a.s.).)  Then $\{A\phi_2 \textrm{ is singular}\}$ implies that some column of $A\phi_2$ must be a linear combination of the rest columns.  Without loss of generality, assume $Au_1 =c_2Au_2 + c_3Au_3+ ... +c_{M-N}Au_{M-N}$ for some $c_2, c_3, ..., c_{M-N}$.  Given $A, u_2, u_3, ..., u_{M-N}$, this would constrain $u_1$ to $M-(M-N)+(M-N-1)=M-1$ dimensions, as there are $(M-N)$ linear constraints and $(M-N-1)$ free variables.  However, $H_{22}$ consists of i.i.d. elements drawn from a continuous distribution, so by symmetry its null space (and hence $u_1, u_2, ..., u_{M-N}$) must spread over $M$ dimensions.  Moreover, given $A, u_2, u_3, ...,$ and $u_{M-N}$, it is not hard to see that $u_1$ is still continuously distributed in the $M$-dimensional space.  Therefore, $\mathcal P\{u_1 \in \: (M-1)\textrm{ dimensional subspace}\mid A, u_2, u_3, ..., u_{M-N}\}=0$, and hence $\mathcal P\{Au_1 \in \mathrm{Col}[Au_2\:Au_3\: ...\: Au_{M-N}]\}=0$.  It then follows easily that $\mathcal P\{A\phi_2 \textrm{ is singular}\}=0$.

\section{Proof of (\ref{eq:6_3})} \label {App_C}
%auto-ignore

\begin{fact} \label{fact_1}
If $X$ is a Gaussian random vector distributed according to $\mathcal N(0, (kP)I_M)$ for some positive constant $k$, $H$ is a $N \times M$ matrix, and $Z \sim \mathcal N(0, I_N),$ then $\mathcal D\{h(HX+Z)\} = \mathrm{rank}(H).$
\end{fact}

We show $\mathcal D\{h(Y_1\mid S_1, U_1)\} =\langle N, M-N, N\rangle$ (a.s.) here, using the simple fact above.  Other terms in Section \ref{Thm2} and other places can be proved along similar lines.  First of all, letting $V_1\triangleq \psi_1 D_1+\phi_1 C_1$ denote the random vector carrying the private messages at Tx1, we have
\begin{align} \begin{split} \label{eq:C_1}
\mathcal D&\{h(Y_1\mid S_1, U_1)\} \\
& =\mathcal D\left\{ p_{cd}h(H_{11}V_1+H_{12}X_2+Z_1)+p_{\overline cd}h(H_{11}V_1+Z_1)+p_{c\overline d}h(H_{12}X_2+Z_1) +(1-p_c-p_d+p_{cd})h(Z_1)\right\} \\
& \overset{(a)}=\mathcal D\left\{ p_{cd}h(H_{11}\psi_1 D_1+H_{12}X_2+Z_1)+p_{\overline cd}h(H_{11}\psi_1 D_1+Z_1)+p_{c\overline d}h(H_{12}X_2+Z_1) \right\}
\end{split} \end{align}
where (a) follows because $H_{11}\phi_1=0$ and the noise term can be dropped in DoF computation.  Note that 
$$ H_{11}\psi_1 D_1+H_{12}X_2 = H_{11}\psi_1 D_1+H_{12}\phi_2 C_2+ H_{12}U_2 
= \left[ H_{11} \: H_{12}\phi_2 \: H_{12} \right] \left[ \begin{array}{c} D_1 \\ C_2 \\ U_2 \end{array}\right],$$
so we must have $\mathcal D\left\{h(H_{11}\psi_1 D_1+H_{12}X_2+Z_1)\right\} \le N.$  On the other hand, we are also lower bounded by 
$\mathcal D\left\{h(H_{11}\psi_1 D_1+H_{12}X_2+Z_1)\right\}  \ge \mathcal D\left\{h(H_{12}U_2+Z_1)\right\} =  N$ (a.s.).   Hence we conclude that $\mathcal D\big\{h(H_{11}\psi_1 D_1 +H_{12}X_2 +Z_1)\big\} =N$ (a.s.).  For the same reason, obvioulsy $\mathcal D\big\{h(H_{12}X_2 +Z_1)\big\} =N$ (a.s.).  Finally, it follows easily from Fact \ref{fact_1} that $\mathcal D\big\{h(H_{11}\psi_1 D_1+Z_1)\big\}=\mathrm{rank}(H_{11}\psi_1)=M-N$ (a.s.).  So the proof is completed.

\section{Proof of Proposition \ref{prop:HKIA_LB_2}: HKIA lower bound for the Type II channel} \label {App_D}
%auto-ignore

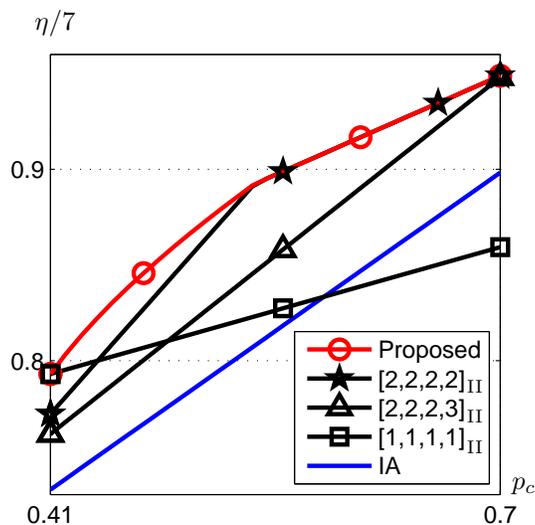
\begin{figure}[b]
\centering
\begin{tikzpicture}
\node[anchor=south west,inner sep=0] at (0,0) {\includegraphics[width=7cm]{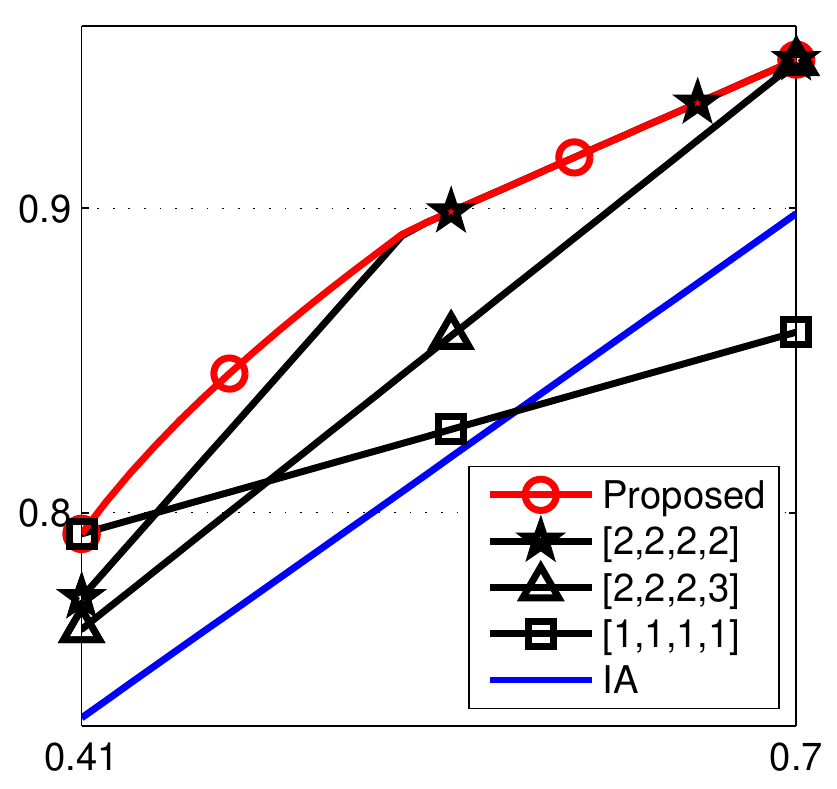}};
\node [right] at (6.7,0.65) {$p_c$};
\node [above] at (0.75,6.5) {$\eta/7$};
\node [right] at (6.1,1.97) {$_\mathrm{II}$};
\node [right] at (6.1,1.57) {$_\mathrm{II}$};
\node [right] at (6.1,1.17) {$_\mathrm{II}$};
\end{tikzpicture}
\caption{HKIA schemes for the $7 \times 6$ bursty channel with $p_d=0.7, p_{d|c}=0.9, \frac{p_d}{1+ p_{d|c}}<p_c\le p_d$}
\label{fig:HKIA_LB_7x6}
\end{figure}

\subsubsection{Extending the HKIA scheme for arbitrary $M$}
The key issue in extending the HKIA scheme in Section \ref{Thm3_A} to the Type II channel with arbitrary $M=3k+q$ (i.e. $k=\lfloor \frac{M}{3} \rfloor$ and $q\in\{0,1,2\}$) is: What changes should be made to the INBF scheme in Fig. \ref{fig:HKIA_LB_INBF}(b) when $M/3$ is not an integer?  Since the HKIA lower bound in Section \ref{Achi_D} is partially tight for $N/M > 2/3$ in Regime 2, naturally we would continue to include its INBF scheme (Fig. \ref{fig:Type_II_BF}(b)) as we did in Section \ref{Thm3_A}.  Similarly, the usual INBF scheme for the non-bursty channel \cite{MMK08} can be employed to enhance the lower bound when $p_c$ becomes large, following the rationale in Section \ref{Thm3_A}.  To better appreciate these two HKIA schemes, we apply them on the $7 \times 6$ channel and illustrate the achieved DoF in Fig. \ref{fig:HKIA_LB_7x6} as $[1,1,1,1]_\mathrm{II}$ and $[2,2,2,3]_\mathrm{II}$.  Note that the $[2,2,2,3]_\mathrm{II}$ HKIA, which incorporates the usual INBF for the non-bursty channel, again outperforms IA and complements the $[1,1,1,1]_\mathrm{II}$ HKIA when $p_c$ is large.

Somewhat surprisingly, however, although the $[2,2,2,3]_\mathrm{II}$ INBF scheme is the best known scheme for the non-bursty $7 \times 6$ channel, it is not the best one to use here.  To see this, consider an HKIA with the $[2,2,2,2]_\mathrm{II}$ INBF which is similar to Fig. \ref{fig:HKIA_LB_INBF}(b), except the dimensions of the $\psi$ and $\phi$ INBF matrices are now $7 \times 2$ each, and the $\hat H_{ij}$ represents 5 rows of the respective $H_{ij}$ channel matrix.   As shown in the figure, the $[2,2,2,2]_\mathrm{II}$ HKIA achieves a strictly higher DoF than the $[2,2,2,3]_\mathrm{II}$ HKIA, although the $[2,2,2,2]_\mathrm{II}$ INBF is clearly suboptimal when the channel is non-bursty.
As an astute reader may have thought about, this seemingly counter-intuitive result is in fact reasonable if we examine the utilization efficiency of received signal dimensions as we did in Section \ref{Disc_A}, c.f. Fig. \ref{fig:HKIA_INBF_Efficiency}(b).  Specifically, compared to the $[2,2,2,2]_\mathrm{II}$ INBF, the $[2,2,2,3]_\mathrm{II}$ is less efficient in utilizing the Rx signal dimensions because it consumes one more Rx dimension but that benefits only the Tx2-Rx2 link while causing more interference to the other three links.  By freeing up this dimension and giving it to the public messages, $[2,2,2,2]_\mathrm{II}$ HKIA is understandably more efficient.  Our proposed HKIA scheme therefore blends the $[2,2,2,2]_\mathrm{II}$ and $[1,1,1,1]_\mathrm{II}$ INBF, and the achievable DoF on the $7 \times 6$ channel is shown in Fig. \ref{fig:HKIA_LB_7x6}.

The general HKIA scheme for arbitrary $M$ blends the $\big[(M-N),(M-N),(M-N),(M-N)\big]_\mathrm{II}$ and the $\left[k,k,k,k\right]_\mathrm{II}$ INBF, similarly to what was done in Section \ref{Thm3_A}.   The $\big[(M-N),(M-N),(M-N),(M-N)\big]_\mathrm{II}$ INBF is just the one illustrated in Fig. \ref{fig:Type_II_BF}(b), while the $\left[ k,k,k,k \right]_\mathrm{II}$ INBF is similar to Fig. \ref{fig:HKIA_LB_INBF}(b), except the $\psi$ and $\phi$ matrices are now $M \times k$ matrices, and the $\hat H_{ij}$ denotes $M-k$ rows of $H_{ij}$.  Specifically, the random coding in (\ref{eq:6_1}) and (\ref{eq:6_1a}) now has
\begin{equation} \label{eq:D_1}
D_1, C_1, C_2, D_2 \sim \mathcal N\left(0, \left[\begin{array}{cc} \frac{1}{4(M-N)}P^bI_{M-N} & 0\\ 0 & \frac{1}{4A}P^aI_A\end{array}\right]\right) \quad (\mathrm{i.i.d.}),
\end{equation}
where $A=k-(M-N)$, and $a, b\in[0,1]$.  (It is easy to verify that $A \ge 0$.)  

\subsubsection{Computation of the achievable sum DoF}

The analysis of the achievable DoF follows the same procedure for the HKIA scheme in Section \ref{Achi_D_3}.  First, the following private DoF is clearly achievable:
\begin{equation} \label{eq:D_2}
\eta_\mathrm{priv} \le 2g(a,b)\langle 2,1,1 \rangle,
\end{equation}
where $g(x, y)\triangleq  x A+y(M-N)$ is a function introduced to simplify the expressions in this section.  To compute the public DoF, we first re-evaluate (\ref{eq:6_3}) and (\ref{eq:6_4}), and it is not hard to verify that they now become 
\begin{align} \begin{split} \label{eq:D_3}
\mathcal D\{h(Y_1\mid S_1)\}				&=\langle N, N, N\rangle, \\
\mathcal D\{h(Y_1\mid S_1, U_1)\}			&=\langle N, g(2a,b), N\rangle, \\
\mathcal D\{h(Y_1\mid S_1, U_2)\}			&=\langle N, N, g(2a,b)\rangle, \textrm{ and}\\
\mathcal D\{h(Y_1\mid S_1, U_1, U_2)\}	&=\langle g(3a,2b), g(2a,b), g(2a,b)\rangle,
\end{split} \end{align}
where $g(3a,2b)=g(2a,2b)+g(a,0)$ in the last equality.  $g(2a,2b)$ is due to $D_1$ and $C_2$, and $g(a,0)$ is due to the residual interference of $C_1$ and $D_2$ (aligned in $A$ dimensions at $Y_1$.)  By symmetry, $\mathcal D\{h(Y_2\mid S_2)\}, \mathcal D\{h(Y_2\mid S_2, U_2)\}, \mathcal D\{h(Y_2\mid S_2, U_1)\},$ and $\mathcal D\{h(Y_2\mid S_2, U_1, U_2)\}$ can again be obtained readily.  With these computed, (\ref{eq:6_2}) in turn evaluates to
\begin{align} \begin{split} \label{eq:D_4}
\eta_{01}, \: \eta_{02} &\le \langle N,0,N \rangle - \langle g(3a,2b), 0, g(2a,b)\rangle, \textrm{ and} \\
\eta_{01}+\eta_{02} &\le \langle N,N,N \rangle - \langle g(3a,2b), g(2a,b), g(2a,b)\rangle.
\end{split} \end{align}

It then follows that
\begin{equation} \label{eq:D_5}
\eta_\mathrm{pub} \le \min\Big(2\Big[\langle N,0,N \rangle - \langle g(3a,2b), 0, g(2a,b)\rangle\Big], \quad
\langle N,N,N \rangle - \langle g(3a,2b), g(2a,b), g(2a,b)\rangle \Big),
\end{equation}
is achievable, and hence the total achievable DoF ($\eta_\mathrm{total} = \eta_\mathrm{priv}+\eta_\mathrm{pub}$) is
\begin{equation} \label{eq:D_6}
\eta_\mathrm{total} \le \min\Big(2\Big[\langle N,0,N \rangle + \langle g(-a,0), g(a,b), g(-a,0)\rangle\Big], \quad
\langle N,N,N \rangle + \langle g(a,2b), g(0,b), g(0,b)\rangle \Big).
\end{equation}

\subsubsection{Optimization of `$a$' and `$b$'}

Note that 
\begin{align} \begin{split} \label{eq:D_7}
\langle g(-a,0), g(a,b), g(-a,0)\rangle &= -aA\langle 1,-1,1\rangle+b(M-N)\langle 0, 1, 0\rangle, \textrm{ and} \\
\langle g(a,2b), g(0,b), g(0,b)\rangle &= aA\langle 1,0,0\rangle+b(M-N)\langle 2, 1, 1\rangle.
\end{split} \end{align}
Since the coefficients of `$b$' is nonnegative in both terms, we should maximize `$b$' and its optimal value is $b=1$.  Substituting $b=1$ into (\ref{eq:D_7}) and (\ref{eq:D_6}), the total achievable DoF ($\eta_\mathrm{total}$) becomes
\begin{equation} \label{eq:D_8}
\min\Big(2\Big[\langle N,M-N,N \rangle - aA\langle 1,-1,1\rangle\Big], \quad
\langle 2M-N,M,M \rangle + aA\langle 1,0,0\rangle \Big).
\end{equation}
Recall that we are evaluating the DoF lower bound in Regime 2.  From (\ref{eq:D_8}), it is straightforward to show that if $\ETS{3N-2M}{M-2N}{2N-M} > 0$ or equivalently $p_c/p_d > 1/(1+\alpha p_{d|c})$, we have
\begin{equation} \label{eq:D_9}
a_\mathrm{opt} = \left\{\begin{aligned}[rl] 
\frac{\ET{3N-2M}{M-2N}{2N-M}}{A\ET{3}{-2}{2}}, & \quad \frac{p_c}{p_d} \le \frac{1}{1+\beta_2\: p_{d\mid c}}, \\
1, & \quad \frac{p_c}{p_d} > \frac{1}{1+\beta_2\: p_{d\mid c}},
\end{aligned} \right.
\end{equation}
where $\beta_2=\frac{q}{k+q}$.  Substituting (\ref{eq:D_9}) into (\ref{eq:D_8}) and simplifying the expressions, it is not hard to verify that
\begin{equation} \label{eq:D_10}
\eta_\mathrm{total} \le \left\{ \begin{aligned}[rl]
M\left( \ET{1}{1}{1}+\ET{1}{0}{0}\frac{\ET{1}{-1}{1}}{\ET{3}{-2}{2}} \right),  & \quad \frac{p_c}{p_d} \le  \frac 1{1+\beta_2\: p_{d|c}}\\
M\ET{1}{1}{1}+k\ET{1}{0}{0},  & \quad \frac{p_c}{p_d} > \frac 1{1+\beta_2\: p_{d|c}}, \\
\end{aligned} \right.
\end{equation}
is achievable, where we have used the identity $\ET{1}{1}{1}+\ET{1}{0}{0}\frac{\ET{1}{-1}{1}}{\ET{3}{-2}{2}}=\ET{2}{1}{1}-\ET{1}{0}{0}\frac{\ET{2}{-1}{1}}{\ET{3}{-2}{2}}$.  Thus we have proved Proposition \ref{prop:HKIA_LB_2}.

\begin{remark}
Since the following three inequalities are equivalent:
\begin{equation*}
\ETS{3N-2M}{M-2N}{2N-M} > 0
\quad \equiv \quad \frac{p_c}{p_d} > \frac{1}{1+\alpha p_{d|c}}
\quad \equiv \quad \frac{N}{M} > \frac{\ETS{2}{-1}{1}}{\ETS{3}{-2}{2}},
\end{equation*}
the sum DoF lower bound in Proposition \ref{prop:HKIA_LB_2} can alternatively be viewed to hold when $\frac{\ET{2}{-1}{1}}{\ET{3}{-2}{2}} < \frac{N}{M} \le 1$ in Regime 2.  (Note that $\frac{2}{3}\le \frac{\ET{2}{-1}{1}}{\ET{3}{-2}{2}}$.)  
\end{remark}
\begin{remark} Moreover, when $\frac{p_c}{p_d} \le \frac{1}{1+\alpha p_{d|c}}$, it is easily verified that the optimal value of $a_\mathrm{opt}$ is zero, and $\eta_\mathrm{total}$ reduces to $2\ET{N}{M-N}{N}$, which is just the DoF achieved by the HKIA in Section \ref{Achi_D_2} when $\frac{N}{M} > \frac{2}{3}$!  Therefore, the HKIA scheme for $\frac{N}{M} > \frac{2}{3}$ in Section \ref{Achi_D_2} may be viewed as a special case of the scheme here.
\end{remark}

\section{Proof of Proposition \ref{prop:HKIA_LB_1}: HKIA lower bound for the Type I channel} \label {App_E}
%auto-ignore

\begin{figure}[b]
\centering
\begin{tikzpicture}
\node[anchor=south west,inner sep=0] at (0,0) {\includegraphics[width=7cm]{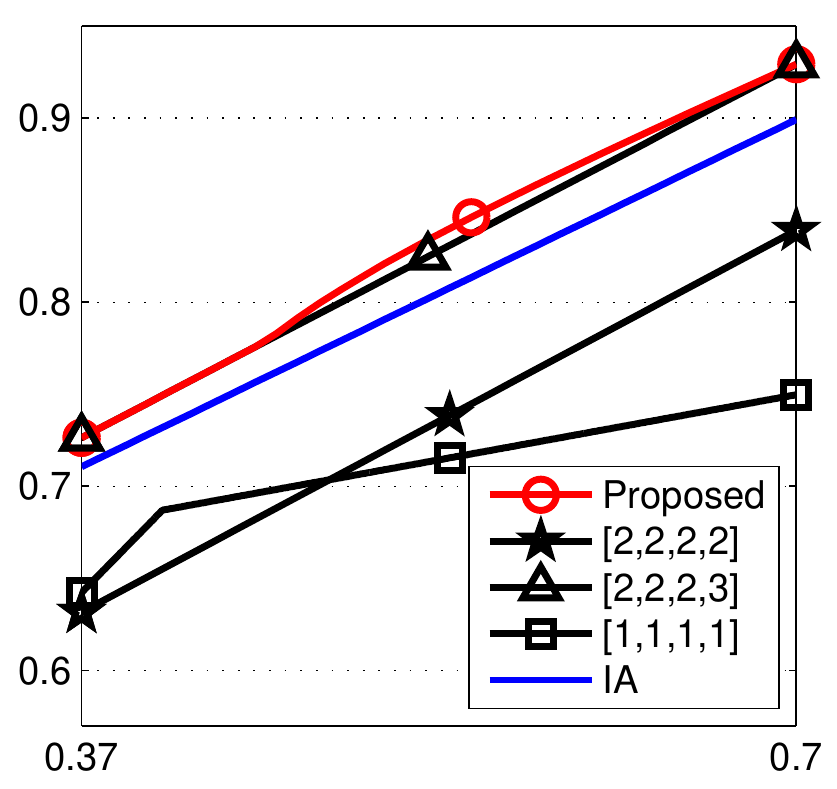}};
\node [right] at (6.7,0.65) {$p_c$};
\node [above] at (0.75,6.5) {$\eta/7$};
\node [right] at (6.1,1.97) {$_\mathrm{I}$};
\node [right] at (6.1,1.57) {$_\mathrm{I}$};
\node [right] at (6.1,1.17) {$_\mathrm{I}$};
%\node [right] at (5.95,1.78) {$_\mathrm{I}$};
%\node [right] at (5.95,1.41) {$_\mathrm{I}$};
%\node [right] at (5.95,1.04) {$_\mathrm{I}$};
\end{tikzpicture}
\caption{HKIA schemes for the $6 \times 7$ bursty channel with $p_d=0.7, p_{d|c}=0.9, \frac{p_d}{1+ p_{d|c}}<p_c\le p_d$}
\label{fig:HKIA_LB_6x7}
\end{figure}
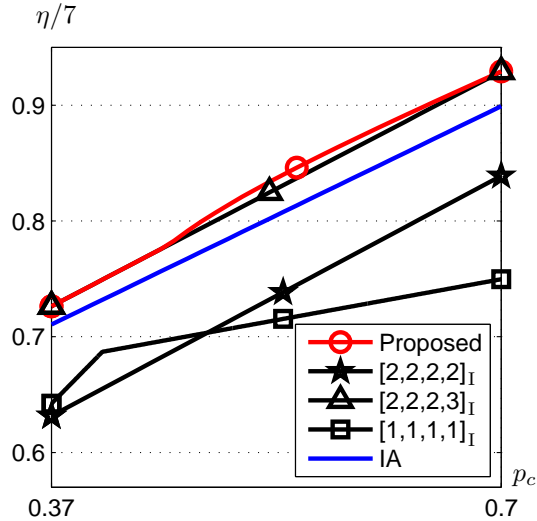

\setcounter{subsubsection}{0}
\subsubsection{Extending the HKIA scheme for arbitrary $N$}

Not surprisingly, when $M/N>2/3$ and $N$ is not a multiple of 3, the HKIA scheme for the Type I channel is not reciprocal to that for the Type II channel as described in Appendix \ref{App_D}, due to the non-reciprocity of the HKIA scheme as discussed in Section \ref{Disc_A}.  For example, the $[1,1,1,1]_\mathrm{I}, [2,2,2,3]_\mathrm{I},$ and $[2,2,2,2]_\mathrm{I}$ HKIA DoF lower bound on the $6 \times 7$ channel are illustrated in Fig. \ref{fig:HKIA_LB_6x7}.  It is evident in the figure that the $[2,2,2,3]_\mathrm{I}$ HKIA, which incorporates usual INBF scheme for the non-bursty channel \cite{MMK08}, delivers a strictly higher DoF than the other two HKIA schemes, again signifying the fact that introducing public messages on the Type I channel is not as effective as on the Type II channel.  In this case, maximizing the dimensions used for orthogonal private message beamforming is more beneficial.  Therefore we derive our lower bound by taking the $[2,2,2,3]_\mathrm{I}$ HKIA scheme and fine-tuning the power of the private INBF signals as was done in Section \ref{Thm3_B}.  (No blending of HKIA schemes.)  This proposed lower bound is also illustrated in Fig. \ref{fig:HKIA_LB_6x7}.

For the general Type I channel with $N=3k+q$ where $k$ is an integer and $q\in\{0,1, 2\}$, the proposed HKIA scheme employs the $[k,k,k,k]_\mathrm{I}$, the $[k,k,k,(k+1)]_\mathrm{I}$ or the $[(k+1),k,k,(k+1)]_\mathrm{I}$ INBF scheme, which is the usual INBF scheme for the non-bursty channel, when $q=0, 1$ or $2$, respectively.  The decoding procedure and the analysis of the the achievable DoF again follow the lines in Section \ref{Achi_D} verbatim.  We include it below for the sake of completeness.  Note that it takes more effort to compute the achievable DoF on the Type I channel, as the form of the INBF scheme varies with $q$ and can be asymmetric.  We distinguish three cases.  

\subsubsection{Proof of Proposition \ref{prop:HKIA_LB_1} when $q=0$}

In this case, we have the $\big[\frac{N}{3},\frac{N}{3},\frac{N}{3},\frac{N}{3}\big]_\mathrm{I}$ HKIA shown in Section \ref{Thm3_B_2}.  (Note that $N/3=k$ when $q=0$.)  So the following private DoF is clearly achievable:
\begin{equation} \label{eq:E_15}
\eta_\mathrm{priv} \le 2ak(p_c+p_d) = ak\ET{4}{2}{2}.
\end{equation}
As for the public DoF, with the new distributions of random variables, (\ref{eq:6_3}) can be verified to be
\begin{align*} 
\mathcal D\{h(Y_1\mid S_1)\}				&=\ET{N}{M}{M}, \\
\mathcal D\{h(Y_1\mid S_1, U_1)\}		&=\ET{M+a(N-M)}{2ak}{M}, \\
\mathcal D\{h(Y_1\mid S_1, U_2)\}		&=\ET{M+a(N-M)}{M}{2ak}, \textrm{ and}\\
\mathcal D\{h(Y_1\mid S_1, U_1, U_2)\}	&=\ET{aN}{2ak}{2ak}.  \mpelabel{eq:7_14}
\end{align*}
$\mathcal D\{h(Y_2\mid S_2)\}, \mathcal D\{h(Y_2\mid S_2, U_2)\}, \mathcal D\{h(Y_2\mid S_2, U_1)\},$ and $\mathcal D\{h(Y_2\mid S_2, U_1, U_2)\}$ are obtained immediately by symmetry.  Hence, by straightforward computation, the achievable public DoF now becomes
\begin{equation} \label{eq:E_16}
\eta_\mathrm{pub} \le \min\left(2\left[\ET{M}{0}{M} - a\ET{M}{0}{2k}\right], \quad
\ET{N}{M}{M} - ak\ET{3}{2}{2} \right),
\end{equation}
and the total achievable DoF is
\begin{equation} \label{eq:E_17}
\eta_\mathrm{total} \le \min\left(2\left[\ET{M}{0}{M} - a\ET{M-2k}{-k}{k} \right], \quad
\ET{N}{M}{M} + ak\ET{1}{0}{0} \right).
\end{equation}

Now the remaining task is to optimize `$a$' so that $\eta_\mathrm{total}$ is maximized.  We distinguish two cases:  If $\big\langle M-2k, -k, k \big\rangle \le 0$, then clearly we should maximize `$a$', i.e. $a_\mathrm{opt}=1$.  Otherwise, the optimal `$a$' is obtained by equating $2\left[\ET{M}{0}{M} - a\ET{M-2k}{-k}{k}\right]$ and $\ET{N}{M}{M} + ak\ET{1}{0}{0}$.  Therefore, noting that $\big\langle M-2k, -k, k \big\rangle \le 0$ is equivalent to $p_c/p_d \le 1/\left(1+\frac{M-2k}{k}p_{d\mid c}\right)$, we have
\begin{equation} \label{eq:E_18}
a_\mathrm{opt} = \left\{\begin{aligned}[rl] 
1, & \quad \frac{p_c}{p_d} \le \frac{1}{1+\left(\frac{M-2k}{k}\right)p_{d\mid c}} \\ 
\frac{\ET{2M-N}{-M}{M}}{\ET{2M-N}{-2k}{2k}}, & \quad \frac{p_c}{p_d} > \frac{1}{1+\left(\frac{M-2k}{k}\right)p_{d\mid c}} 
\end{aligned} \right.
\end{equation}
Substituting (\ref{eq:E_18}) into (\ref{eq:E_17}) and simplifying the expressions, we then conclude that the following total DoF is achievable:
\begin{equation}\label{eq:E_19}
\eta_\mathrm{total} \le \left\{ \begin{aligned}[rl]
2\ET{2k}{k}{M-k},  & \quad \frac{p_c}{p_d} \le  \frac 1{1+\left(\frac{M-2k}{k}\right)p_{d|c}}\\
\ET{N}{M}{M}+\ET{k}{0}{0} \frac{\ET{2M-N}{-M}{M}}{\ET{2M-N}{-2k}{2k}},  & \quad \frac{p_c}{p_d} > \frac 1{1+\left(\frac{M-2k}{k}\right)p_{d|c}}. \\
\end{aligned} \right.
\end{equation}

\subsubsection{Proof of Proposition \ref{prop:HKIA_LB_1} when $q=1$}

When $q=1$, the usual $[k,k,k,(k+1)]_\mathrm{I}$ INBF for the non-bursty channel \cite{MMK08} is used in the HKIA, so the $D_1, C_1, C_2,$ and $D_2$ random vectors in (\ref{eq:7_11}) and (\ref{eq:7_12}) now have the following distribution
\begin{align} \begin{split} \label{eq:E_1}
D_1, C_1, C_2  &\sim \mathcal{N}\left(0,  \frac{P^a}{4k}I_{k}\right) \textrm{, and} \\
D_2 &\sim \mathcal{N}\left(0,  \frac{P^a}{4(k+1)}I_{k+1}\right), 
\end{split} \end{align}
where $a\in [0,1]$.  The achievable private DoF is hence easily seen to be
\begin{equation} \label{eq:E_2}
\eta_\mathrm{priv} \le a\ET{4k+1}{2k+1}{2k}.
\end{equation}

The determination of the achievable public DoF takes a little more effort due to the asymmetry between $D_1$ and $D_2$, but it still parallels the steps in Section \ref{Thm3_B}.  So again the following differential entropies of $Y_1$ are evaluated:
\begin{align} \begin{split} \label{eq:E_3}
\mathcal D\{h(Y_1\mid S_1)\}				&=\ET{N}{M}{M}, \\
\mathcal D\{h(Y_1\mid S_1, U_1)\}			&=\ET{M+(N-M)a}{2ka}{M}, \\
\mathcal D\{h(Y_1\mid S_1, U_2)\}			&=\ET{M+(N-M)a}{M}{(2k+1)a}, \textrm{ and}\\
\mathcal D\{h(Y_1\mid S_1, U_1, U_2)\}	&=\ET{Na}{2ka}{(2k+1)a}.
\end{split} \end{align}
The differential entropies of $Y_2$ is not symmetric to those of $Y_1$ but can be verified to be
\begin{align*}
\mathcal D\{h(Y_2\mid S_2)\}				&=\ET{N}{M}{M}, \\
\mathcal D\{h(Y_2\mid S_2, U_2)\}			&=\ET{M+(N-M)a}{(2k+1)a}{M}, \\
\mathcal D\{h(Y_2\mid S_2, U_1)\}			&=\ET{M+(N-M)a}{M}{2ka}, \textrm{ and}\\
\mathcal D\{h(Y_2\mid S_2, U_1, U_2)\}	&=\ET{Na}{(2k+1)a}{2ka}.  \mpelabel{eq:E_4}
\end{align*}
From (\ref{eq:E_3}) and (\ref{eq:E_4}), it is straightforward to verify that the achievable public DoF is
\begin{equation} \label{eq:E_5}
\eta_\mathrm{pub} \le \min\Big( \ET{2M}{0}{2M} - a\ET{2M}{0}{4k+1}, \quad 
\ET{N}{M}{M} -a\ET{N}{2k+1}{2k} \Big),
\end{equation}

It follows from (\ref{eq:E_2}) and (\ref{eq:E_5}) that the total achievable DoF is
\begin{equation} \label{eq:E_6}
\eta_\mathrm{total} \le \min\Big( \ET{2M}{0}{2M} -  a\ET{2M-4k-1}{-2k-1}{2k+1},  \quad
\ET{N}{M}{M} + a\ET{k}{0}{0} \Big).
\end{equation}
The optimal value of `$a$' can then be shown to be
\begin{equation} \label{eq:E_7}
a_\mathrm{opt} = \left\{\begin{aligned}[rl] 
1, & \quad \frac{p_c}{p_d} \le \frac{1}{1+\left(\frac{2M-4k-1}{2k+1}\right)p_{d\mid c}}, \\
\frac{\ET{2M-N}{-M}{M}}{\ET{2M-N}{-2k-1}{2k+1}}, & \quad \frac{p_c}{p_d} > \frac{1}{1+\left(\frac{2M-4k-1}{2k+1}\right)p_{d\mid c}},
\end{aligned} \right.
\end{equation}
which leads to the following achievable DoF:
\begin{equation} \label{eq:E_8}
\eta_\mathrm{total} \le \left\{ \begin{aligned}[rl]
\ET{4k+1}{2k+1}{2M-2k-1}, & \quad \frac{p_c}{p_d} \le \frac{1}{1+\left(\frac{2M-4k-1}{2k+1}\right)p_{d\mid c}}, \\
\ET{N}{M}{M}+\ET{k}{0}{0}\frac{\ET{2M-N}{-M}{M}}{\ET{2M-N}{-2k-1}{2k+1}}, & \quad \frac{p_c}{p_d} > \frac{1}{1+\left(\frac{2M-4k-1}{2k+1}\right)p_{d\mid c}}.
\end{aligned} \right.
\end{equation}

\subsubsection{Proof of Proposition \ref{prop:HKIA_LB_1} when $q=2$}

When $q=2$, the HKIA scheme incorporates the usual $[(k+1),k,k,(k+1)]_\mathrm{I}$ INBF for the non-bursty channel \cite{MMK08}.  In this case, $D_1, C_1, C_2,$ and $D_2$ random vectors in (\ref{eq:7_11}) and (\ref{eq:7_12}) now take the following distribution
\begin{align} \begin{split} \label{eq:E_1}
C_1, C_2  &\sim \mathcal{N}\left(0,  \frac{P^a}{4k}I_{k}\right) \textrm{, and} \\
D_1, D_2 &\sim \mathcal{N}\left(0,  \frac{P^a}{4(k+1)}I_{k+1}\right), 
\end{split} \end{align}
where $a\in [0,1]$.  The achievable private DoF is hence 
\begin{equation} \label{eq:E_9}
\eta_\mathrm{priv} \le 2a\ET{2k+1}{k+1}{k}.
\end{equation}
To evaluate the achievable public DoF, we first compute
\begin{align} \begin{split} \label{eq:E_10}
\mathcal D\{h(Y_1\mid S_1)\}				&=\ET{N}{M}{M}, \\
\mathcal D\{h(Y_1\mid S_1, U_1)\}			&=\ET{M+(N-M)a}{(2k+1)a}{M}, \\
\mathcal D\{h(Y_1\mid S_1, U_2)\}			&=\ET{M+(N-M)a}{M}{(2k+1)a}, \textrm{ and}\\
\mathcal D\{h(Y_1\mid S_1, U_1, U_2)\}	&=\ET{Na}{(2k+1)a}{(2k+1)a}.
\end{split} \end{align}
The differential entropies of $Y_2$ is obtained readily due to symmetry.  Therefore, the achievable public DoF is 
\begin{equation} \label{eq:E_11}
\eta_\mathrm{pub} \le \min\Big( 2\ET{M}{0}{M} - 2a\ET{M}{0}{2k+1}, \quad 
\ET{N}{M}{M} -a\ET{N}{2k+1}{2k+1} \Big),
\end{equation}
and the total achievable DoF is
\begin{equation} \label{eq:E_12}
\eta_\mathrm{total} \le \min\Big( 2\ET{M}{0}{M} - 2a\ET{M-2k-1}{-k-1}{k+1}, \quad 
\ET{N}{M}{M} + a\ET{k}{1}{-1} \Big).
\end{equation}

Straightforward optimization of `$a$' then leads to
\begin{equation} \label{eq:E_13}
a_\mathrm{opt} = \left\{\begin{aligned}[rl] 
1, & \quad \frac{p_c}{p_d} \le \frac{1}{1+\left(\frac{2M-4k-2}{2k+2}\right)p_{d\mid c}}, \\
\frac{\ET{2M-N}{-M}{M}}{\ET{2M-N}{-2k-1}{2k+1}}, & \quad \frac{p_c}{p_d} > \frac{1}{1+\left(\frac{2M-4k-2}{2k+2}\right)p_{d\mid c}},
\end{aligned} \right.
\end{equation}
and the achievable DoF is
\begin{equation} \label{eq:E_14}
\eta_\mathrm{total} \le \left\{ \begin{aligned}[rl]
\ET{4k+2}{2k+2}{2M-2k-2}, & \quad \frac{p_c}{p_d} \le \frac{1}{1+\left(\frac{2M-4k-2}{2k+2}\right)p_{d\mid c}}, \\
\ET{N}{M}{M}+\ET{k}{1}{-1}\frac{\ET{2M-N}{-M}{M}}{\ET{2M-N}{-2k-1}{2k+1}}, & \quad \frac{p_c}{p_d} > \frac{1}{1+\left(\frac{2M-4k-2}{2k+2}\right)p_{d\mid c}}.
\end{aligned} \right.
\end{equation}
Finally, noting that (\ref{eq:E_18}), (\ref{eq:E_19}), (\ref{eq:E_7}), (\ref{eq:E_8}), (\ref{eq:E_13}) and (\ref{eq:E_14}) are equivalent to Proposition \ref{prop:HKIA_LB_1}, we have completed its proof.

%\section*{Acknowledgment}

\end{document}